\documentclass[aps,pre,reprint,superscriptaddress,longbibliography,floatfix]{revtex4-2}

\usepackage[T1]{fontenc}
\usepackage[utf8]{inputenc}
\usepackage{lmodern}
\usepackage{amsmath,amssymb,bm,mathtools}
\usepackage{graphicx}
\usepackage{booktabs}
\usepackage{microtype}
\usepackage{xcolor}
\usepackage[colorlinks=true,allcolors=blue!55!black]{hyperref}
\hypersetup{
pdftitle={Dataset Complexity Shapes Finite-Distance Loss Geometry in Neural Networks}, pdfauthor={Jaeyong Bae and Hawoong Jeong}, pdfsubject={Dataset-dependent local entropy around neural-network solutions}, pdfkeywords={neural networks, statistical physics, dataset complexity, local entropy, sequential Monte Carlo}
}

\graphicspath{{figures/}}

\newcommand{\dd}{\,\mathrm d}
\newcommand{\E}{\mathbb E}
\newcommand{\R}{\mathbb R}
\newcommand{\1}{\mathbf 1}
\newcommand{\CMS}{\mathcal C_{\mathrm{MS}}}

\begin{document}

\title{Dataset Complexity Shapes Finite-Distance Loss Geometry in Neural Networks}

\author{Jaeyong Bae}
\affiliation{Department of Physics, Korea Advanced Institute of Science and Technology, Daejeon, Korea}
\author{Hawoong Jeong}
\affiliation{Department of Physics, Korea Advanced Institute of Science and Technology, Daejeon, Korea}
\affiliation{Center of Complex Systems, Korea Advanced Institute of Science and Technology, Daejeon, Korea}

\begin{abstract}
Finite datasets can share the same size and low-order statistics while differing strongly in structural complexity. We connect this dataset complexity to loss-landscape geometry by pairing local label mixing across neighborhood scales with local entropy around trained neural-network solutions. Adapted from the Franz--Parisi construction in spin-glass theory, local entropy measures the effective volume of low-loss, solution-like parameter configurations at each distance from a reference. We estimate it in finite networks using adaptive sequential Monte Carlo. In a controlled synthetic sweep, greater dataset complexity produces a larger decrease in local entropy near the reference. Farther away, its radial derivative becomes weak and nearly common across conditions. Dataset complexity therefore changes where the effective solution volume contracts, rather than making it decrease uniformly faster. Experiments on real image data show the same qualitative trend, with label randomization further amplifying the effect. These results show that dataset structure shapes how low-loss neighborhoods are organized across finite distances from trained solutions.
\end{abstract}

\maketitle

\section{Introduction}
\label{sec:introduction}

Understanding how collective behavior emerges from many interacting components is a central goal across many fields of science.  Artificial neural networks can also be viewed from this perspective.  They are complex systems built from many simple elements, and their behavior emerges from how those elements interact during learning.

Despite their remarkable success, the principles governing how neural networks learn and generalize remain only partly understood.  Many theoretical studies therefore begin with simplified models that preserve the ingredients believed to be essential.  Statistical physics provides a natural framework for this approach by describing learning through collective variables and the geometry of high-dimensional spaces.

Neural-network learning is shaped jointly by the training data, the architecture and its learnable parameters, and the loss that guides their evolution.  During training, the dataset remains fixed while the parameters change through optimization.  In statistical-mechanical language, the data can therefore be treated as quenched disorder and the parameters as degrees of freedom coupled through a common energy-like function.  This viewpoint has been used to study learning transitions, generalization, and the organization of solutions in perceptrons and other high-dimensional models \cite{Gardner1988SpaceInteractions,Seung1992StatisticalMechanics,Bahri2020StatisticalMechanicsDeepLearning}.

Much of the analytical progress based on this viewpoint relies on simplified input ensembles.  Gaussian inputs and teacher--student constructions often make disorder averages tractable and reduce the learning problem to a small number of macroscopic variables.  Hidden-manifold, random-feature, and Gaussian-mixture models extend this framework to structured settings \cite{Goldt2020HiddenManifold,Gerace2020RandomFeaturesHiddenManifold,Mignacco2020DMFTGaussianMixtureSGD,Loureiro2022GenericFeatureMapsTeacherStudent}. Under suitable conditions, universality results further show that some non-Gaussian or standardized Gaussian-mixture inputs behave similarly to Gaussian inputs \cite{Bae2026GaussianUniversality,Gerace2024GaussianUniversality}. These results help identify which aspects of learning are mainly controlled by low-order statistics of the input distribution.

An actual finite dataset, however, can contain structure that is not fully described by these statistics.  Two datasets may have the same number of samples, the same class balance, and similar input means and covariances, while their labels are arranged in very different ways.  Different labels may be strongly mixed among nearby samples or form large, well-separated regions. The resulting learning problems may therefore differ even when the dataset size and low-order input statistics are similar.  The organization of the dataset should thus be considered together with its size and distribution \cite{HoBasu2002ComplexityMeasures,Lorena2019ClassificationComplexitySurvey}.

In many theoretical descriptions, the difficulty of a learning problem is summarized by model size or by a load such as the number of examples per trainable parameter.  These quantities are essential, but they do not say how the labels are arranged within a realized dataset.  Here we use \emph{dataset complexity} for this second notion of how strongly examples with different labels are interwoven across local scales.  It can change while the sample count, parameter count, and class balance remain fixed.

In this work, we ask whether a more complex organization of the dataset is reflected in the local loss landscape of the neural network.  On the data side, we quantify label mixing over several neighborhood scales.  The resulting statistic is normalized against random label mixing and serves as our operational measure of dataset complexity.  It is small when nearby samples usually share a label and increases as different labels become locally interwoven.

On the parameter side, the loss at a single trained solution does not describe how other good configurations are organized around it, so a neighborhood-scale picture of the loss landscape is important.  Hessian spectra provide a one-point probe of local curvature, while associated notions of sharpness can depend on parameterization \cite{Sagun2017HessianOverparam,Dinh2017SharpMinima}.  By construction, such one-point information does not describe how the geometry changes over finite displacements.

Finite-distance studies partly address this limitation by examining the landscape along selected paths or low-dimensional structures, revealing mode connectivity and other nonlocal organization \cite{Garipov2018LossSurfaces,FortJastrzebski2019LargeScale,Dold2025PathsAmbientSpaces}. Because these constructions are conditioned on particular paths or subspaces, however, they do not by themselves characterize the broader neighborhood around a reference.

Franz--Parisi potentials and local entropy take a complementary view by characterizing the abundance of low-loss configurations around a reference at fixed overlap or within a finite neighborhood, thereby summarizing more than a single prescribed path \cite{Franz1995RecipesMetastable,Baldassi2016LocalEntropy,Baldassi2020ShapingLandscape,Musso2021PartialLocalEntropy,Baldassi2021UnveilingStructure,Baldassi2023TypicalAtypical,Chaudhari2017EntropySGD}. Complementary empirical and computational studies have examined local-to-global loss-landscape structure, the global density of states, and finite local-volume estimates around trained solutions \cite{Yang2021TaxonomizingLandscape,Mele2025DensityStates,Huang2020GeneralizationVisualizations,ScherlisBelrose2025LocalVolume}. Here we ask whether dataset complexity shapes local entropy around a trained reference.

\begin{figure}[t]
  \centering
  \includegraphics[width=\columnwidth]
    {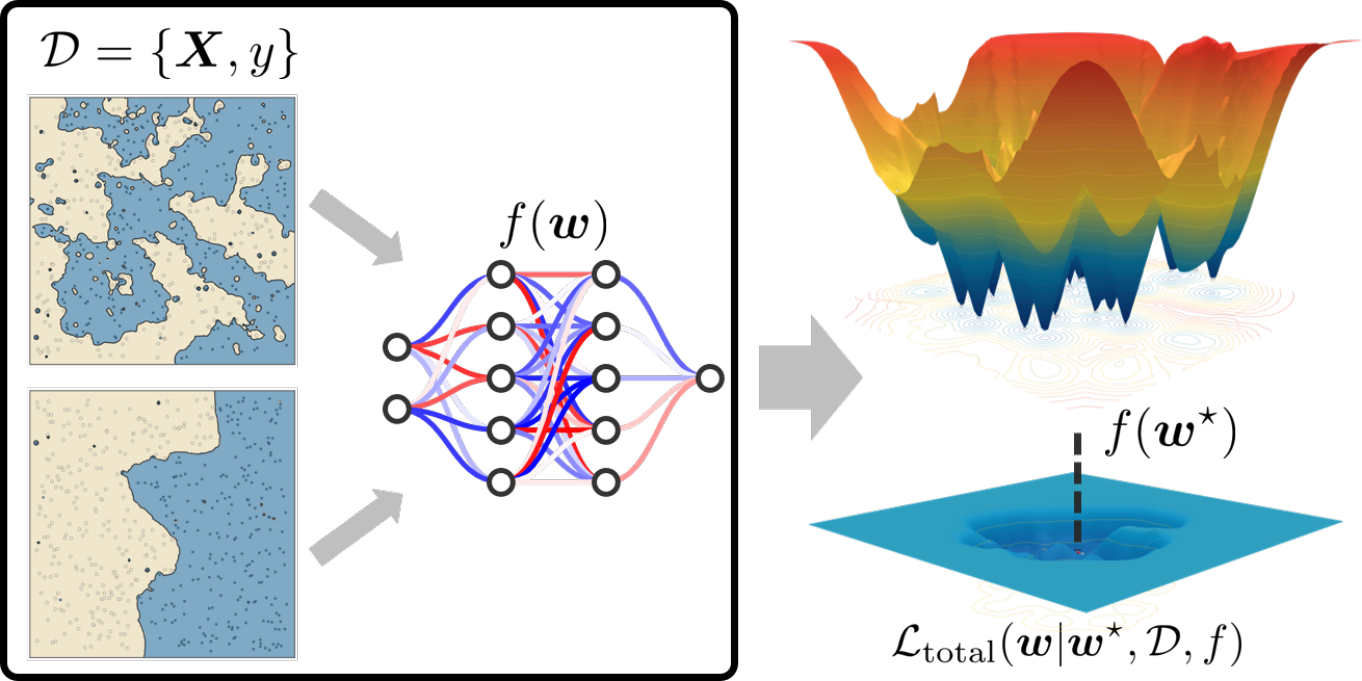}
\caption{ Schematic overview of our study.  We examine how dataset complexity affects finite-distance loss geometry around trained neural-network solutions. }
  \label{fig:dataset-to-landscape-schematic}
\end{figure}

We connect these two sides by comparing dataset complexity with local entropy centered on a trained reference.  We use local entropy and its radial derivative to examine how this effective solution volume changes with distance. Because this quantity is difficult to evaluate beyond analytically tractable models, we introduce an importance-sampling method for finite empirical neural networks.  Fig.~\ref{fig:dataset-to-landscape-schematic} provides a schematic overview of our study, in which we examine how finite-distance loss geometry changes as dataset complexity varies.

Across controlled synthetic data and MNIST experiments, stronger local mixing of the labels is accompanied by a stronger change in the surrounding local entropy.  This change is not uniform over distance.  The radial derivative first becomes more negative and then returns toward zero, showing that the low-loss volume contracts most strongly over an inner distance range. Low-loss configurations therefore reorganize differently across distance ranges around the trained reference.  Because this contraction is concentrated within a finite distance range, a single local notion of flatness cannot summarize it.

The Methods introduce the dataset-complexity measure, the Franz--Parisi-inspired local entropy, and the sampling method used to estimate it.  In the Results, we first compare the estimator with a replica-symmetric calculation for a Gaussian perceptron.  We then examine how dataset complexity is reflected in local entropy across controlled synthetic examples and MNIST.

\section{Methods}
\label{sec:methods}
\subsection{Experimental systems and learning models}
\label{sec:experimental-systems}

We examine three experimental settings that range from controlled synthetic patterns to structured image data.  A two-dimensional synthetic family allows the label arrangement to be varied systematically and visualized directly. Two complementary MNIST settings test whether the link between dataset structure and local entropy persists in higher-dimensional, naturally structured data.  The observables in data and parameter space are introduced in the following subsection.

\subsubsection{Synthetic dataset}
\label{subsec:synthetic-system}

We first construct a two-dimensional synthetic family in which the spatial organization of binary labels can be controlled directly.  For each dataset, \(n=512\) points are drawn uniformly from \([-1,1]^2\) and connected by a mutual ten-nearest-neighbor graph \(\mathcal E\).  The labels are initialized with equal numbers of \(+1\) and \(-1\).

To reorganize the labels without changing the class balance, we use Kawasaki exchange dynamics with the ferromagnetic energy
\begin{equation}
  E_{\mathrm{Ising}}(\boldsymbol y)
  =
  -\sum_{(i,j)\in\mathcal E}y_i y_j .
\end{equation}
At each update, two opposite labels are exchanged with probability \(\min\{1,\exp[-\beta_{\mathrm{data}}\Delta E_{\mathrm{Ising}}]\}\). Small values of \(\beta_{\mathrm{data}}\) produce locally interwoven labels, whereas larger values favor extended same-label domains.  We consider 18 values \(\beta_{\mathrm{data}}=0.05,0.07,\ldots,0.39\), generate 60 independent datasets for each value, and retain the label configurations after 2000 sweeps.

\subsubsection{MNIST dataset constructions}
\label{subsec:mnist-systems}

To examine whether the same behavior persists in structured image data, we also perform experiments on MNIST.  The original \(28\times28\) images are reduced to \(10\times10\) pixels by box averaging, and each input coordinate is standardized using the corresponding training set.  Each binary task contains 512 training images and is evaluated on ten dataset realizations.

Unlike the synthetic system, MNIST has no direct control parameter that continuously changes the organization of its labels.  We therefore consider two complementary constructions.  The first starts from a balanced odd--even classification task and flips each label with probability \(\eta\).  We use \[ \eta\in\{0,0.05,0.15,0.25,0.5\}. \] The case \(\eta=0\) preserves the original odd--even labels, while \(\eta=0.5\) removes their relation to the images in expectation.  Within each paired dataset, the image coordinates are shared across all values of \(\eta\).  A common set of uniform random variates makes the flipped labels nested as \(\eta\) increases.

The second construction considers binary classification between individual digit pairs.  Some pairs are clearly separated in image space, whereas others share similar shapes and are more strongly interwoven.  For every one of the 45 pairs, we calculated the multiscale complexity \(\CMS\), defined in the next subsection, on ten balanced datasets per pair.  We ranked the pairs by their mean \(\CMS\) across these datasets and fixed the 12 tasks at ranks \(1,5,9,\ldots,45\).  For each selected task, we then calculated local entropy on those datasets, using ten trained references per task and dataset.  Thus the task selection and landscape comparison use the same datasets. Unlike the label-noise construction, differences among the digit-pair tasks arise from the natural structure of the digit classes rather than from artificial label perturbations.

\subsubsection{Network models and regularized loss}
\label{subsec:learning-models}

For both the synthetic and MNIST experiments, we use a fully connected feed-forward network with two hidden layers and a scalar output.  For an input \(\boldsymbol x\in\mathbb R^d\), the network is written as
\begin{align}
  \boldsymbol h^{(1)}
  &=
  \tanh\!\left(
    W^{(1)}\boldsymbol x+\boldsymbol b^{(1)}
  \right),\\
  \boldsymbol h^{(2)}
  &=
  \tanh\!\left(
    W^{(2)}\boldsymbol h^{(1)}+\boldsymbol b^{(2)}
  \right),\\
  f_{\boldsymbol w}(\boldsymbol x)
  &=
  \boldsymbol a^{\mathsf T}\boldsymbol h^{(2)}+b .
  \label{eq:network-definition}
\end{align}
Here \(\boldsymbol w\) denotes the collection of all weights and biases, while \(\boldsymbol a\) is the output-layer weight vector.  The scalar \(f_{\boldsymbol w}(\boldsymbol x)\) is a signed classification score, and the predicted label is \(\operatorname{sgn}[f_{\boldsymbol w}(\boldsymbol x)]\). For the two-dimensional synthetic data, both hidden layers contain 48 units. For the downsampled MNIST data, both contain 20 units.  The synthetic and MNIST models have \(P=2545\) and \(P=2461\) trainable parameters, respectively.

For a binary dataset \(\mathcal D=\{(\boldsymbol x_i,y_i)\}_{i=1}^{n}\), with \(y_i\in\{-1,+1\}\), the prediction-dependent part of the loss is the mean logistic loss,
\begin{equation}
  \overline L_{\mathrm{CE}}(\boldsymbol w;\mathcal D)
  =
  \frac{1}{n}\sum_{i=1}^{n}
  \log\!\left[
    1+\exp\!\left(
      -y_i f_{\boldsymbol w}(\boldsymbol x_i)
    \right)
  \right].
  \label{eq:mean-logistic-loss}
\end{equation}
This loss depends on the signed margin \(y_i f_{\boldsymbol w}(\boldsymbol x_i)\), becoming small when a sample is classified correctly with a large margin.

The regularized learning loss is
\begin{equation}
  \mathcal L_{\mathrm{reg}}(\boldsymbol w;\mathcal D)
  =
  \gamma\,\overline L_{\mathrm{CE}}(\boldsymbol w;\mathcal D)
  +
  \lambda\frac{\|\boldsymbol w\|_2^2}{2P}.
  \label{eq:dnn-shell-energy}
\end{equation}
Dividing by \(P\) turns \(\|\boldsymbol w\|_2^2\) into the mean squared parameter magnitude, up to the factor \(1/2\), and keeps the regularization on a comparable scale across network sizes.  We use \(\gamma=1\) and \(\lambda=0.01\) for both network families.

\subsection{Dataset complexity and local entropy}
\label{sec:operational-observables}

Our goal is to examine whether changes in dataset structure are reflected in the loss landscape.  The experimental systems above allow us to vary or compare the organization of their labels.  We now define an observable for this organization and a reference-conditioned local entropy in parameter space.

On the data side, motivated by the view that harder classification problems tend to involve more intricate, potentially glassy decision boundaries, we measure local label mixing across neighborhood scales.
On the parameter side, we measure the effective solution volume at each
distance from a trained reference using local entropy adapted from the
Franz--Parisi construction.

\subsubsection{Multiscale dataset complexity}
\label{subsec:dataset-complexity}

Let \(\mathcal D=\{(\boldsymbol x_i,y_i)\}_{i=1}^{n}\), with \(y_i\in\{-1,+1\}\), and let \(N_j(i)\) be the \(j\)th Euclidean nearest neighbor of sample \(i\) among the other standardized training inputs.  For a neighborhood containing the first \(k\) neighbors, we define the directed label-mismatch fraction
\begin{equation}
  C_k(\mathcal D)
  =
  \frac{1}{nk}
  \sum_{i=1}^{n}
  \sum_{j=1}^{k}
  \mathbf 1
  \!\left[
    y_i\ne y_{N_j(i)}
  \right].
  \label{eq:knn-label-mismatch}
\end{equation}
The quantity \(C_k\) is small when neighboring samples tend to share a label and increases as different labels become locally interwoven.

Equivalently, with \(A_{ij}^{(k)}=\mathbf 1\{j\in\{N_1(i),\ldots,N_k(i)\}\}\),
\begin{equation}
  \sum_{i,j}
  A_{ij}^{(k)}
  |y_i-y_j|
  =
  2nk C_k(\mathcal D).
  \label{eq:cnn-gtv}
\end{equation}
Taken together, the nearest-neighbor construction and this edgewise identity place \(C_k\) at the intersection of neighborhood- and multiscale classification-complexity measures and graph total-variation formulations \cite{Smith2014InstanceComplexity,Singh2003MultiresolutionComplexity,Acena2026DynamicDisagreeing,Shuman2013EmergingGraphSignal,Ortega2018GraphSignalProcessing}. We also normalize by the mismatch expected under random label mixing at the realized class counts.  If class \(c\) occurs \(n_c\) times, this probability is
\begin{equation}
  q_{\mathcal D}
  = \sum_c\frac{n_c}{n}
  \frac{(n-n_c)
  }{
    (n-1)
  }.
  \label{eq:chance-label-mismatch}
\end{equation}
Our dataset-complexity measure is
\begin{equation}
  \mathcal C_{\mathrm{MS}}(\mathcal D)
  =
  \frac{1}{k_{\max}q_{\mathcal D}}
  \sum_{k=1}^{k_{\max}}
  C_k(\mathcal D),
  \qquad
  k_{\max}
  =
  \left\lfloor
    \sqrt{n-1}
  \right\rfloor .
  \label{eq:multiscale-dataset-complexity}
\end{equation}
We choose \(k_{\max}=\lfloor\sqrt{n-1}\rfloor\), so the widest neighborhood grows sublinearly with \(n\).  Here \(n-1\) is the number of possible neighbors of each sample.  All datasets have \(n=512\), and hence \(k_{\max}=22\).  Larger \(\CMS\) indicates stronger local label mixing. Random mixing at the same class counts gives \(\CMS=1\) in expectation.

\subsubsection{Local entropy as a measure of loss-landscape complexity}
\label{subsec:hard-soft-shell}

The loss at one trained parameter vector says little about how many other good parameter settings lie nearby.  Directly mapping this high-dimensional neighborhood is generally impractical.  We therefore use local entropy to measure the effective solution volume at a fixed distance from the reference. We first state its strict zero-error form and then the continuous form used for finite networks.

We denote a retained trained reference by \(\widetilde{\boldsymbol w}\).  We retain references with zero training classification error---often called interpolating solutions---so that
\begin{equation}
  y_i
  f_{\widetilde{\boldsymbol w}}(\boldsymbol x_i)
  >
  0
  \qquad
  \text{for every }i.
  \label{eq:exact-reference-condition}
\end{equation}

Using the Heaviside function \(\Theta(z)=1\) for \(z>0\) and zero otherwise, we define the interpolation indicator
\begin{equation}
  \mathcal I_{\mathrm{int}}(\boldsymbol w)
  =
  \prod_{i=1}^{n}
  \Theta\!\left[
    y_i f_{\boldsymbol w}(\boldsymbol x_i)
  \right].
\end{equation}
At the normalized distance
\begin{equation}
  r
  =
  \frac{
    \|
      \boldsymbol w-\widetilde{\boldsymbol w}
    \|_2
  }{
    \sqrt P
  },
\end{equation}
we use the radial constraint
\begin{equation}
  \delta_r
  (\boldsymbol w\mid\widetilde{\boldsymbol w})
  \equiv
  \delta\!\left(
    Pr^2
    -
    \|
      \boldsymbol w-\widetilde{\boldsymbol w}
    \|_2^2
  \right).
  \label{eq:radial-shell-constraint}
\end{equation}
The corresponding hard local entropy is
\begin{equation}
  \phi_{\mathrm h}
  (r\mid\widetilde{\boldsymbol w},\mathcal D)
  =
  \frac{1}{P}
  \log
  \int_{\mathbb R^P}
  \mathrm d^P w\,
  \delta_r
  (\boldsymbol w\mid\widetilde{\boldsymbol w})
  \mathcal I_{\mathrm{int}}(\boldsymbol w).
  \label{eq:hard-local-entropy}
\end{equation}
High-dimensional solution volumes can grow or shrink exponentially with \(P\).  Dividing their logarithm by \(P\) gives an entropy density that remains on an interpretable scale as the number of trainable parameters changes.

The local entropy depends on the dataset and on the chosen reference.  Let \(p_{\mathrm{ref}}(\widetilde{\boldsymbol w}\mid\mathcal D)\) be the normalized density of retained references for a fixed dataset.  We use the shorthand
\begin{equation}
  \mathbb E_{\widetilde{\boldsymbol w}}
  \left[
    B(\widetilde{\boldsymbol w},\mathcal D)
  \right]
  \equiv
  \int_{\mathbb R^P}\mathrm d^P\widetilde w\,
  p_{\mathrm{ref}}
  (\widetilde{\boldsymbol w}\mid\mathcal D)
  B(\widetilde{\boldsymbol w},\mathcal D).
  \label{eq:reference-expectation-shorthand}
\end{equation}
At finite \(P\), the ensemble-averaged hard local entropy is
\begin{equation}
  \Phi_{\mathrm h}^{(P)}(r)
  =
  \mathbb E_{\mathcal D}
  \mathbb E_{\widetilde{\boldsymbol w}}
  \left[
    \phi_{\mathrm h}
    (r\mid\widetilde{\boldsymbol w},\mathcal D)
  \right].
  \label{eq:averaged-hard-local-entropy}
\end{equation}
In the thermodynamic construction, \(p_{\mathrm{ref}}\) represents a prescribed ensemble of exact solutions.  In the finite neural-network experiments, it is the empirical ensemble of retained trained references satisfying Eq.~\eqref{eq:exact-reference-condition}. With the relevant intensive control parameters fixed, the thermodynamic limit defines the Franz--Parisi local entropy density \cite{Franz1995RecipesMetastable,Baldassi2016LocalEntropy},
\begin{equation}
  \Phi_{\mathrm{FP}}(r)
  =
  \lim_{P\to\infty}
  \Phi_{\mathrm h}^{(P)}(r).
  \label{eq:fp-local-entropy}
\end{equation}

The hard construction is natural for classification models that permit an analytic treatment.  In a finite neural network, a strict count of exact solutions would sharply separate otherwise similar parameter settings.  We instead let every direction contribute continuously, with lower regularized loss contributing more.  The shell inverse temperature \(\beta_{\mathrm{sh}}\) controls the strength of this preference.  We use \(\beta_{\mathrm{sh}}=100\) and the Boltzmann factor \(\exp[-\beta_{\mathrm{sh}}\mathcal L_{\mathrm{reg}}]\).  The full radial-shell partition function is
\begin{align}
  \Omega_{\beta}
  (r\mid\widetilde{\boldsymbol w},\mathcal D)
  &=
  \int_{\mathbb R^P}
  \mathrm d^P w\,
  \delta_r
  (\boldsymbol w\mid\widetilde{\boldsymbol w})
  \notag\\
  &\quad\times
  \exp\!\left[
    -\beta_{\mathrm{sh}}
    \mathcal L_{\mathrm{reg}}
    (\boldsymbol w;\mathcal D)
  \right].
  \label{eq:soft-shell-partition}
\end{align}
The ensemble average of this normalized log volume defines the soft local entropy,
\begin{equation}
  \Phi_{\beta}^{(P)}(r)
  =
  \mathbb E_{\mathcal D}
  \mathbb E_{\widetilde{\boldsymbol w}}
  \left[
    \frac{1}{P}
    \log
    \Omega_{\beta}
    (r\mid\widetilde{\boldsymbol w},\mathcal D)
  \right].
  \label{eq:soft-local-entropy}
\end{equation}

To distinguish the geometric growth of a shell from its loss-dependent organization, write
\begin{equation}
  \rho
  =
  \|
    \boldsymbol w-\widetilde{\boldsymbol w}
  \|_2,
  \qquad
  \mathrm d^P w
  =
  \rho^{P-1}
  \mathrm d\rho\,
  \mathrm d\omega ,
\end{equation}
where \(\mathrm d\omega\) is the angular surface element.  The radial constraint fixes \(\rho=\sqrt P\,r\), and its Jacobian leaves the geometric factor \(\rho^{P-2}/2\).

For any function \(B\), define the normalized fixed-shell average
\begin{equation}
  \left\langle
    B(\boldsymbol w)
  \right\rangle_{
    r,\widetilde{\boldsymbol w}
  }
  \equiv
  \frac{
    \displaystyle
    \int_{\mathbb R^P}
    \mathrm d^P w\,
    \delta_r
    (\boldsymbol w\mid\widetilde{\boldsymbol w})
    B(\boldsymbol w)
  }{
    \displaystyle
    \int_{\mathbb R^P}
    \mathrm d^P w\,
    \delta_r
    (\boldsymbol w\mid\widetilde{\boldsymbol w})
  }.
  \label{eq:exact-shell-average}
\end{equation}
This normalization leaves an average over angular directions at the fixed distance \(r\).  The partition function then factorizes exactly,
\begin{align}
  \Omega_{\beta}
  (r\mid\widetilde{\boldsymbol w},\mathcal D)
  &=
  \frac{S_{P-1}}{2}
  \left(
    \sqrt P\,r
  \right)^{P-2}
  \notag\\
  &\quad\times
  \left\langle
    \exp\!\left[
      -\beta_{\mathrm{sh}}
      \mathcal L_{\mathrm{reg}}
      (\boldsymbol w;\mathcal D)
    \right]
  \right\rangle_{
    r,\widetilde{\boldsymbol w}
  } .
  \label{eq:soft-shell-factorization}
\end{align}
Here \(S_{P-1}\) is the surface area of the unit \((P-1)\)-sphere.

The local entropy for a particular dataset and reference is
\begin{equation}
  \phi_{\mathrm E,\beta}^{(P)}
  (r\mid\widetilde{\boldsymbol w},\mathcal D)
  =
  \frac{1}{P}
  \log
  \left\langle
    \exp\!\left[
      -\beta_{\mathrm{sh}}
      \mathcal L_{\mathrm{reg}}
      (\boldsymbol w;\mathcal D)
    \right]
  \right\rangle_{
    r,\widetilde{\boldsymbol w}
  } .
  \label{eq:energetic-shell-contribution}
\end{equation}
The Boltzmann factor gives greater weight to lower-loss configurations, which often correspond to candidates with higher training accuracy.  Throughout, we therefore use effective solution volume as an intuitive description of this local-entropy quantity. Its ensemble average is
\begin{equation}
  \Phi_{\mathrm E,\beta}^{(P)}(r)
  =
  \mathbb E_{\mathcal D}
  \mathbb E_{\widetilde{\boldsymbol w}}
  \left[
    \phi_{\mathrm E,\beta}^{(P)}
    (r\mid\widetilde{\boldsymbol w},\mathcal D)
  \right].
  \label{eq:averaged-energetic-local-entropy}
\end{equation}
The geometric contribution is
\begin{equation}
  \mathcal G_S^{(P)}(r)
  =
  \frac{1}{P}
  \log\!\left[
    \frac{S_{P-1}}{2}
    \left(
      \sqrt P\,r
    \right)^{P-2}
  \right].
  \label{eq:entropic-shell-contribution}
\end{equation}
Consequently, the full radial-shell local entropy decomposes as
\begin{equation}
  \Phi_{\beta}^{(P)}(r)
  =
  \mathcal G_S^{(P)}(r)
  +
  \Phi_{\mathrm E,\beta}^{(P)}(r).
  \label{eq:local-entropy-decomposition}
\end{equation}

The geometric term depends only on \(P\) and \(r\), not on the dataset or the trained reference.  All differences among datasets at fixed architecture therefore enter through \(\phi_{\mathrm E,\beta}^{(P)}\).  In the empirical neural networks, we report the centered local entropy and its radial derivative,
\begin{align}
  \Delta\Phi_{\mathrm E,\beta}^{(P)}(r)
  &=
  \mathbb E_{\mathcal D}
  \mathbb E_{\widetilde{\boldsymbol w}}
  \left[
    \phi_{\mathrm E,\beta}^{(P)}
    (r\mid\widetilde{\boldsymbol w},\mathcal D)
    -
    \phi_{\mathrm E,\beta}^{(P)}
    (r_0\mid\widetilde{\boldsymbol w},\mathcal D)
  \right],
  \notag\\
  g_{\mathrm E,\beta}^{(P)}(r)
  &=
  \mathbb E_{\mathcal D}
  \mathbb E_{\widetilde{\boldsymbol w}}
  \left[
    \frac{\partial}{\partial r}
    \phi_{\mathrm E,\beta}^{(P)}
    (r\mid\widetilde{\boldsymbol w},\mathcal D)
  \right].
  \label{eq:energetic-shell-response}
\end{align}
For the perceptron benchmark, the finite-\(N\) estimator and the replica calculation evaluate the same full radial-shell local entropy.

\subsection{Importance sampling for local entropy}
\label{sec:path-sampler}

At a fixed radius, only a small set of directions may retain low regularized loss, so uniform sampling can miss most of the relevant parameter configurations.  The angular dependence of the \(L_2\) term can be included in the importance distribution.  We then use adaptive sequential Monte Carlo (SMC) to approximate local entropy.

\subsubsection{Importance sampling on the fixed shell}
\label{subsec:fixed-shell-importance}

For a fixed dataset \(\mathcal D\) and trained reference \(\widetilde{\boldsymbol w}\), the reference-conditioned local entropy in Eq.~\eqref{eq:energetic-shell-contribution} is
\begin{equation}
  \phi_{\mathrm E,\beta}^{(P)}
  (r\mid\widetilde{\boldsymbol w},\mathcal D)
  =
  \frac{1}{P}
  \log
  \left\langle
    e^{
      -\beta_{\mathrm{sh}}
      \mathcal L_{\mathrm{reg}}(\boldsymbol w;\mathcal D)
    }
  \right\rangle_{r,\widetilde{\boldsymbol w}} .
  \label{eq:conditional-energetic-contribution}
\end{equation}
Here the bracket denotes the normalized angular average at fixed radius.

At a fixed normalized radius \(r\), a point on the parameter-space shell is written as
\begin{equation}
  \boldsymbol w_r(\boldsymbol u)
  =
  \widetilde{\boldsymbol w}
  +
  \sqrt P\,r\,\boldsymbol u,
  \qquad
  \boldsymbol u\in\mathbb S^{P-1},
  \qquad
  \|\boldsymbol u\|_2=1.
  \label{eq:fixed-shell-parameterization}
\end{equation}
We write \(\langle\cdot\rangle_{\sigma}\) for the average with respect to the normalized uniform probability measure on the unit sphere \(\mathbb S^{P-1}\).  Along the shell, define
\begin{align}
  C_r(\boldsymbol u)
  &=
  \gamma\,\overline L_{\mathrm{CE}}
  \bigl(
    \boldsymbol w_r(\boldsymbol u);
    \mathcal D
  \bigr),
  \label{eq:radial-ce-definition}\\
  R_r(\boldsymbol u)
  &=
  \lambda
  \frac{
    \|\boldsymbol w_r(\boldsymbol u)\|_2^2
  }{2P}.
  \label{eq:radial-l2-definition}
\end{align}
Substituting these two terms gives
\begin{equation}
  \phi_{\mathrm E,\beta}^{(P)}
  (r\mid\widetilde{\boldsymbol w},\mathcal D)
  =
  \frac{1}{P}
  \log
  \left\langle
    e^{
      -\beta_{\mathrm{sh}}
      [C_r(\boldsymbol u)+R_r(\boldsymbol u)]
    }
  \right\rangle_{\sigma}.
  \label{eq:ge-uniform-angular-form}
\end{equation}

The simplest approximation of Eq.~\eqref{eq:ge-uniform-angular-form} is uniform Monte Carlo sampling. Directions \(\boldsymbol u\) are drawn uniformly from the sphere, and the angular average is replaced by a finite sample average.  This approach becomes inefficient when the weighted measure is concentrated in a narrow region of the shell, since most sampled directions then contribute negligibly.

HMC trajectories can be useful for exploring individual low-loss pathways \cite{Zambon2025SamplingSpace}, but may move between separated angular regions slowly in the shell considered here.  We therefore first isolate the angular dependence that is known exactly.

The angular dependence of the quadratic regularization term can be obtained explicitly.  On the fixed shell,
\begin{equation}
  \beta_{\mathrm{sh}}R_r(\boldsymbol u)
  =
  \beta_{\mathrm{sh}}\lambda
  \frac{\|\widetilde{\boldsymbol w}\|_2^2}{2P}
  +
  \frac{
    \beta_{\mathrm{sh}}\lambda r^2
  }{2}
  +
  \beta_{\mathrm{sh}}\lambda r
  \frac{
    \widetilde{\boldsymbol w}\!\cdot\boldsymbol u
  }{\sqrt P}.
  \label{eq:l2-shell-expansion}
\end{equation}
We introduce
\begin{align}
  \Lambda_r
  &=
  \beta_{\mathrm{sh}}\lambda
  \frac{\|\widetilde{\boldsymbol w}\|_2^2}{2P}
  +
  \frac{
    \beta_{\mathrm{sh}}\lambda r^2
  }{2},
  \label{eq:radial-l2-constant}\\
  \boldsymbol\mu_r
  &=
  -\frac{
    \widetilde{\boldsymbol w}
  }{
    \|\widetilde{\boldsymbol w}\|_2
  },
  \qquad
  \kappa_r
  =
  \beta_{\mathrm{sh}}\lambda r
  \frac{
    \|\widetilde{\boldsymbol w}\|_2
  }{
    \sqrt P
  }.
  \label{eq:vmf-parameters}
\end{align}
For \(r\|\widetilde{\boldsymbol w}\|_2>0\), this gives
\begin{equation}
  \beta_{\mathrm{sh}}R_r(\boldsymbol u)
  =
  \Lambda_r
  -
  \kappa_r
  \boldsymbol\mu_r\!\cdot\boldsymbol u.
  \label{eq:compact-l2-decomposition}
\end{equation}
Therefore,
\begin{align}
  \phi_{\mathrm E,\beta}^{(P)}
  (r\mid\widetilde{\boldsymbol w},\mathcal D)
  &=
  -\frac{\Lambda_r}{P}
  \notag\\
  &\quad+
  \frac{1}{P}
  \log
  \left\langle
    e^{\kappa_r\boldsymbol\mu_r\cdot\boldsymbol u}
    e^{
      -\beta_{\mathrm{sh}}C_r(\boldsymbol u)
    }
  \right\rangle_{\sigma}.
  \label{eq:ge-expanded-angular-form}
\end{align}

The factor \(e^{\kappa_r\boldsymbol\mu_r\cdot\boldsymbol u}\) is the angular bias generated by the quadratic regularization.  It is largest at \(\boldsymbol u=\boldsymbol\mu_r\), the direction that minimizes \(\|\boldsymbol w_r(\boldsymbol u)\|_2\) among points on the fixed shell. Instead of leaving this known bias inside a uniform Monte Carlo weight, we absorb it into the sampling measure.

We therefore define
\begin{align}
  q_r(\boldsymbol u)
  &=
  \frac{
    e^{\kappa_r\boldsymbol\mu_r\cdot\boldsymbol u}
  }{
    M_P(\kappa_r)
  },
  \label{eq:l2-vmf-proposal}\\
  M_P(\kappa)
  &=
  \left\langle
    e^{\kappa\boldsymbol\mu_r\cdot\boldsymbol u}
  \right\rangle_{\sigma}.
  \label{eq:vmf-normalization}
\end{align}
The density \(q_r\), defined relative to the normalized uniform surface measure, is the von Mises--Fisher distribution with mean direction \(\boldsymbol\mu_r\) and concentration \(\kappa_r\).  By rotational invariance, \(M_P(\kappa)\) depends only on \(\kappa\) and is available analytically.

Using \(q_r\), the remaining loss-dependent factor is
\begin{equation}
  \mathcal Z_L(r)
  =
  \mathbb E_{\boldsymbol u\sim q_r}
  \left[
    e^{
      -\beta_{\mathrm{sh}}C_r(\boldsymbol u)
    }
  \right].
  \label{eq:loss-partition-factor}
\end{equation}
The angular average in Eq.~\eqref{eq:ge-expanded-angular-form} then satisfies
\begin{equation}
  \left\langle
    e^{\kappa_r\boldsymbol\mu_r\cdot\boldsymbol u}
    e^{
      -\beta_{\mathrm{sh}}C_r(\boldsymbol u)
    }
  \right\rangle_{\sigma}
  =
  M_P(\kappa_r)\,
  \mathcal Z_L(r).
  \label{eq:importance-change-of-measure}
\end{equation}
Hence,
\begin{align}
  \phi_{\mathrm E,\beta}^{(P)}
  (r\mid\widetilde{\boldsymbol w},\mathcal D)
  &=
  -\frac{\Lambda_r}{P}
  +
  \frac{\log M_P(\kappa_r)}{P}
  \notag\\
  &\quad+
  \frac{\log\mathcal Z_L(r)}{P}.
  \label{eq:ge-importance-decomposition}
\end{align}
The first two terms are available analytically.  The numerical calculation is therefore reduced to estimating the residual loss-partition factor \(\mathcal Z_L(r)\) from samples drawn initially from \(q_r\). We estimate this factor numerically with adaptive sequential Monte Carlo, which introduces the loss weighting through a sequence of tempered particle distributions \cite{DelMoral2006SMCSamplers,Zhou2016AdaptiveSMC}.  The tempering, resampling, and mutation steps are described in Appendix~\ref{app:sampler-details}.

\section{Results}
\label{sec:results}
\subsection{A controlled perceptron benchmark}
\label{sec:results-perceptron}

Before turning to nonlinear networks, we test the fixed-radius SMC estimator in a perceptron whose thermodynamic-limit local entropy can be calculated independently.  The benchmark asks whether the numerically estimated finite-size local entropy approaches the independently computed replica-symmetric prediction as the number of trainable weights increases.

\begin{figure}[t]
  \centering
  \includegraphics[width=\columnwidth]
    {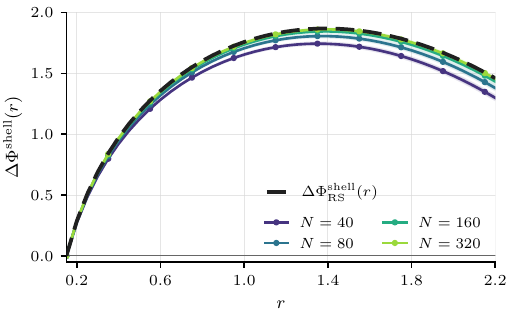}
\caption{ Full radial-shell local entropy in the controlled perceptron benchmark. The dashed black curve shows the independently calculated thermodynamic-limit RS result \(\Delta\Phi_{\mathrm{RS}}^{\mathrm{shell}}(r)\).  Colored curves show the SMC estimates \(\Delta\Phi_N^{\mathrm{shell}}(r)\) for \(N\in\{40,80,160,320\}\) trainable weights.  Shaded bands denote one standard error across datasets after averaging 10 references within each of 10 independent datasets.  Each radius uses \(2^{15}\) particles.}
  \label{fig:perceptron-benchmark}
\end{figure}

We consider a Gaussian perceptron with
\begin{equation}
  \mathcal D_N
  =
  \left\{
    (\boldsymbol x^\mu,y^\mu)
  \right\}_{\mu=1}^{M},
  \qquad
  M=\lfloor\alpha N\rfloor,
\end{equation}
and scalar output
\begin{equation}
  f_{\boldsymbol w}(\boldsymbol x^\mu)
  =
  \frac{
    \boldsymbol x^\mu\cdot\boldsymbol w
  }{\sqrt N}.
  \label{eq:benchmark-perceptron}
\end{equation}
Here the number of trainable parameters is \(P=N\).  Absorbing the labels into the patterns, \(\boldsymbol\xi^\mu=y^\mu\boldsymbol x^\mu\), we take \(\boldsymbol\xi^\mu\sim\mathcal N(\boldsymbol 0,I_N)\) and define the signed field
\begin{equation}
  h^\mu(\boldsymbol w)
  =
  \frac{
    \boldsymbol\xi^\mu\cdot\boldsymbol w
  }{\sqrt N}.
\end{equation}
The single-pattern logistic loss is
\begin{equation}
  \ell(h)=\log(1+e^{-h}),
\end{equation}
and the corresponding mean prediction loss is
\begin{equation}
  \overline L_{\mathrm{CE}}^{\mathrm{perc}}
  (\boldsymbol w;\mathcal D_N)
  =
  \frac{1}{M}
  \sum_{\mu=1}^{M}
  \ell\!\left[h^\mu(\boldsymbol w)\right].
\end{equation}

To match the notation used for the neural networks, we write the regularized perceptron loss as
\begin{equation}
  \mathcal L_{\mathrm{reg}}^{\mathrm p}
  (\boldsymbol w;\mathcal D_N)
  =
  \gamma_{\mathrm p}\,\overline L_{\mathrm{CE}}^{\mathrm{perc}}
  (\boldsymbol w;\mathcal D_N)
  +
  \lambda_{\mathrm p}
  \frac{\|\boldsymbol w\|_2^2}{2P} .
  \label{eq:benchmark-general-energy}
\end{equation}
In the local-entropy calculation, the shell weight is \(\exp[-\beta_{\mathrm p}\mathcal L_{\mathrm{reg}}^{\mathrm p}]\). With \(P=N\), we use \(\gamma_{\mathrm p}=M\), \(\lambda_{\mathrm p}=P\), and \(\beta_{\mathrm p}=1\).  Equation~\eqref{eq:benchmark-general-energy} then equals \(M\overline L_{\mathrm{CE}}^{\mathrm{perc}} +\|\boldsymbol w\|_2^2/2\).  At fixed \(\alpha=M/N\), both terms scale linearly with \(N\), as required by the analytic calculation.

We first characterize the reference ensemble.  Exact reference solutions are assigned the weight
\begin{equation}
  W_{\mathrm{ref}}
  (\widetilde{\boldsymbol w};\mathcal D_N)
  =
  \exp\!\left[
    -\frac{\lambda_{\mathrm{ref}}}{2}
    \|\widetilde{\boldsymbol w}\|_2^2
  \right]
  \prod_{\mu=1}^{M}
  \Theta\!\left[
    h^\mu(\widetilde{\boldsymbol w})
  \right].
  \label{eq:benchmark-reference-weight}
\end{equation}
We set \(\lambda_{\mathrm{ref}}=1\) in this reference weight.  It controls the typical norm of the exact solutions used as shell centers, separately from the soft logistic weighting of the surrounding shell. The corresponding partition function and normalized reference density are
\begin{align}
  Z_{\mathrm{ref}}^0(\mathcal D_N)
  &=
  \int_{\mathbb R^N}
  \mathrm d^N\widetilde{\boldsymbol w}\,
  W_{\mathrm{ref}}
  (\widetilde{\boldsymbol w};\mathcal D_N),
  \\
  p_{\mathrm{ref}}^0
  (\widetilde{\boldsymbol w}\mid\mathcal D_N)
  &=
  \frac{
    W_{\mathrm{ref}}
    (\widetilde{\boldsymbol w};\mathcal D_N)
  }{
    Z_{\mathrm{ref}}^0(\mathcal D_N)
  }
  .
  \label{eq:benchmark-reference-measure}
\end{align}

The replica-symmetric calculation summarizes the reference ensemble by two extremized order parameters, the squared norm per parameter and the typical overlap between reference solutions.  Their saddle-point equations and the retained branch are given in Appendix~\ref{app:replica-saddle}.  The reference measure contains only exact classifiers, whereas shell configurations are weighted by the soft logistic loss in Eq.~\eqref{eq:benchmark-general-energy}.  Given \(p_{\mathrm{ref}}^0\), averaging the loss-weighted shell volume over datasets and hard references gives the finite-\(N\) local entropy
\begin{equation}
  \Phi_N(r)
  =
  \mathbb E_{\mathcal D_N}
  \mathbb E_{
    \widetilde{\boldsymbol w}
    \sim p_{\mathrm{ref}}^0(\cdot\mid\mathcal D_N)
  }
  \left[
    \frac{1}{N}
    \log
    \Omega
    (r\mid\widetilde{\boldsymbol w},\mathcal D_N)
  \right].
  \label{eq:benchmark-finite-local-entropy}
\end{equation}
The inner expectation averages over the exact solutions used as shell centers.  For each center, the loss-weighted volume of parameter configurations at distance \(\sqrt N r\) is
\begin{align}
  \Omega
  (r\mid\widetilde{\boldsymbol w},\mathcal D_N)
  &=
  \int_{\mathbb R^N}
    \mathrm d^Nw\,
    \delta\!\left(
      Nr^2
      -
      \|\boldsymbol w-\widetilde{\boldsymbol w}\|_2^2
    \right)
  \notag\\
  &\quad\times
    \exp\!\left[
      -\beta_{\mathrm p}
      \mathcal L_{\mathrm{reg}}^{\mathrm p}
      (\boldsymbol w;\mathcal D_N)
    \right].
  \label{eq:benchmark-shell-mass}
\end{align}
Candidates with lower regularized logistic loss contribute more strongly.  The remaining step is to calculate the thermodynamic-limit local entropy of configurations on this shell.  The replica-symmetric calculation introduces three additional order parameters.  Writing
\begin{equation}
  \boldsymbol w^\gamma
  =
  \sqrt Q\,\boldsymbol u^\gamma,
  \qquad
  \|\boldsymbol u^\gamma\|_2^2=N,
\end{equation}
\(Q\) is the squared norm per parameter of a shell configuration, and
\begin{equation}
  \frac{
    \boldsymbol u^\gamma
    \cdot
    \boldsymbol u^\delta
  }{N}
  =
  \delta_{\gamma\delta}
  +
  (1-\delta_{\gamma\delta})p
  \label{eq:benchmark-shell-overlap}
\end{equation}
defines the overlap \(p\) between two shell replicas.  Let \(a\) index the reference replicas and \(\gamma\) the shell replicas.  Let \(\widetilde{\boldsymbol v}^{\,a}\) denote the normalized direction of reference replica \(a\), with \(\|\widetilde{\boldsymbol v}^{\,a}\|_2^2=N\).  The copy \(a=1\) serves as the center of the shell, and its mixed overlap with shell replica \(\gamma\) is
\begin{equation}
  \frac{
    \widetilde{\boldsymbol v}^{\,a}
    \cdot
    \boldsymbol u^\gamma
  }{N}
  =
  c\,\delta_{a1}
  +
  t(1-\delta_{a1}).
  \label{eq:benchmark-mixed-overlap}
\end{equation}
The overlap for \(a=1\) is fixed by the prescribed distance,
\begin{equation}
  c
  =
  c_d(Q,r)
  =
  \frac{
    Q+\widetilde Q_\star-r^2
  }{
    2\sqrt{Q\widetilde Q_\star}
  }.
  \label{eq:benchmark-distance-cosine}
\end{equation}
Under replica symmetry, all nonselected reference replicas are equivalent, so \(t\) is the overlap between a shell configuration and any copy \(a\ne1\). The overlap with \(a=1\) is fixed by \(r\). In this way, \(Q\) records the typical size of a candidate on the shell, \(p\) records how strongly two such candidates align, and \(t\) records their shared relation to the reference ensemble.  The prescribed distance supplies the remaining overlap \(c\). After taking the two replica limits, the local entropy separates into a radial term, a conditional directional entropy, and a one-pattern energetic term,
\begin{align}
  \mathcal F_{\mathrm{RS}}(Q,p,t;r)
  &=
  f_{\mathrm{rad}}^{\mathrm{shell}}(Q)
  +
  G_{\mathrm S}^{\mathrm{shell}}(p,t;c,q_{\mathrm{ref}})
  \notag\\
  &\quad+
  \alpha
  G_{\mathrm E}^{\mathrm{shell}}(Q,p,t;r).
  \label{eq:benchmark-shell-functional}
\end{align}
The three terms respectively record the typical weight scale, the amount of directional space compatible with the overlaps, and the loss-based weighting of classification fields.  Their competition determines how much solution-like parameter volume remains at the prescribed distance. Here \(c=c_d(Q,r)\).  With the coefficient choices above, the radial term is
\begin{equation}
  f_{\mathrm{rad}}^{\mathrm{shell}}(Q)
  =
  \frac{1}{2}\log Q
  -
  \frac{1}{2}Q .
  \label{eq:benchmark-shell-radial}
\end{equation}
The directional contribution follows from the Schur complement of the reference block.  With
\begin{equation}
  B(c,t)
  =
  (1-c^2)(1-2q_{\mathrm{ref}})
  +
  q_{\mathrm{ref}}^2
  -
  2q_{\mathrm{ref}}ct
  +
  t^2,
  \label{eq:benchmark-shell-b}
\end{equation}
it is
\begin{equation}
  G_{\mathrm S}^{\mathrm{shell}}
  =
  \frac{B(c,t)}{
    2(1-p)(1-q_{\mathrm{ref}})^2
  }
  +
  \frac{1}{2}\log\!\left[2\pi(1-p)\right].
  \label{eq:benchmark-shell-entropic}
\end{equation}

To write the energetic term, let
\begin{equation}
  A
  =
  p-\frac{t^2}{q_{\mathrm{ref}}}
  -\frac{(c-t)^2}{1-q_{\mathrm{ref}}}
  \label{eq:benchmark-feasibility-remainder}
\end{equation}
and
\begin{equation}
  g
  =
  \frac{t}{\sqrt{q_{\mathrm{ref}}}}z_0
  +
  \frac{c-t}{\sqrt{1-q_{\mathrm{ref}}}}z_1
  +
  \sqrt A\,z_2
  +
  \sqrt{1-p}\,z_3 .
  \label{eq:benchmark-conditioned-field}
\end{equation}
With \(Dz=e^{-z^2/2}\mathrm dz/\sqrt{2\pi}\) and \(H(x)=\int_x^\infty Dz\), the selected reference satisfies its hard classification constraint.  The shell field is weighted by the soft logistic loss, giving
\begin{widetext}
\begin{align}
  G_{\mathrm E}^{\mathrm{shell}}(Q,p,t;r)
  &=
  \int Dz_0\,
  \frac{1}{
    H\!\left(
      -\sqrt{
        \frac{q_{\mathrm{ref}}}{1-q_{\mathrm{ref}}}
      }\,z_0
    \right)
  }
  \int Dz_1\,
  \Theta\!\left(
    \sqrt{q_{\mathrm{ref}}}\,z_0
    +
    \sqrt{1-q_{\mathrm{ref}}}\,z_1
  \right)
  \notag\\[-2pt]
  &\quad\times
  \int Dz_2\,
  \log\int Dz_3\,
  \exp\!\left[
    -\beta_{\mathrm p}
    \ell\!\left(\sqrt Q\,g\right)
  \right].
  \label{eq:benchmark-shell-energetic}
\end{align}
\end{widetext}

The covariance representation requires \(A\ge0\) and \(p<1\).  The replica-symmetric prediction is the constrained stationary value
\begin{equation}
  \Phi_{\mathrm{RS}}(r)
  =
  \underset{Q,p,t;\,A\ge0,\,p<1}{\operatorname{extr}}
  \mathcal F_{\mathrm{RS}}(Q,p,t;r)
  ,
  \label{eq:benchmark-final-rs}
\end{equation}
with
\begin{equation}
  \left(
    \sqrt{\widetilde Q_\star}-r
  \right)^2
  <
  Q
  <
  \left(
    \sqrt{\widetilde Q_\star}+r
  \right)^2 .
  \label{eq:benchmark-q-domain}
\end{equation}
We evaluate the variational free entropy by numerically solving the constrained RS saddle-point problem.  The reparameterization and numerical checks are given in Appendix~\ref{app:replica-saddle}.  The detailed replica derivation is provided in the Supplementary Material.

For the analytic benchmark, we set \(\alpha=0.1\) and consider the common distance range
\begin{equation}
  0.15\le r\le2.20,
  \qquad
  \Delta r=0.05 .
\end{equation}
We compare the full radial-shell local entropy because its geometric growth makes the finite-\(N\) and thermodynamic-limit curves particularly easy to distinguish.  With \(r_0=0.15\), the analytic result is
\begin{equation}
  \Delta\Phi_{\mathrm{RS}}^{\mathrm{shell}}(r)
  =
  \Phi_{\mathrm{RS}}(r)
  -
  \Phi_{\mathrm{RS}}(r_0).
  \label{eq:benchmark-rs-full-shell-profile}
\end{equation}
The finite-\(N\) curve is estimated with the same angular expectation used for the neural networks.  For one dataset and reference,
\begin{equation}
  \phi_{\mathrm E,\beta}^{(N),\mathrm{perc}}
  (r\mid\widetilde{\boldsymbol w},\mathcal D_N)
  =
  \frac{1}{N}
  \log
  \left\langle
    e^{-\beta_{\mathrm p}
    \mathcal L_{\mathrm{reg}}^{\mathrm p}
    (\boldsymbol w;\mathcal D_N)}
  \right\rangle_{r,\widetilde{\boldsymbol w}}.
  \label{eq:benchmark-perceptron-angular-profile}
\end{equation}
We center this quantity before averaging over datasets and references,
\begin{align}
  \Delta\Phi_{\mathrm E,\beta}^{(N),\mathrm{perc}}(r)
  &=
  \mathbb E_{\mathcal D_N}
  \mathbb E_{
    \widetilde{\boldsymbol w}
    \sim p_{\mathrm{ref}}^0(\cdot\mid\mathcal D_N)
  }
  \left[
    \phi_{\mathrm E,\beta}^{(N),\mathrm{perc}}
    (r\mid\widetilde{\boldsymbol w},\mathcal D_N)
  \right.
  \notag\\[-2pt]
  &\hspace{19mm}\left.
    -
    \phi_{\mathrm E,\beta}^{(N),\mathrm{perc}}
    (r_0\mid\widetilde{\boldsymbol w},\mathcal D_N)
  \right].
  \label{eq:benchmark-centered-perceptron-angular-profile}
\end{align}
We then restore the finite-dimensional shell-area term,
\begin{equation}
  \Delta\Phi_N^{\mathrm{shell}}(r)
  =
  \Delta\Phi_{\mathrm E,\beta}^{(N),\mathrm{perc}}(r)
  +
  \frac{N-2}{N}
  \log\!\left(\frac{r}{r_0}\right).
  \label{eq:benchmark-finite-full-shell-profile}
\end{equation}
The coefficient \((N-2)/N\) is the exact finite-dimensional geometric factor. We evaluate \(N\in\{40,80,160,320\}\) on this common radial grid.  Each size uses 10 datasets and 10 hard references per dataset.  At each radius, we use \(2^{15}\) particles.

Fig.~\ref{fig:perceptron-benchmark} directly compares the two curves defined above.  The finite-size SMC curve, \(\Delta\Phi_N^{\mathrm{shell}}(r)\), approaches the independently calculated RS curve, \(\Delta\Phi_{\mathrm{RS}}^{\mathrm{shell}}(r)\), as the number of perceptron weights increases.  This agreement provides a further check that our sampling procedure accurately approximates the local entropy in this benchmark.

\subsection{Controlled synthetic label organization}
\label{sec:results-synthetic}

Having benchmarked the shell estimator in a controlled analytical setting, we now ask how the organization of a finite dataset affects the local entropy landscape around a trained neural network solution.  We begin with the synthetic system, where the label arrangement can be changed systematically.

\begin{figure}[!t]
  \centering
  \includegraphics[width=\columnwidth]
    {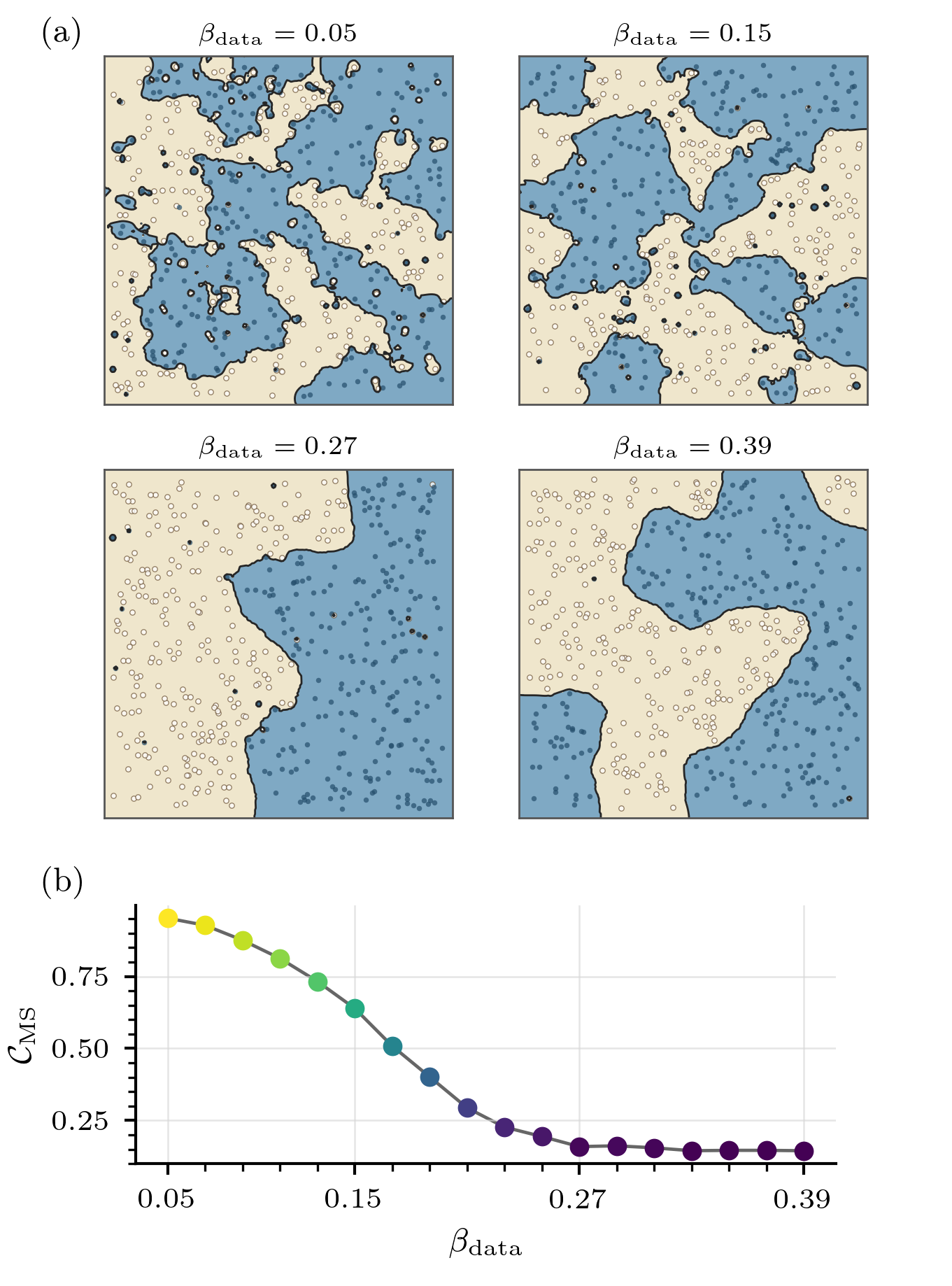}
\caption{Controlled organization of the synthetic labels. (a) Four representative observed configurations spanning the controlled sweep.  The colored regions and black interfaces visualize the organization of the sampled labels.  (b) Mean \(\mathcal C_{\mathrm{MS}}\) at every \(\beta_{\mathrm{data}}\), with standard errors across 60 independent datasets.}
  \label{fig:synthetic-label-control}
\end{figure}

The generator parameter \(\beta_{\mathrm{data}}\) controls the tendency of neighboring labels to align during the Kawasaki dynamics. Fig.~\ref{fig:synthetic-label-control} shows representative label configurations.  At small \(\beta_{\mathrm{data}}\), the two labels remain locally interwoven and form an irregular interface throughout the sample.  As \(\beta_{\mathrm{data}}\) increases, same-label regions grow and the interface becomes progressively smoother.

We quantify this change using the multiscale dataset complexity \(\mathcal C_{\mathrm{MS}}\).  As shown in Fig.~\ref{fig:synthetic-label-control}(b), \(\mathcal C_{\mathrm{MS}}\) generally decreases with \(\beta_{\mathrm{data}}\) and reaches a shallow plateau at the ordered end of the sweep.  The generator therefore supplies a graded family of label organizations for the landscape comparison.

For each dataset, we train networks from fresh random initializations with the same regularized loss later used to weight the shell.  Adam stops at the first endpoint that classifies every training sample correctly.  Unresolved attempts continue with L-BFGS under the same loss.  We retain the first ten distinct zero-error endpoints as references and estimate the local entropy around each one.  With the common empirical baseline \(r_0=0.01\), we report
\begin{equation}
  \begin{aligned}
  \Delta\Phi_{\mathrm E}(r)
  &=
  \mathbb E_{\mathcal D}
  \mathbb E_{\widetilde{\boldsymbol w}}
  \left[
    \phi_{\mathrm E}
    (r\mid\widetilde{\boldsymbol w},\mathcal D)
    -
    \phi_{\mathrm E}
    (r_0\mid\widetilde{\boldsymbol w},\mathcal D)
  \right],
  \\
  g_{\mathrm E}(r)
  &=
  \frac{\mathrm d}{\mathrm dr}
  \Delta\Phi_{\mathrm E}(r).
  \end{aligned}
  \label{eq:synthetic-centered-energetic-profile}
\end{equation}

\begin{figure*}[t]
  \centering
  \includegraphics[width=\textwidth]
    {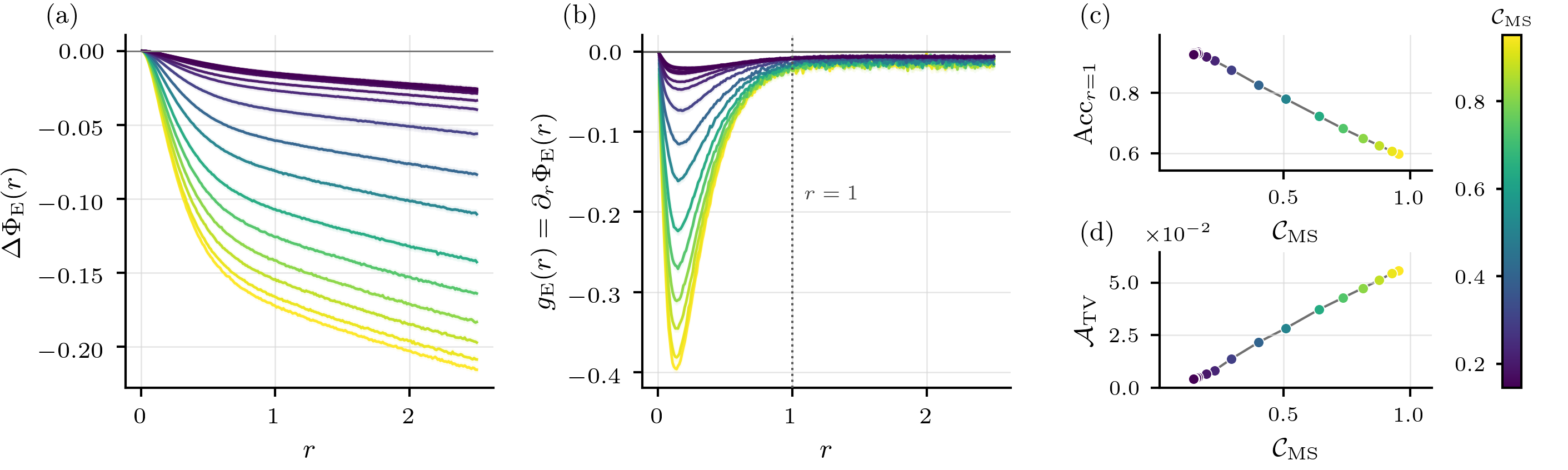}
\caption{Local entropy across the controlled synthetic sweep. (a) Centered local entropy \(\Delta\Phi_{\mathrm E}(r)\).  More negative values correspond to a smaller effective solution volume relative to \(r_0\).  (b) Its radial derivative \(g_{\mathrm E}(r)\).  The dotted line marks \(r=1\).  (c) Particle-weighted, sample-wise training accuracy among the configurations contributing to the \(r=1\) shell, plotted against \(\mathcal C_{\mathrm{MS}}\).  (d) Total radial variation \(\mathcal A_{\mathrm{TV}}\) plotted against \(\mathcal C_{\mathrm{MS}}\).  Curves show means at each \(\beta_{\mathrm{data}}\) after averaging ten references within each of 60 independent datasets.  Shaded bands and error bars denote standard errors across datasets.  Color denotes \(\mathcal C_{\mathrm{MS}}\).}
  \label{fig:synthetic-shell-response}
\end{figure*}

In Fig.~\ref{fig:synthetic-shell-response}(a), \(\Delta\Phi_{\mathrm E}(r)\) decreases more rapidly for datasets with larger \(\mathcal C_{\mathrm{MS}}\).  Their effective solution volume therefore narrows sooner as the distance from the trained reference increases.  For datasets with smaller \(\mathcal C_{\mathrm{MS}}\), the decrease is weaker over the same radial range.

Fig.~\ref{fig:synthetic-shell-response}(b) shows where this difference is produced.  At the high-complexity end of the sweep, \(g_{\mathrm E}(r)\) has a deep negative minimum in the inner radial region.  The minimum becomes progressively shallower as \(\mathcal C_{\mathrm{MS}}\) decreases.  At large \(r\), the derivatives approach a common weakly negative value, leaving the stronger inner contraction as a persistent offset in \(\Delta\Phi_{\mathrm E}\).  Dataset complexity changes both the strength and the radial location of the main contraction.  It does more than rescale one common decay curve.

The same shell samples provide a functional check. Fig.~\ref{fig:synthetic-shell-response}(c) shows that particle-weighted training accuracy at \(r=1\) falls to \(0.598\) at the highest-complexity endpoint, compared with about \(0.93\) in the low-complexity conditions. Thus the outer derivatives converge even though the shell accuracies remain markedly different.  Most of the difference in effective solution volume has already accumulated in the inner region.

For a compact scalar summary, define for each retained reference \(j\) the shell-temperature-normalized derivative and its total radial variation by
\begin{equation}
  \begin{aligned}
  u_j(r)
  &=
  \frac{
    g_{\mathrm E,j}(r)
    -
    g_{\mathrm E,j}(r_0)
  }{
    \beta_{\mathrm{sh}}
  },
  \\
  \mathcal A_{\mathrm{TV},j}
  &=
  \int_{r_0}^{1}
  \left|
    \frac{\mathrm du_j(r)}{\mathrm dr}
  \right|
  \mathrm dr.
  \end{aligned}
  \label{eq:local-entropy-glassness}
\end{equation}
We evaluate the integral on the measured grid as \(\sum_i|u_j(r_{i+1})-u_j(r_i)|\), and report \(\mathcal A_{\mathrm{TV}}\) by averaging within datasets and then across datasets. As shown in Fig.~\ref{fig:synthetic-shell-response}(d), \(\mathcal A_{\mathrm{TV}}\) follows the broad graded change across the synthetic sweep.

\subsection{Label organization and local entropy in MNIST}
\label{sec:results-mnist}

The synthetic system provides direct control of label organization, but its two-dimensional geometry is deliberately simple.  We next ask whether the same relation appears in structured image data.  We use two complementary MNIST comparisons.  The label-noise comparison changes labels while keeping the image coordinates fixed, whereas the digit-pair comparison uses naturally different binary classification tasks.  The first provides the cleaner control of label organization.  The second tests whether the relation persists when tasks differ naturally.

\subsubsection{Label-noise sweep with fixed inputs}
\label{subsec:results-mnist-noise}

\begin{figure}[!b]
  \centering
  \includegraphics[width=\columnwidth]{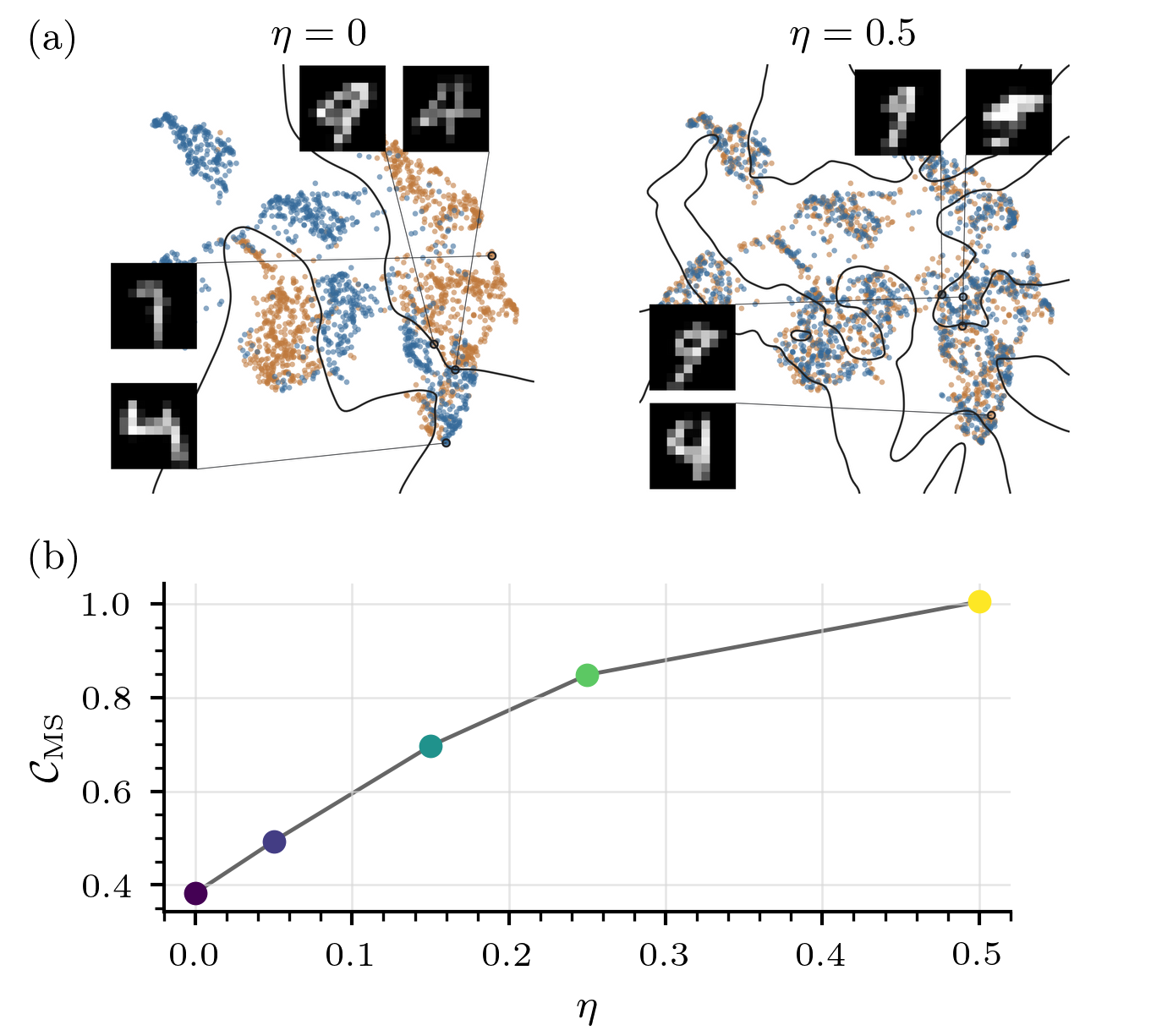}
\caption{Dataset organization in the MNIST label-noise sweep. (a) Endpoint visualizations in shared UMAP coordinates for the clean endpoint (\(\eta=0\)) and randomized endpoint (\(\eta=0.5\)). Colors show the observed labels, and black contours summarize their organization in the embedding.  UMAP is used only for visualization. (b) Mean \(\mathcal C_{\mathrm{MS}}\) at each noise probability, with standard errors across ten paired datasets.  The complexity measure is calculated in the standardized 100-dimensional input space.}
  \label{fig:mnist-label-noise-control}
  \label{fig:mnist-label-noise-complexity}
\end{figure}

We first consider the balanced odd--even task and flip labels with probability \(\eta\).  Within each of ten paired datasets, the coordinates are shared across all noise levels and the flip masks are nested. Increasing \(\eta\) therefore changes the local label organization without changing the input geometry of that dataset.  The measured dataset complexity \(\mathcal C_{\mathrm{MS}}\) increases accordingly, as shown in Fig.~\ref{fig:mnist-label-noise-complexity}(b).

Fig.~\ref{fig:mnist-label-noise-control} visualizes the two endpoints in the same embedding.  Clean odd--even labels form locally organized regions, whereas the randomized labels are strongly mixed.

\begin{figure*}[t]
  \centering
  \includegraphics[width=\textwidth]{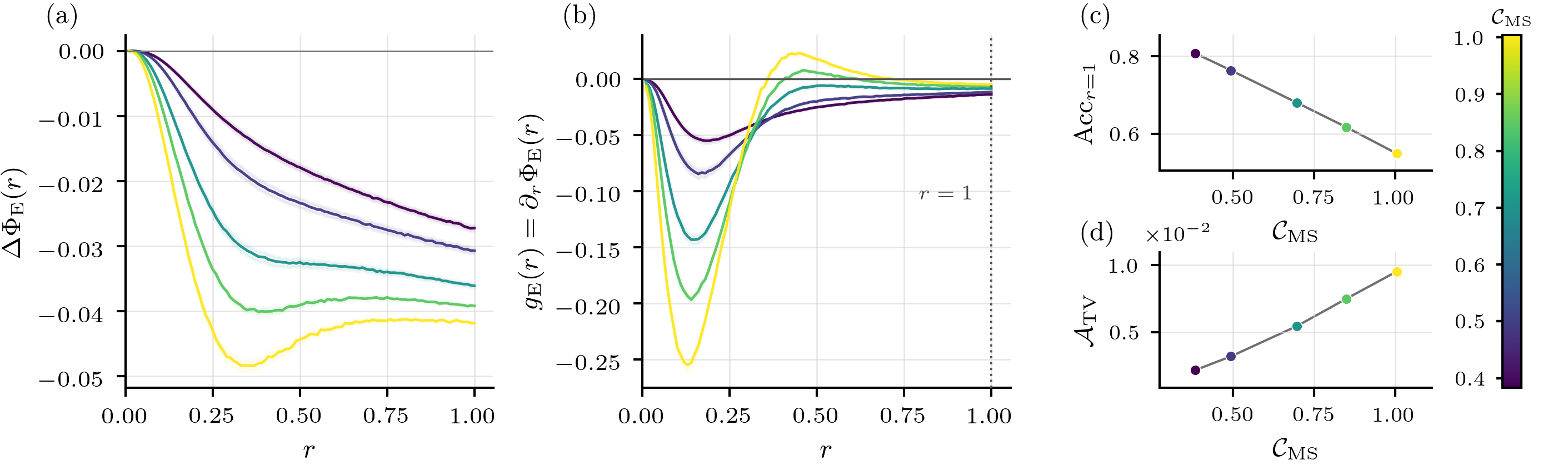}
\caption{Local entropy under MNIST label noise with fixed inputs. (a) Centered local entropy \(\Delta\Phi_{\mathrm E}(r)\). More negative values correspond to a smaller effective solution volume relative to \(r_0\).  (b) Its radial derivative \(g_{\mathrm E}(r)\).  The dotted line marks \(r=1\).  (c) Particle-weighted, sample-wise training accuracy among the configurations contributing to the \(r=1\) shell, plotted against \(\mathcal C_{\mathrm{MS}}\).  (d) Total radial variation \(\mathcal A_{\mathrm{TV}}\) plotted against \(\mathcal C_{\mathrm{MS}}\).  Curves average ten retained references within each of ten paired datasets.  Shaded bands and error bars denote standard errors across datasets.  Color denotes \(\mathcal C_{\mathrm{MS}}\).}
  \label{fig:mnist-label-noise}
\end{figure*}

In Fig.~\ref{fig:mnist-label-noise}(a), increasing label noise makes \(\Delta\Phi_{\mathrm E}(r)\) fall more sharply just outside the trained reference.  The effective solution volume therefore becomes concentrated in a narrower neighborhood of the solution.

Fig.~\ref{fig:mnist-label-noise}(b) shows the same change through the radial derivative \(g_{\mathrm E}(r)\).  Its negative inner minimum becomes deeper as \(\mathcal C_{\mathrm{MS}}\) increases.  At the largest noise levels, \(g_{\mathrm E}(r)\) crosses zero over an intermediate radial range.  The effective solution volume then increases locally with radius after contracting sharply at smaller \(r\).

Fig.~\ref{fig:mnist-label-noise}(c) shows the particle-weighted training accuracy of the \(r=1\) shell.  The outer rise of \(g_{\mathrm E}\) toward and across zero therefore reflects a change in the radial organization of the effective solution volume rather than recovery of the classification function.  The monotonic increase of \(\mathcal A_{\mathrm{TV}}\) with \(\eta\) in Fig.~\ref{fig:mnist-label-noise}(d) summarizes the full change in radial shape.

The clean and randomized endpoints show the largest radial contrast, including a much deeper inner minimum and an intermediate range with \(g_{\mathrm E}(r)>0\) under random labels.  Because the retained zero-training-error references were not certified stationary points of the regularized loss, we also compared the endpoints using a separate symmetrized finite-distance loss response that cancels the residual first-order tilt.  The symmetrized response rose in both conditions, but earlier and by a much larger amount under random labels.  This independent check supports the conclusion that residual first-order tilt alone cannot explain the clean--randomized contrast.  Details are given in Appendix~\ref{app:finite-distance-audit}.

\subsubsection{Naturally different digit-pair tasks}
\label{subsec:results-mnist-pairs}

We next consider binary classification between individual MNIST digit pairs. Unlike the label-noise sweep with fixed inputs, changing the digit pair also changes the distribution and geometry of the inputs.  This provides a natural comparison in which experimental control is weaker.

We calculated \(\mathcal C_{\mathrm{MS}}\) for all 45 digit pairs on ten balanced datasets per pair and ranked the pairs by their mean across those datasets.  The selection rule used \(\mathcal C_{\mathrm{MS}}\) alone, rather than any local-entropy outcome, and retained the pairs at ranks \(1,5,9,\ldots,45\).  For this fixed set of twelve tasks, we evaluated local entropy on the same datasets, using ten trained references per task and dataset.  Thus pair selection and the reported complexity values use the same \(\mathcal C_{\mathrm{MS}}\) definition and the same datasets.

\begin{figure}[!t]
  \centering
  \includegraphics[width=\columnwidth]{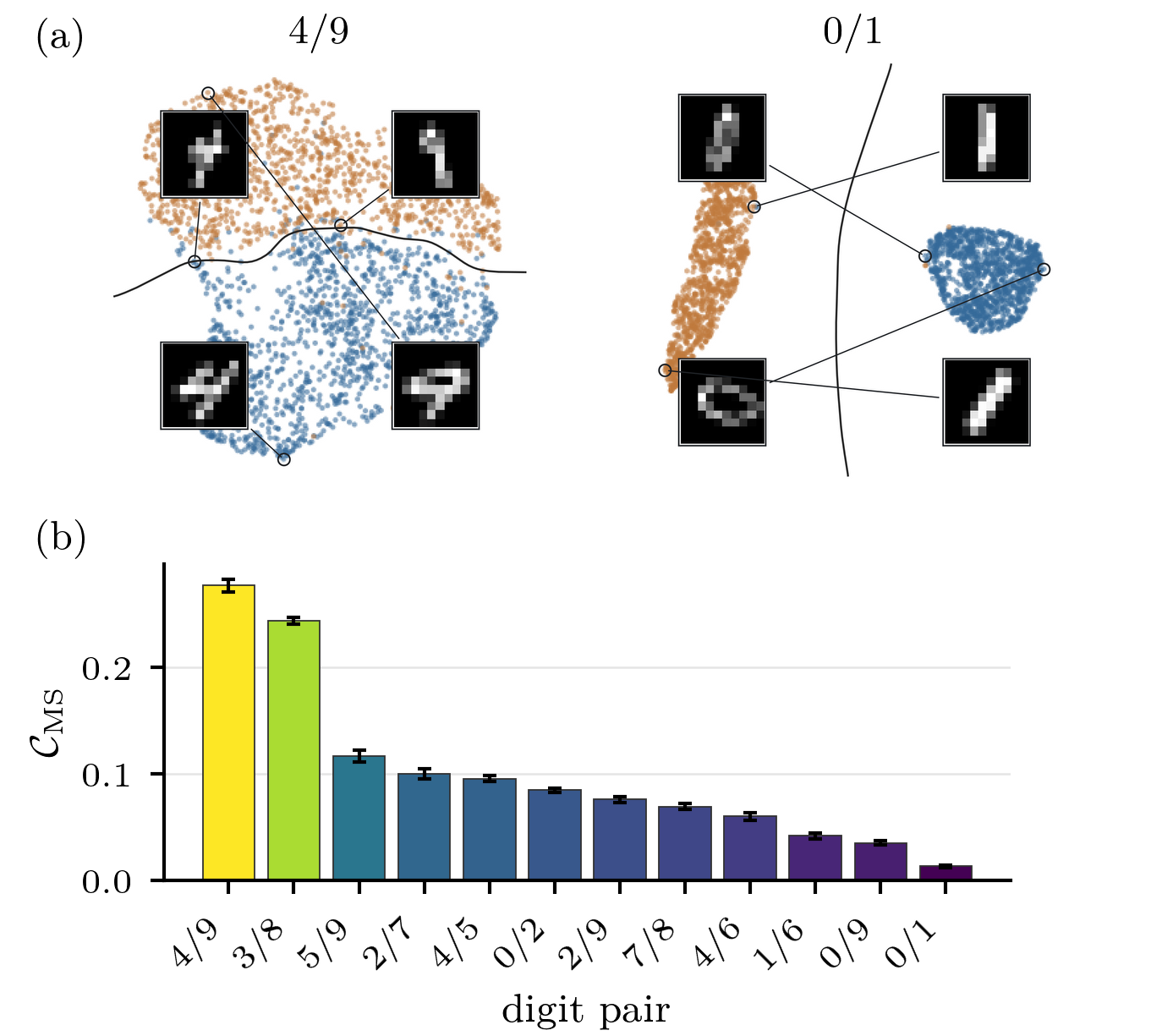}
\caption{Dataset organization in the selected MNIST digit-pair tasks. (a) UMAP visualizations of the lower-complexity \(0/1\) and higher-complexity \(4/9\) endpoints.  (b) Mean \(\mathcal C_{\mathrm{MS}}\) for the pairs at ranks \(1,5,9,\ldots,45\) after ranking all 45 tasks by their mean across ten balanced datasets per pair.  Error bars denote standard errors across those datasets.}
  \label{fig:mnist-digit-selection}
  \label{fig:mnist-digit-complexity}
\end{figure}

Fig.~\ref{fig:mnist-digit-selection}(a) illustrates the two endpoint pairs, and panel (b) reports \(\mathcal C_{\mathrm{MS}}\) for all 12 selected pairs. The mean complexity values range from approximately \(0.013\) to \(0.277\). This range is narrower and lower than the ranges generated by both controlled sweeps, so the digit-pair comparison probes a milder regime of dataset organization.

\begin{figure*}[t]
  \centering
  \includegraphics[width=\textwidth]{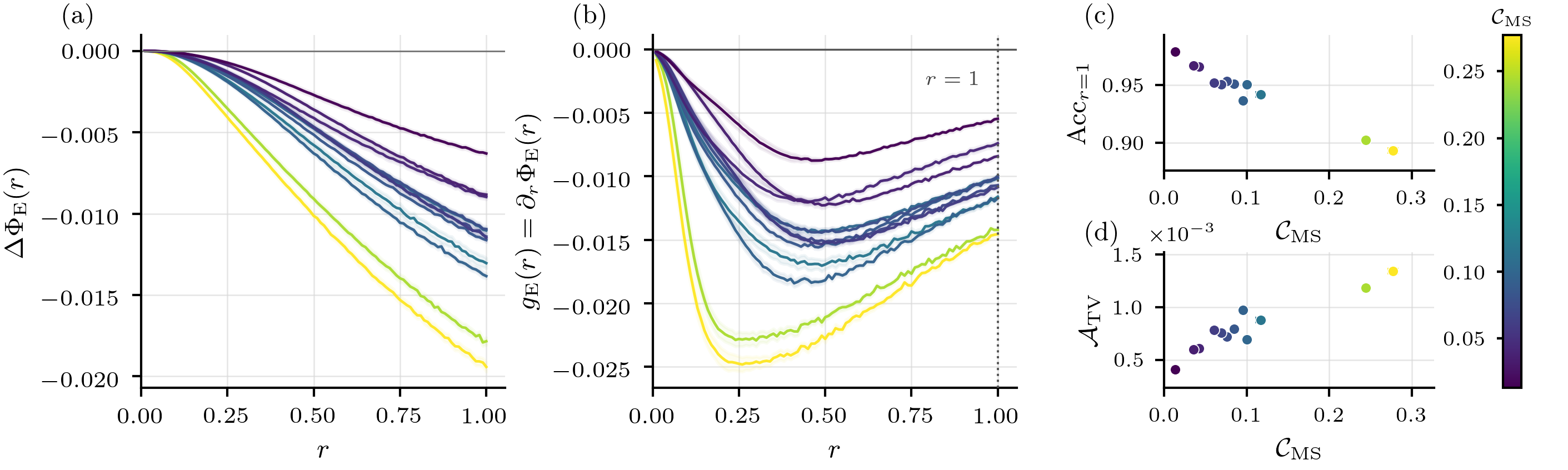}
\caption{Local entropy across the selected MNIST digit-pair tasks. (a) Centered local entropy \(\Delta\Phi_{\mathrm E}(r)\). (b) Its radial derivative \(g_{\mathrm E}(r)\).  The dotted line marks \(r=1\).  (c) Particle-weighted, sample-wise training accuracy among the configurations contributing to the \(r=1\) shell, plotted against \(\mathcal C_{\mathrm{MS}}\).  (d) Total radial variation \(\mathcal A_{\mathrm{TV}}\) plotted against \(\mathcal C_{\mathrm{MS}}\).  Curves average ten references within each of ten balanced datasets per task.  Shaded bands and error bars denote standard errors across datasets, and color denotes mean \(\mathcal C_{\mathrm{MS}}\).}
  \label{fig:mnist-digit-shell}
\end{figure*}

Fig.~\ref{fig:mnist-digit-shell} nevertheless shows the same broad tendency across the naturally different tasks.  For pairs with larger \(\mathcal C_{\mathrm{MS}}\), \(\Delta\Phi_{\mathrm E}(r)\) falls more rapidly in the inner radial range, and their effective solution volume narrows sooner around the trained reference.

At larger \(r\), the \(g_{\mathrm E}(r)\) curves move closer to zero and to one another.  The effective solution volume is still smaller than it was near the reference, but its additional decrease per radial step has become weaker. Several intermediate tasks overlap or exchange order, while the contrast between the two endpoints remains clear.

At \(r=1\), particle-weighted, sample-wise training accuracy is \(0.979\) for the low-complexity \(0/1\) task and \(0.893\) for the high-complexity \(4/9\) task.  The effective-volume separation is therefore observed while both sets of shell configurations remain highly accurate.

\section{Discussion}
\label{sec:discussion}

\subsection{Finite-distance hardening}

Across all three empirical comparisons, \(g_{\mathrm E}(r)\) develops a negative inner minimum and then relaxes toward a weaker value.  Higher dataset complexity generally moves the strongest contraction of effective solution volume closer to the reference and makes it sharper, rather than uniformly rescaling the same radial profile.  Figure~\ref{fig:hardening-geometry-composite}(a) isolates this shape through \(h(r)=-g_{\mathrm E}(r)/r\).  A positive \(h\) simply means that effective solution volume contracts with radius, as expected near a trained solution.  The nontrivial feature is that this normalized contraction initially becomes stronger with distance before turning downward.

The contribution from a fixed Hessian is quadratic in the displacement, so we use the squared-radius coordinate \(\tau=r^2\).  Write \(\boldsymbol w_\tau(\boldsymbol u)=\widetilde{\boldsymbol w} +\sqrt{P\tau}\,\boldsymbol u\) and define
\begin{equation}
  \mathcal A(\tau,\boldsymbol u)
  =
  \beta_{\mathrm{sh}}
  \left[
    \mathcal L_{\mathrm{reg}}
    (\boldsymbol w_\tau(\boldsymbol u);\mathcal D)
    -
    \mathcal L_{\mathrm{reg}}
    (\widetilde{\boldsymbol w};\mathcal D)
  \right].
  \label{eq:discussion-directional-excess-action}
\end{equation}
Along the same direction, define
\begin{equation}
  \kappa(\tau,\boldsymbol u)
  =
  \frac{2}{P}\partial_\tau\mathcal A(\tau,\boldsymbol u)
  =
  \frac{1}{Pr}\partial_r\mathcal A(\tau,\boldsymbol u).
  \label{eq:discussion-directional-next-step-slope}
\end{equation}
At a given radius, \(\mathcal A\) is the scaled loss change reached along \(\boldsymbol u\), and \(\kappa\) is its outward slope in the same direction. A large positive \(\kappa\) means that the next outward step raises the loss rapidly.  A smaller \(\kappa\) means a weaker increase, or a decrease when \(\kappa<0\).

Directions with smaller \(\mathcal A\) carry greater weight in local entropy. Let \(\langle\cdot\rangle_\tau\) denote this weighted directional average, with relative weights proportional to \(e^{-\mathcal A}\).  Exact differentiation gives
\begin{equation}
  h(r)
  =
  -\frac{g_{\mathrm E}(r)}{r}
  =
  \mathbb E_{\mathcal D}
  \mathbb E_{\widetilde{\boldsymbol w}}
  \left[
    \langle\kappa\rangle_\tau
  \right].
  \label{eq:discussion-normalized-radial-derivative}
\end{equation}
Thus \(h\) is the weighted mean of \(\kappa\) among the directions currently contributing to effective solution volume.  Differentiating this same mean separates its change exactly into two simultaneous terms.
\begin{equation}
  \partial_\tau h
  =
  \mathbb E_{\mathcal D}
  \mathbb E_{\widetilde{\boldsymbol w}}
  \left[
    \langle\partial_\tau\kappa\rangle_\tau
    -\frac{P}{2}\operatorname{Var}_\tau(\kappa)
  \right].
  \label{eq:discussion-exact-hardening-balance}
\end{equation}
The first term tracks changes in the slopes of the currently weighted directions.  For the next outward step, a direction with smaller \(\kappa\) is more favorable because its loss rises less.  The nonpositive variance term records the resulting shift of weight toward such directions as the shell expands.  A rise in \(h\) is therefore nontrivial.  It means that the average outward response becomes steeper even while the weighting moves toward directions expected to be easier.

Equation~\eqref{eq:discussion-exact-hardening-balance} now tests whether the rise can be explained using only curvature at the reference.  For comparison with the usual local-Hessian description \cite{Sagun2017HessianOverparam}, we retain only the second-order Taylor term around a stationary reference, \[ \Delta\mathcal L_{\mathrm{quad}}(r,\boldsymbol u) = \frac{1}{2} \bigl(\sqrt{P}\,r\boldsymbol u\bigr)^{\mathsf T} H \bigl(\sqrt{P}\,r\boldsymbol u\bigr) = \frac{Pr^2}{2}\boldsymbol u^{\mathsf T}H\boldsymbol u, \] where \(H=\nabla_{\boldsymbol w}^{2} \mathcal L_{\mathrm{reg}}(\widetilde{\boldsymbol w};\mathcal D)\) is the Hessian at the reference.  For this baseline, \[ \mathcal A_{\mathrm{quad}}(\tau,\boldsymbol u) = \frac{\beta_{\mathrm{sh}}P\tau}{2} \boldsymbol u^{\mathsf T}H\boldsymbol u, \qquad \kappa_{\mathrm{quad}}(\boldsymbol u) = \beta_{\mathrm{sh}} \boldsymbol u^{\mathsf T}H\boldsymbol u. \] Hence \(\partial_\tau\kappa_{\mathrm{quad}}=0\).  If \(H\) is isotropic, the variance also vanishes and \(h\) is constant.  If \(H\) is anisotropic, the variance term makes \(h\) nonincreasing.  Dataset complexity may sharpen or distort the Hessian at the reference, but if that quadratic form remains fixed with radius, \(h\) cannot rise.  The observed complexity-dependent hardening therefore cannot be explained by local Hessian shaping alone.

\subsection{A possible geometric interpretation}

One plausible microscopic hypothesis is constraint crowding.  Close to the reference, many directions can remain comparably solution-like.  As the shell moves outward, restrictions associated with different examples and margins become costly at different distances.  If a more complex dataset brings many of these restrictions into conflict over a narrower displacement range, a large family of solution-like directions can disappear almost at once, leaving only a smaller corridor.  In Eq.~\eqref{eq:discussion-exact-hardening-balance}, this would appear as an increase of \(\kappa\) across many directions that still contribute to effective solution volume, making \(\langle\partial_\tau\kappa\rangle_\tau\) positive.  When this increase exceeds the opposing variance term, \(h\) rises and \(g_{\mathrm E}\) makes its sharp inner downward excursion.  Farther from the reference, the surviving corridors favor directions for which another outward step adds less loss. Newly encountered restrictions may also be spread over a broader radial range. The contraction can then weaken beyond the turning range.

\begin{figure*}[!t]
  \centering
  \includegraphics[width=0.99\textwidth]{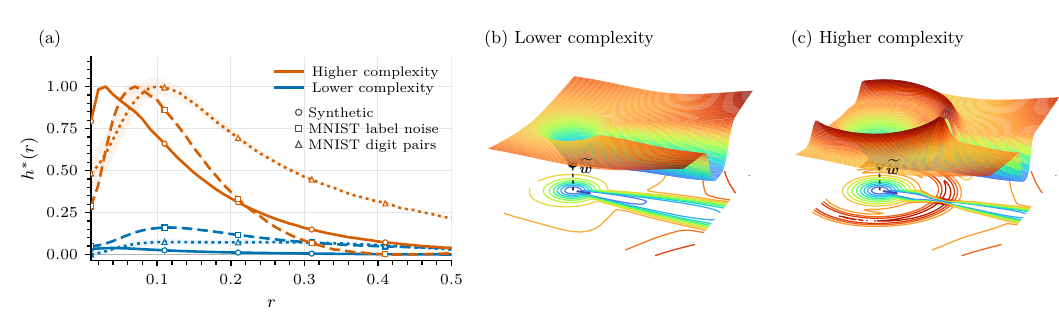}
\caption{ Dimensionless rescaled radial response and a schematic corridor interpretation. (a) The horizontal axis is shell radius \(r\), and the vertical axis is \(h^*=(h-h_{\min})/(h_{\max}-h_{\min})\), with \(h=-g_{\mathrm E}/r\). Each higher/lower pair uses joint extrema over the displayed \(r\) range and the same range for both standard-error bands.  This scaling compares radial shape and turning location, not absolute response magnitude. Panels (b) and (c) illustrate the constraint-crowding hypothesis.  The higher-complexity sketch places the main bottleneck closer to the reference and concentrates the loss of solution-like directions into a narrower radial interval. }
  \label{fig:hardening-geometry-composite}
\end{figure*}

For intuition, panels (b) and (c) of Fig.~\ref{fig:hardening-geometry-composite} turn this hypothesis into a corridor sketch.  The corridor represents directions that remain solution-like as the radius increases.  In the higher-complexity case, many such directions are lost over an earlier and narrower interval, producing the sharper bottleneck.

\subsection{Random limits and natural data}

Random-label or random-pattern systems are often used to make theoretical analyses tractable.  With the inputs fixed, however, our randomization changes the strength and radial scale of the inner bottleneck and also creates a sequence of signs in \(g_{\mathrm E}\) not observed in the selected natural digit-pair tasks.  Randomization can therefore introduce an unintended qualitative change in landscape geometry, rather than merely extending natural data to a uniformly harder regime.  A result derived in a random limit may describe this special geometry created by randomization.

Conversely, the much milder digit-pair regime raises a broader possibility. Part of the empirical success of modern neural networks may come from organization already present in real inputs and labels, which leaves a broader range of accurate solution candidates around a trained reference.  Whether this extends beyond the selected tasks, architecture, representation, and parameter metric remains open.

\section{Conclusion}
\label{sec:conclusion}

We examined how the structural complexity of a finite dataset appears in the finite-distance loss geometry around a trained neural-network solution.  We paired label mixing across scales with local entropy around each trained reference and estimated the latter using adaptive SMC.

The controlled synthetic study showed the clearest relation.  Greater label mixing produced a deeper inner contraction of effective solution volume, followed by a weaker outer response.  Holding the MNIST inputs fixed while randomizing their labels strengthened this bottleneck and moved it toward the reference.  Strong randomization also made \(g_{\mathrm E}\) change from negative to positive and then back to negative.  The natural digit-pair tasks occupied a milder regime whose main differences were concentrated in the inner radial range.

Further analysis shows that the observed nontrivial response reflects a radial reorganization of the loss landscape, with the loss response strengthening within the contributing directions faster than local entropy shifts weight toward easier ones.  A stationary fixed-Hessian Taylor baseline cannot generate the measured inner rise followed by a turning range.

Taken together, the results extend the usual flat--sharp description from one reference point to a sequence of radial scales.  Fitting a more complex dataset does not lead only to a scalar change from a flat to a sharp solution; it changes where radial contraction is concentrated and, in the strong random-label regime, can change the qualitative outer response.

Future work can use the same radial observables in paired interventions that separate changes induced by data organization from those induced by model architecture.  One can hold the data and training protocol fixed while changing an architectural feature, such as residual connections, or hold the architecture fixed while reorganizing the labels or representation.  Such comparisons can test which design choices preserve a broad effective solution volume, shift the bottleneck location, or change the outer reorganization.  We expect these paired finite-distance comparisons to support a more global and mechanistic understanding of how neural-network design shapes usable solution neighborhoods.

\begin{acknowledgments}
This work was supported by the Basic Science Research Program through the National Research Foundation of Korea (NRF Grant No. RS-2025-00514776).
\end{acknowledgments}

\appendix
\section{Adaptive SMC and controls}
\label{app:sampler-details}

\begin{table*}[t]
  \begin{center}
  \caption{Sampling controls used for the reported local-entropy estimates.  Every row uses
  direct-transition testing, systematic resampling, and a vMF random-walk
  concentration \(\kappa_{\mathrm{move}}\) equal to 80 times the parameter dimension.}
  \label{tab:app-secondary-smc-settings}
  \vspace{4pt}

  \begin{ruledtabular}
  \begin{tabular}{lcccc}
    System
    & \(\rho_\mathrm{step}\)
    & \(\rho_\mathrm{pool}\)
    & Min. \(\Delta t\)
    & MH sweeps\\
    \hline
    Synthetic DNN
    & 0.75 & 0.75 & \(10^{-4}\)  & 1\\
    MNIST label noise
    & 0.95 & 0.50 & \(10^{-4}\)  & 2\\
    MNIST digit pairs
    & 0.95 & 0.50 & \(10^{-4}\) & 2\\
    Perceptron
    & 0.85 & 0.85 & \(5\times10^{-5}\) & 2
  \end{tabular}
  \end{ruledtabular}
  \end{center}
\end{table*}

This section describes the adaptive sampler and records the numerical controls used for the reported local-entropy estimates.  References are first averaged within each dataset, and datasets are the independent units used for uncertainty estimates.

\subsection{Adaptive tempered particle construction}
\label{subsec:tempered-pool}

The remaining numerical quantity in Eq.~\eqref{eq:ge-importance-decomposition} is
\begin{equation}
  \mathcal Z_L(r)
  =
  \int_{\mathbb S^{P-1}}
  \mathrm d\sigma(\boldsymbol u)\,
  q_r(\boldsymbol u)
  e^{
    -\beta_{\mathrm{sh}}C_r(\boldsymbol u)
  } .
  \label{eq:loss-partition-integral}
\end{equation}
Although this is formally an expectation under \(q_r\), its direct finite-sample estimate can be dominated by a small number of directions with low loss.  We therefore replace the single transition from the unweighted case to the full loss weight by a sequence
\begin{equation}
  0=t_0<t_1<\cdots<t_K=1.
  \label{eq:temperature-sequence}
\end{equation}
Along this path, define
\begin{equation}
  \mathcal Z_L(t;r)
  =
  \int_{\mathbb S^{P-1}}
  \mathrm d\sigma(\boldsymbol u)\,
  q_r(\boldsymbol u)
  e^{
    -t\beta_{\mathrm{sh}}C_r(\boldsymbol u)
  } .
  \label{eq:tempered-loss-partition}
\end{equation}
Because \(\mathcal Z_L(0;r)=1\) and \(\mathcal Z_L(1;r)=\mathcal Z_L(r)\), we can compute the desired value as a product of smaller changes along this path \cite{Neal2001AnnealedImportance},
\begin{align}
  \mathcal Z_L(r)
  &=
  \prod_{k=0}^{K-1}
  \frac{
    \mathcal Z_L(t_{k+1};r)
  }{
    \mathcal Z_L(t_k;r)
  }
  \notag\\
  &=
  \prod_{k=0}^{K-1}
  \mathbb E_{\boldsymbol u\sim\pi_{t_k}}
  \left[
    e^{
      -\Delta t_k
      \beta_{\mathrm{sh}}C_r(\boldsymbol u)
    }
  \right],
  \label{eq:tempered-product-identity}
\end{align}
where
\begin{equation}
  \Delta t_k=t_{k+1}-t_k
\end{equation}
and
\begin{equation}
  \pi_t(\boldsymbol u)
  =
  \frac{
    q_r(\boldsymbol u)
    e^{
      -t\beta_{\mathrm{sh}}C_r(\boldsymbol u)
    }
  }{
    \mathcal Z_L(t;r)
  }
  \label{eq:tempered-target}
\end{equation}
is the normalized angular distribution at tempering value \(t\).

We approximate each \(\pi_{t_k}\) with a weighted particle set
\begin{equation}
  \mathcal P_k
  =
  \left\{
    \left(
      \boldsymbol u_i^{(k)},
      p_i^{(k)}
    \right)
  \right\}_{i=1}^{N_{\mathrm p}},
  \qquad
  \sum_i p_i^{(k)}=1,
  \label{eq:weighted-particle-pool}
\end{equation}
and use
\begin{equation}
  \mathbb E_{\pi_{t_k}}[F(\boldsymbol u)]
  \simeq
  \sum_i
  p_i^{(k)}
  F(\boldsymbol u_i^{(k)}).
  \label{eq:discrete-pool-approximation}
\end{equation}
Here \(p_i^{(k)}\) is the normalized probability assigned to particle \(\boldsymbol u_i^{(k)}\). The particle approximation gives
\begin{equation}
  \widehat{\mathcal Z}_L(r)
  =
  \prod_{k=0}^{K-1}
  \left[
    \sum_i
    p_i^{(k)}
    e^{
      -\Delta t_k\beta_{\mathrm{sh}}
      C_r(\boldsymbol u_i^{(k)})
    }
  \right].
  \label{eq:tempered-normalizer-estimate}
\end{equation}

At each \(t_k\), we reweight the particles, use overlap thresholds to set the tempering increment and resampling decision, and apply a Metropolis mutation \cite{DelMoral2006SMCSamplers,Zhou2016AdaptiveSMC}. The initial set is sampled from the known distribution \(q_r\),
\begin{equation}
  \boldsymbol u_i^{(0)}
  \overset{\mathrm{iid}}{\sim}q_r,
  \qquad
  p_i^{(0)}=\frac{1}{N_{\mathrm p}}.
  \label{eq:initial-particle-pool}
\end{equation}
For a tentative next temperature \(s\in(t_k,1]\), reweighting the current particle locations gives
\begin{equation}
  \widetilde p_i^{(k)}(s)
  =
  \frac{
    p_i^{(k)}
    e^{
      -(s-t_k)\beta_{\mathrm{sh}}
      C_r(\boldsymbol u_i^{(k)})
    }
  }{
    \displaystyle
    \sum_j
    p_j^{(k)}
    e^{
      -(s-t_k)\beta_{\mathrm{sh}}
      C_r(\boldsymbol u_j^{(k)})
    }
  }.
  \label{eq:tentative-particle-probability}
\end{equation}
These probabilities form the discrete approximation of \(\pi_s\) that would be obtained without changing the current particle locations.

To quantify whether this reweighted pool remains sufficiently well represented, we use
\begin{equation}
  \mathcal O
  (\boldsymbol b\Vert\boldsymbol a)
  =
  \left(
    \sum_i\frac{b_i^2}{a_i}
  \right)^{-1}
  =
  \frac{
    1
  }{
    1+\chi^2
    (\boldsymbol b\Vert\boldsymbol a)
  },
  \label{eq:particle-overlap}
\end{equation}
where
\begin{equation}
  \chi^2
  (\boldsymbol b\Vert\boldsymbol a)
  =
  \sum_i
  \frac{
    (b_i-a_i)^2
  }{
    a_i
  }
  \label{eq:particle-chi-square}
\end{equation}
is the Pearson \(\chi^2\) divergence.  The overlap is one when the probability vectors coincide and decreases as \(\boldsymbol b\) becomes concentrated relative to \(\boldsymbol a\).  With the current particle probabilities as the baseline, it is the normalized conditional effective sample size used to control an intermediate SMC transition.

We choose the largest tentative temperature for which the reweighted current locations retain the prescribed overlap,
\begin{multline}
  t_{k+1}
  =
  \max
  \Bigl\{
    s\in(t_k,1]:
    \mathcal O
    \bigl(
      \widetilde{\boldsymbol p}^{(k)}(s)
      \Vert
      \boldsymbol p^{(k)}
    \bigr)
    \ge
    \rho_{\mathrm{step}}
  \Bigr\}.
  \label{eq:adaptive-temperature-selection}
\end{multline}
If a proposed value of \(s\) would assign appreciable probability to only a few of the current particle locations, the temperature increment is reduced.  The factor inside the corresponding bracket in Eq.~\eqref{eq:tempered-normalizer-estimate} is recorded before any resampling.

After accepting \(t_{k+1}\), we set
\begin{equation}
  \widetilde p_i^{(k+1)}
  =
  \widetilde p_i^{(k)}(t_{k+1}).
\end{equation}
Even when every individual transition satisfies the step-overlap condition, the particle probabilities carried across several transitions can gradually become uneven.  We therefore also compare the accumulated weights with the equal-weight vector \(e_i=1/N_{\mathrm p}\),
\begin{align}
  \mathcal O_{\mathrm{pool}}^{(k+1)}
  &=
  \mathcal O
  \left(
    \widetilde{\boldsymbol p}^{(k+1)}
    \Vert
    \boldsymbol e
  \right)
  \notag\\
  &=
  \frac{
    1
  }{
    N_{\mathrm p}
    \sum_i
    \left(
      \widetilde p_i^{(k+1)}
    \right)^2
  }.
  \label{eq:pool-overlap}
\end{align}
This is the fraction of the nominal particle number represented by the current weights.  If
\begin{equation}
  \mathcal O_{\mathrm{pool}}^{(k+1)}
  <
  \rho_{\mathrm{pool}},
  \label{eq:pool-resampling-condition}
\end{equation}
we apply systematic resampling according to \(\widetilde{\boldsymbol p}^{(k+1)}\) and reset all probabilities to \(1/N_{\mathrm p}\).  Otherwise, the particle locations and their updated probabilities are retained.  The step overlap therefore controls whether the current locations can represent the next target, while the pool overlap determines whether the accumulated weighted representation should be reconstructed.

After each accepted temperature transition, we apply a local Metropolis--Hastings mutation, irrespective of whether resampling was performed.  Denote the particle locations after the optional resampling by \(\overline{\boldsymbol u}_i\).  We draw a proposal from
\begin{equation}
  \boldsymbol v_i
  \sim
  \operatorname{vMF}
  \left(
    \overline{\boldsymbol u}_i,
    \kappa_{\mathrm{move}}
  \right).
  \label{eq:local-vmf-mutation}
\end{equation}
The vMF variates are generated with Wood's sampler \cite{Wood1994SimulationVMF}. Up to an additive constant, the log density of the tempered target is
\begin{equation}
  \Psi_t(\boldsymbol u)
  =
  \log q_r(\boldsymbol u)
  -
  t\beta_{\mathrm{sh}}C_r(\boldsymbol u).
  \label{eq:tempered-log-target}
\end{equation}
The symmetric proposal is accepted with probability
\begin{equation}
  \alpha_i
  =
  \min
  \left\{
    1,\,
    e^{
      \Psi_{t_{k+1}}(\boldsymbol v_i)
      -
      \Psi_{t_{k+1}}
      (\overline{\boldsymbol u}_i)
    }
  \right\}.
  \label{eq:compact-mh-acceptance}
\end{equation}
The mutation leaves \(\pi_{t_{k+1}}\) invariant and does not modify the particle probabilities.  It provides local exploration of the new tempered distribution and can restore diversity among copies produced by resampling.

We repeat these steps until \(t_K=1\).  Equation~\eqref{eq:tempered-normalizer-estimate} then gives \(\widehat{\mathcal Z}_L(r)\).  Substitution into Eq.~\eqref{eq:ge-importance-decomposition} gives \(\widehat{\phi}_{\mathrm E,\beta}^{(P)} (r\mid\widetilde{\boldsymbol w},\mathcal D)\).

\subsection{Numerical controls}

Table~\ref{tab:app-secondary-smc-settings} collects the adaptive controls used by the samplers in the reported calculations.

For each independently seeded pool \(s\in\{1,2\}\), Eq.~\eqref{eq:tempered-normalizer-estimate} gives a numerical estimate \(\widehat Z_s\equiv\widehat{\mathcal Z}_{L,s}(r)\) of the loss-partition factor in Eq.~\eqref{eq:loss-partition-factor}.  To reduce sensitivity to one atypical realization of the sampling path, we use
\begin{equation}
  \widehat Z
  =
  \frac{\widehat Z_1+\widehat Z_2}{2}.
  \label{eq:app-two-pool-normalizer}
\end{equation}
To calculate an expectation from the terminal particles, such as the radial response in Eq.~\eqref{eq:direct-radial-response}, we combine the two pool estimates using the same normalizers,
\begin{equation}
  \widehat A
  =
  \frac{
    \widehat Z_1\widehat A_1+\widehat Z_2\widehat A_2
  }{
    \widehat Z_1+\widehat Z_2
  }.
  \label{eq:app-two-pool-terminal}
\end{equation}

The DNN calculations use two independently seeded pools of 512 particles.  Their normalizers and terminal expectations are combined using Eqs.~\eqref{eq:app-two-pool-normalizer} and \eqref{eq:app-two-pool-terminal}.  At each \(r\), particles are initialized from the vMF importance distribution in Eq.~\eqref{eq:l2-vmf-proposal} and evolved using the controls in Table~\ref{tab:app-secondary-smc-settings}.

\subsection{Radial derivative estimator}
\label{app:radial-score-estimator}

We evaluate \(g_{\mathrm E}(r)\) directly from the terminal SMC particles rather than from finite differences between neighboring grid points.  Since \(\partial_r\boldsymbol w_r(\boldsymbol u)=\sqrt P\,\boldsymbol u\),
\begin{equation}
  g_{\mathrm E,\beta}^{(P)}
  (r\mid\widetilde{\boldsymbol w},\mathcal D)
  =
  -\frac{\beta_{\mathrm{sh}}}{P}
  \mathbb E_{\pi_1}
  \left[
    \nabla_{\boldsymbol w}\mathcal L_{\mathrm{reg}}
    \bigl(\boldsymbol w_r(\boldsymbol u);\mathcal D\bigr)
    \mathbin{\cdot}\sqrt P\,\boldsymbol u
  \right].
  \label{eq:direct-radial-response}
\end{equation}
Here \(\pi_1\) is the final loss-weighted angular distribution in Eq.~\eqref{eq:tempered-target}.  Thus \(g_{\mathrm E}(r)\) is evaluated from the particles at radius \(r\), without differentiating an interpolated local entropy curve.  We check that all evaluated directional derivatives are finite.  Statistical uncertainty is computed across dataset replicates.

\section{Perceptron benchmark saddle}
\label{app:replica}
\label{app:replica-saddle}

The main text defines the measure over exact references.  It also states the replica-symmetric shell functional and its constrained extremization in Eqs.~\eqref{eq:benchmark-shell-functional}--\eqref{eq:benchmark-final-rs}. Here we give the reference saddle and the numerical parameterization used to select the shell extremum.  The detailed replica construction is provided in the Supplementary Material.

\subsection{Reference saddle}

Using the replica method, we introduce formal copies of the reference.  The index \(a\) labels these copies, which serve as bookkeeping variables for the normalized reference average.  The reference sector is described by its squared norm per parameter,
\begin{equation}
  \widetilde Q
  =
  \frac{
    \|\widetilde{\boldsymbol w}\|_2^2
  }{N},
\end{equation}
and by the overlap \(q_{\mathrm{ref}}\) between two replicated reference directions.  We separate the norm and direction of copy \(a\) by writing \(\widetilde{\boldsymbol w}^{\,a} =\sqrt{\widetilde Q}\,\widetilde{\boldsymbol v}^{\,a}\), with \(\|\widetilde{\boldsymbol v}^{\,a}\|_2^2=N\).  Thus \(\widetilde{\boldsymbol v}^{\,a}\) is the normalized direction of that copy, and the replica-symmetric ansatz is
\begin{equation}
  \frac{
    \widetilde{\boldsymbol v}^{\,a}
    \cdot
    \widetilde{\boldsymbol v}^{\,b}
  }{N}
  =
  \delta_{ab}
  +
  (1-\delta_{ab})q_{\mathrm{ref}} .
\end{equation}

Under this ansatz, the reference free-entropy density is
\begin{equation}
  \phi_{\mathrm{ref}}^{\mathrm{RS}}
  =
  \underset{
    \widetilde Q>0,\;0<q_{\mathrm{ref}}<1
  }{\operatorname{extr}}
  \mathcal F_{\mathrm{ref}}^{\mathrm{RS}}
  (\widetilde Q,q_{\mathrm{ref}}),
  \label{eq:benchmark-reference-rs}
\end{equation}
where
\begin{equation}
  \mathcal F_{\mathrm{ref}}^{\mathrm{RS}}
  (\widetilde Q,q_{\mathrm{ref}})
  =
  f_{\mathrm{rad}}^{\mathrm{ref}}(\widetilde Q)
  +
  G_{\mathrm S}^{\mathrm{ref}}(q_{\mathrm{ref}})
  +
  \alpha G_{\mathrm E}^{\mathrm{ref}}(q_{\mathrm{ref}}).
  \label{eq:benchmark-reference-functional}
\end{equation}
Thus the many exact reference solutions are summarized by two macroscopic quantities, their squared norm \(\widetilde Q\) and the typical directional similarity \(q_{\mathrm{ref}}\) between two solutions drawn from the same reference ensemble. The radial contribution is
\begin{equation}
  f_{\mathrm{rad}}^{\mathrm{ref}}(\widetilde Q)
  =
  \frac{1}{2}\log\widetilde Q
  -
  \frac{\lambda_{\mathrm{ref}}}{2}\widetilde Q,
  \label{eq:benchmark-reference-radial}
\end{equation}
the directional entropic contribution is
\begin{equation}
  G_{\mathrm S}^{\mathrm{ref}}(q_{\mathrm{ref}})
  =
  \frac{1}{2(1-q_{\mathrm{ref}})}
  +
  \frac{1}{2}\log(1-q_{\mathrm{ref}})
  +
  \frac{1}{2}\log(2\pi),
  \label{eq:benchmark-reference-entropic}
\end{equation}
and the contribution from the exact classification constraints is
\begin{equation}
  G_{\mathrm E}^{\mathrm{ref}}(q_{\mathrm{ref}})
  =
  \int Dz\,
  \log
  H\!\left(
    -\sqrt{
      \frac{q_{\mathrm{ref}}}{1-q_{\mathrm{ref}}}
    }\,z
  \right),
  \label{eq:benchmark-reference-energetic}
\end{equation}
where \(Dz=e^{-z^2/2}\mathrm dz/\sqrt{2\pi}\) and \(H(x)=\int_x^\infty Dz\). The three contributions in Eq.~\eqref{eq:benchmark-reference-functional} respectively describe the radial scale of the reference weights, the available angular parameter space, and the restriction imposed by exact classification of the random patterns.

We denote by \((\widetilde Q_\star,q_{\mathrm{ref}})\) the selected stationary point satisfying
\begin{align}
  \left.
  \partial_{\widetilde Q}
  \mathcal F_{\mathrm{ref}}^{\mathrm{RS}}(\widetilde Q,q)
  \right|_{\widetilde Q=\widetilde Q_\star,\,q=q_{\mathrm{ref}}}
  &=0,
  \notag\\
  \left.
  \partial_q
  \mathcal F_{\mathrm{ref}}^{\mathrm{RS}}(\widetilde Q,q)
  \right|_{\widetilde Q=\widetilde Q_\star,\,q=q_{\mathrm{ref}}}
  &=0.
  \label{eq:benchmark-reference-saddle}
\end{align}

\subsection{Feasible shell parameterization}

At the selected reference saddle, \(\widetilde Q=\widetilde Q_\star=\lambda_{\mathrm{ref}}^{-1}\) and \(q=q_{\mathrm{ref}}\).  These values are held fixed in the shell calculation. For fixed \(r>0\), the admissible shell norm satisfies
\begin{equation}
  \left(\sqrt{\widetilde Q_\star}-r\right)^2
  <Q<
  \left(\sqrt{\widetilde Q_\star}+r\right)^2,
  \label{eq:app-geometric-q-domain}
\end{equation}
and the selected-reference overlap is
\begin{equation}
  c=c_d(Q,r)
  =
  \frac{Q+\widetilde Q_\star-r^2}
       {2\sqrt{Q\widetilde Q_\star}}.
  \label{eq:app-distance-cosine}
\end{equation}

To enforce the covariance constraints effectively, introduce \(\chi\in(-1,1)\) and \(\zeta\in[0,1)\), and write
\begin{align}
  t(Q,\chi)
  &=q_{\mathrm{ref}}c
  +\chi\sqrt{
    q_{\mathrm{ref}}(1-q_{\mathrm{ref}})(1-c^2)
  },
  \label{eq:app-feasible-t}\\
  p(Q,\chi,\zeta)
  &=c^2+\chi^2(1-c^2)
  +\zeta(1-c^2)(1-\chi^2).
  \label{eq:app-feasible-p}
\end{align}
The feasibility remainder in Eq.~\eqref{eq:benchmark-feasibility-remainder} and the upper overlap margin then become
\begin{align}
  A
  &=\zeta(1-c^2)(1-\chi^2),
  \notag\\
  1-p
  &=(1-\zeta)(1-c^2)(1-\chi^2).
  \label{eq:app-feasible-identities}
\end{align}
Thus \(A\geq0\) and \(p<1\) are satisfied by construction.  The face \(\zeta=0\) gives \(A=0\), whereas \(0<\zeta<1\) covers the feasible interior.

Let \(\mathcal I_r\) denote the interval in Eq.~\eqref{eq:app-geometric-q-domain} and \(\mathcal C=(-1,1)\times[0,1)\).  For each radius, continuation and independent multistart searches locate feasible stationary candidates. The plotted curve uses the retained RS local entropy results,
\begin{equation}
  \begin{aligned}
  \Phi_{\mathrm{RS}}(r)
  &=\max_{Q\in\mathcal I_r}
    \min_{(\chi,\zeta)\in\mathcal C}
  \\
  &\quad
  \mathcal F_{\mathrm{RS}}
  \bigl(Q,p(Q,\chi,\zeta),t(Q,\chi);r\bigr).
  \end{aligned}
  \label{eq:app-physical-saddle}
\end{equation}
Interior solutions have vanishing derivatives in \((Q,\chi,\zeta)\).

\paragraph{Numerical evaluation.}

The analytic curve uses
\begin{align}
  \alpha&=0.1,
  &q_{\mathrm{ref}}&\simeq0.0633,
  \notag\\
  \beta_{\mathrm p}
  &=\lambda_{\mathrm{ref}}
  =\lambda_{\mathrm p}/P=1.
  \label{eq:app-analytic-contract}
\end{align}
We evaluated the constrained saddle from \(r=0.15\) to \(2.20\) in steps of \(0.05\).  At each \(r\), continuation and independent multistart searches recovered the same saddle.  It remained stable under finer quadrature and satisfied the stationary or one-sided conditions and the curvature signs required by Eq.~\eqref{eq:app-physical-saddle}.  These checks establish numerical convergence of the chosen RS extremum.

\paragraph{Finite-size comparison.}

The finite calculation uses \(N\in\{40,80,160,320\}\) and \(r_0=0.15\). For every \(N\), we generated 10 Gaussian-pattern datasets and retained 10 hard references per dataset.  At each \(r\), the angular SMC calculation used \(2^{15}=32768\) particles divided into two independently seeded pools of \(2^{14}\) particles. After first averaging the ten references within each dataset and then averaging equally over datasets, sampled values of \(r\), and \(N\), the mean absolute split difference \( |\log\widehat Z_1-\log\widehat Z_2|/N \) was \(2.23\times10^{-4}\).  Together with the agreement with the independently computed RS result, this close agreement between pools further supports the numerical reliability of our sampling procedure.

\section{Reference and radial settings}
\label{app:reference-radius-settings}

For the synthetic networks, reference searches used Adam with learning rate \(10^{-3}\) for at most 6,000 epochs.  Attempts unresolved by Adam continued with the same regularized loss under L-BFGS for at most 3,000 iterations, with at most 200 fresh attempts per dataset.  Both MNIST protocols used Adam with learning rate \(0.022\) for at most 4,200 steps, followed when needed by strong-Wolfe L-BFGS for at most 50 outer cycles of 20 iterations, with at most 240 fresh attempts per dataset--condition combination.  A reference was accepted at the first inspected checkpoint whose full-training evaluation satisfied Eq.~\eqref{eq:exact-reference-condition}.  For each dataset and condition, the first ten finite, mutually distinct accepted endpoints were retained.  The synthetic shell grid was \(r=0.01,0.02,\ldots,2.50\), and both MNIST grids were \(r=0.01,0.02,\ldots,1.00\), with \(r_0=0.01\).  The radial-variation summary uses the common window \(0.01\le r\le1.00\).

\section{Symmetric check of reference tilt}
\label{app:finite-distance-audit}

The retained MNIST references achieved zero training error but were not certified stationary points of the regularized loss.  A residual gradient could therefore make one side of a direction easier than the other.  To cancel this one-sided effect without moving or retraining the reference, we evaluated equal displacements on both sides of each sampled direction.
\begin{equation}
  \begin{aligned}
  K_r(\boldsymbol u)
  =
  \frac{\beta_{\mathrm{sh}}}{Pr^2}
  \Bigl[
    &\mathcal L_{\mathrm{reg}}
    (\widetilde{\boldsymbol w}+\sqrt P\,r\boldsymbol u)+\mathcal L_{\mathrm{reg}}
    (\widetilde{\boldsymbol w}-\sqrt P\,r\boldsymbol u)
    \\
    &-2\mathcal L_{\mathrm{reg}}
    (\widetilde{\boldsymbol w})
  \Bigr].
  \label{eq:appendix-symmetric-secant}
  \end{aligned}
\end{equation}
Because the losses at \(+\boldsymbol u\) and \(-\boldsymbol u\) are added before averaging, the linear contribution of the reference gradient, and more generally every contribution that changes sign under \(\boldsymbol u\to-\boldsymbol u\), cancels.

We evaluated Eq.~\eqref{eq:appendix-symmetric-secant} on the retained clean \(\eta=0\) and random-label \(\eta=0.5\) MNIST references.  The two conditions used the same ten sets of image coordinates, with ten retained references per condition and dataset.  For each reference, we drew 256 fresh direction pairs and evaluated \(r\in\{0.01,0.03,0.07,0.10,0.12,0.20,0.40\}\).  Directions and references were averaged within each dataset, and the ten datasets were used as the independent units for uncertainty estimates.

Relative to \(r=0.01\), \(K_r\) increased over the inner range in every dataset under both label conditions.  The clean-label response reached a mean maximum of \(K_r-K_{0.01}=0.417\) at \(r=0.20\), whereas the random-label response reached \(3.508\) at the earlier radius \(r=0.12\). Thus finite-distance hardening remains in the symmetric two-sided response and is substantially amplified by label randomization, showing that a residual first-order reference tilt cannot by itself explain the observed inner increase of \(h\).


\newpage
\bibliography{references}

\providecommand{\dd}{\,\mathrm d}
\providecommand{\E}{\mathbb E}
\providecommand{\R}{\mathbb R}
\providecommand{\1}{\mathbf 1}
\providecommand{\Prob}{\mathbb P}
\providecommand{\Ncal}{\mathcal N}
\providecommand{\Dcal}{\mathcal D}
\providecommand{\Tr}{\operatorname{Tr}}
\providecommand{\extr}{\operatorname*{extr}}
\providecommand{\Var}{\operatorname{Var}}
\providecommand{\Cov}{\operatorname{Cov}}
\providecommand{\Qt}{\widetilde Q}
\providecommand{\vt}{\widetilde v}
\providecommand{\wt}{\widetilde{\boldsymbol w}}
\providecommand{\xvec}{\boldsymbol x}
\providecommand{\wvec}{\boldsymbol w}
\providecommand{\uvec}{\boldsymbol u}
\providecommand{\vvec}{\widetilde{\boldsymbol v}}
\providecommand{\htilde}{\widetilde h}
\providecommand{\Htail}{H}
\providecommand{\phig}{\varphi}
\providecommand{\doteqN}{\doteq}

\clearpage
\onecolumngrid
\begingroup
\setcounter{equation}{0}
\setcounter{subsection}{0}
\setcounter{table}{0}
\setcounter{figure}{0}
\renewcommand{\theequation}{S\arabic{equation}}
\renewcommand{\thesubsection}{S\arabic{subsection}}
\renewcommand{\thetable}{S\arabic{table}}
\renewcommand{\thefigure}{S\arabic{figure}}
\renewcommand{\theHequation}{supp.equation.\arabic{equation}}
\renewcommand{\theHsubsection}{supp.subsection.\arabic{subsection}}
\renewcommand{\theHtable}{supp.table.\arabic{table}}
\renewcommand{\theHfigure}{supp.figure.\arabic{figure}}

\begin{center}
  {\Large\bfseries Supplementary Material}\par
\end{center}
\vspace{10pt}


\subsection{Model, reference ensemble, and target local entropy}
\label{app:rs-model_local_entropy}

Let the dataset be
\begin{equation}
\Dcal_N=\{(\xvec^\mu,y^\mu)\}_{\mu=1}^{M},
\qquad
\xvec^\mu\in\R^N,
\qquad
 y^\mu\in\{-1,+1\},
\qquad
M=\lfloor \alpha N\rfloor .
\label{eq:app-rs-data_set_def}
\end{equation}
The perceptron model output is
\begin{equation}
 f_{\wvec}(\xvec^\mu)
 =
 \frac{\xvec^\mu\cdot \wvec}{\sqrt N},
 \qquad
 \wvec\in\R^N ,
\label{eq:app-rs-perceptron_score}
\end{equation}
and the signed margin is
\begin{equation}
 y^\mu f_{\wvec}(\xvec^\mu)
 =
 \frac{y^\mu \xvec^\mu\cdot \wvec}{\sqrt N}.
\label{eq:app-rs-signed_margin_x}
\end{equation}
Define the label-absorbed Gaussian pattern
\begin{equation}
 \boldsymbol\xi^\mu := y^\mu \xvec^\mu,
 \qquad
 \boldsymbol\xi^\mu\sim\Ncal(\boldsymbol 0,I_N),
 \qquad
 \mu=1,\ldots,M .
\label{eq:app-rs-xi_def}
\end{equation}
With this notation the single-pattern field is
\begin{equation}
 h^\mu(\wvec)
 :=
 \frac{\boldsymbol\xi^\mu\cdot \wvec}{\sqrt N}
 =
 y^\mu f_{\wvec}(\xvec^\mu),
\label{eq:app-rs-hmu_def}
\end{equation}
and successful classification of pattern $\mu$ is the event $h^\mu(\wvec)>0$.

To measure soft violations of the classification constraints we use a single-pattern loss.  A canonical choice is logistic cross-entropy,
\begin{equation}
 \ell(h)=\log(1+e^{-h}).
\label{eq:app-rs-logistic_loss_def}
\end{equation}

Here the sum over $M=\lfloor\alpha N\rfloor$ patterns, $\sum_{\mu=1}^{M}\ell(h^\mu(\wvec))$, contributes at order $N$ and therefore competes with the entropic terms in the large-$N$ saddle.

The shell inverse temperature is $\beta_{\mathrm p}$, and $\lambda_{\mathrm{sh}}$ denotes the intensive coefficient of the quadratic shell penalty. For the benchmark evaluated in the main text, $\beta_{\mathrm p}=\lambda_{\mathrm{sh}}=1$.

The reference configuration is drawn from the exact-classification sector, with an $L^2$ prior that makes the radial integral normalizable. The unnormalized reference weight is
\begin{equation}
W_{\rm ref}(\wt;\Dcal_N)=
 \exp\!\left[-\frac{\lambda_{\mathrm{ref}}}{2}\|\wt\|_2^2\right]
 \prod_{\mu=1}^{M}\Theta(h^\mu(\wt)),
\label{eq:app-rs-Wref_def}
\end{equation}
where $\Theta(z)=\1\{z>0\}$.  The reference partition function is
\begin{equation}
 Z_{\rm ref}^0(\Dcal_N)
 :=
 \int_{\R^N}\dd^N\wt\, W_{\rm ref}(\wt;\Dcal_N),
\label{eq:app-rs-Zref_def}
\end{equation}
and the corresponding probability density is
\begin{equation}
 p_{\mathrm{ref}}^0(\wt\mid\Dcal_N)
 :=
 \frac{W_{\rm ref}(\wt;\Dcal_N)}{Z_{\rm ref}^0(\Dcal_N)}.
\label{eq:app-rs-Pref_def}
\end{equation}

Given a selected reference $\wt$, the local object of interest is the Gibbs mass of parameter vectors $\wvec$ lying at squared distance $Nr^2$ from $\wt$.  The shell is hard in distance and soft in loss.  Its mass is
\begin{align}
 \Omega(r\mid\wt,\Dcal_N)
 &:={}
 \int_{\R^N}\dd^N\wvec\,
 \exp\!\left[
 -\beta_{\mathrm p}\sum_{\mu=1}^{M}\ell(h^\mu(\wvec))
 -\frac{\beta_{\mathrm p}\lambda_{\mathrm{sh}}}{2}\|\wvec\|_2^2
 \right]
 \delta\!\left(Nr^2-\|\wvec-\wt\|_2^2\right).
\label{eq:app-rs-Omega_def}
\end{align}
The reference-conditioned soft Franz--Parisi observable considered here is 
\begin{align}
 \Phi(r)
 &:=
 \lim_{N\to\infty}
 \frac1N
 \E_{\Dcal_N}\E_{\wt\sim p_{\mathrm{ref}}^0(\cdot\mid\Dcal_N)}
 \log\Omega(r\mid\wt,\Dcal_N)
\label{eq:app-rs-Phi_def_a}\\
 &=
 \lim_{N\to\infty}
 \frac1N
 \E_{\Dcal_N}\left[
 \frac{1}{Z_{\rm ref}^0(\Dcal_N)}
 \int_{\R^N}\dd^N\wt\,
 W_{\rm ref}(\wt;\Dcal_N)
 \log\Omega(r\mid\wt,\Dcal_N)
 \right].
\label{eq:app-rs-Phi_def_b}
\end{align}
The reference is sampled from a hard exact-solution ensemble, while the shell is weighted by the soft loss $\ell$ and by the $L^2$ shell energy.
\subsection{Replica construction and reference-sector decomposition}
\label{app:rs-replica_full_object}

For $Z_{\rm ref}^0(\Dcal_N)>0$, the reference-normalization factor is represented formally as
\begin{equation}
 \frac{1}{Z_{\rm ref}^0(\Dcal_N)}
 =
 \lim_{n\to0}\left[Z_{\rm ref}^0(\Dcal_N)\right]^{n-1}.
\label{eq:app-rs-replica_inverse_Zref}
\end{equation}
For $\Omega(r\mid\wt,\Dcal_N)>0$,
\begin{equation}
 \log\Omega(r\mid\wt,\Dcal_N)
 =
 \left.\frac{\partial}{\partial s}\Omega(r\mid\wt,\Dcal_N)^s\right|_{s=0}.
\label{eq:app-rs-replica_log_Omega}
\end{equation}
The calculation first treats $n$ and $s$ as positive integers and only afterwards performs the replica continuations $n\to0$ and $s\to0$.  This continuation is the formal replica step.

For integers $n\ge1$ and $s\ge1$, define the replicated partition
\begin{align}
 Z_{n,s}(r)
 &:={}
 \E_{\Dcal_N}
 \int
 \left[\prod_{a=1}^{n}\dd^N\wt^a\right]
 \left[\prod_{a=1}^{n}W_{\rm ref}(\wt^a;\Dcal_N)\right]
 \Omega(r\mid\wt^1,\Dcal_N)^s .
\label{eq:app-rs-Zns_compact}
\end{align}
Expanding \eqref{eq:app-rs-Zns_compact} using \eqref{eq:app-rs-Wref_def} and \eqref{eq:app-rs-Omega_def} gives
\begin{align}
 Z_{n,s}(r)
 &=
 \E_{\Dcal_N}
 \int
 \left[\prod_{a=1}^{n}\dd^N\wt^a\right]
 \left[
 \prod_{a=1}^{n}
 \exp\!\left(-\frac{\lambda_{\mathrm{ref}}}2\|\wt^a\|_2^2\right)
 \prod_{\mu=1}^{M}\Theta(h^\mu(\wt^a))
 \right]
\notag\\
&\quad\times
 \int
 \left[\prod_{\gamma=1}^{s}\dd^N \wvec^\gamma\right]
 \left[
 \prod_{\gamma=1}^{s}
 \exp\!\left(
 -\beta_{\mathrm p}\sum_{\mu=1}^{M}\ell(h^\mu(\wvec^\gamma))
 -\frac{\beta_{\mathrm p}\lambda_{\mathrm{sh}}}2\|\wvec^\gamma\|_2^2
 \right)
 \right]
\notag\\
&\quad\times
 \left[
 \prod_{\gamma=1}^{s}
 \delta\!\left(Nr^2-\|\wvec^\gamma-\wt^1\|_2^2\right)
 \right].
\label{eq:app-rs-Zns_full}
\end{align}
The selected reference in the original conditional average is represented by the replica $a=1$.
The formal local entropy density is then computed by
\begin{equation}
 \Phi(r)
 =
 \lim_{N\to\infty}\frac1N
 \lim_{n\to0}
 \left.\frac{\partial}{\partial s}\log Z_{n,s}(r)\right|_{s=0}.
\label{eq:app-rs-Phi_from_Zns}
\end{equation}
When $s=0$, no shell replicas remain, so the shell block $P$ and mixed block
$T$ are absent and the joint Gram matrix $\mathcal M$ contains only the reference block
$R$.  As a result, \eqref{eq:app-rs-Zns_compact} reduces to the reference-only
replicated object
\begin{equation}
 Z_{n,0}
 =
 \E_{\Dcal_N}\left[\left(Z_{\rm ref}^0(\Dcal_N)\right)^n\right].
\label{eq:app-rs-Zn0_reference_only}
\end{equation}
Since $Z_{n,0}$ is independent of $s$,
\begin{equation}
 \left.\frac{\partial}{\partial s}\log Z_{n,s}(r)\right|_{s=0}
 =
 \left.\frac{\partial}{\partial s}\log\frac{Z_{n,s}(r)}{Z_{n,0}}\right|_{s=0}.
\label{eq:app-rs-conditional_ratio_derivative}
\end{equation}
This identity is often useful when isolating the shell contribution from the reference normalization.

We next isolate the reference-sector quantities that will be passed to the shell calculation.  The reference partition \eqref{eq:app-rs-Zref_def} is
\begin{equation}
 Z_{\rm ref}^0(\Dcal_N)
 =
 \int_{\R^N}\dd^N\wt\,
 \exp\!\left[-\frac{\lambda_{\mathrm{ref}}}2\|\wt\|_2^2\right]
 \prod_{\mu=1}^{M}
 \Theta\!\left(\frac{\boldsymbol\xi^\mu\cdot\wt}{\sqrt N}\right).
\label{eq:app-rs-Zref_expanded}
\end{equation}
Define the radial variable and normalized direction
\begin{equation}
 \Qt:=\frac{\|\wt\|_2^2}{N},
 \qquad
 \wt=\sqrt{\Qt}\,\vvec,
 \qquad
 \|\vvec\|_2^2=N,
 \qquad
 \Qt>0.
\label{eq:app-rs-Qtilde_vtilde_def}
\end{equation}
Let
\begin{equation}
 \rho:=\|\wt\|_2=\sqrt{N\Qt},
 \qquad
 n_{\wt}:=\frac{\wt}{\|\wt\|_2},
 \qquad
 n_{\wt}\in S^{N-1},
 \qquad
 \vvec=\sqrt N\,n_{\wt}.
\label{eq:app-rs-rho_n_def}
\end{equation}
Let $\dd\Omega(n_{\wt})$ denote the angular surface element on the unit sphere $S^{N-1}$.  The Euclidean volume element is
\begin{equation}
 \dd^N\wt
 =
 \rho^{N-1}\dd\rho\,\dd\Omega(n_{\wt}).
\label{eq:app-rs-polar_measure_wtilde}
\end{equation}
From \eqref{eq:app-rs-rho_n_def},
\begin{equation}
 \dd\rho
 =
 \frac{\sqrt N}{2\sqrt{\Qt}}\dd\Qt,
\label{eq:app-rs-drho_dQtilde}
\end{equation}
so
\begin{equation}
 \rho^{N-1}\dd\rho
 =
 \frac{N^{N/2}}2\Qt^{N/2-1}\dd\Qt.
\label{eq:app-rs-rho_jacobian_Qtilde}
\end{equation}
Let $\dd\sigma(\vvec)$ denote the surface element on the sphere $\{\vvec\in\R^N:\|\vvec\|_2^2=N\}$.  Since $\vvec=\sqrt N n_{\wt}$,
\begin{equation}
 \dd\sigma(\vvec)
 =
 N^{(N-1)/2}\dd\Omega(n_{\wt}).
\label{eq:app-rs-dsigma_dOmega}
\end{equation}
Equations \eqref{eq:app-rs-polar_measure_wtilde}--\eqref{eq:app-rs-dsigma_dOmega} give
\begin{equation}
 \dd^N\wt
 =
 \frac{\sqrt N}{2}\Qt^{N/2-1}\dd\Qt\,\dd\sigma(\vvec).
\label{eq:app-rs-dNwtilde_exact_radial}
\end{equation}
At the level of $N^{-1}\log Z$, subexponential factors are omitted.  We write
\begin{equation}
 \dd^N\wt
 \doteqN
 \dd\Qt\,\dd\sigma(\vvec)\,\Qt^{N/2-1}.
\label{eq:app-rs-dNwtilde_exp_order}
\end{equation}

For $\Qt>0$, the zero-margin classification constraints are invariant under positive rescaling,
\begin{equation}
 \Theta\!\left(\frac{\boldsymbol\xi^\mu\cdot\wt}{\sqrt N}\right)
 =
 \Theta\!\left(\sqrt{\Qt}\frac{\boldsymbol\xi^\mu\cdot\vvec}{\sqrt N}\right)
 =
 \Theta\!\left(\frac{\boldsymbol\xi^\mu\cdot\vvec}{\sqrt N}\right).
\label{eq:app-rs-scale_invariance_theta}
\end{equation}
Substituting \eqref{eq:app-rs-dNwtilde_exp_order} and \eqref{eq:app-rs-scale_invariance_theta} into \eqref{eq:app-rs-Zref_expanded},
\begin{equation}
 Z_{\rm ref}^0(\Dcal_N)
 \doteqN
 Z_{\rm rad}^{\rm ref}\,Z_{\rm dir}^{\rm ref}(\Dcal_N),
\label{eq:app-rs-Zref_rad_dir_factor}
\end{equation}
where
\begin{equation}
 Z_{\rm rad}^{\rm ref}
 :=
 \int_{0}^{\infty}\dd\Qt\,
 \Qt^{N/2-1}
 \exp\!\left[-\frac{N\lambda_{\mathrm{ref}}}2\Qt\right],
\label{eq:app-rs-Zrad_ref_def}
\end{equation}
\begin{equation}
 Z_{\rm dir}^{\rm ref}(\Dcal_N)
 :=
 \int_{\|\vvec\|_2^2=N}\dd\sigma(\vvec)
 \prod_{\mu=1}^{M}
 \Theta\!\left(\frac{\boldsymbol\xi^\mu\cdot\vvec}{\sqrt N}\right).
\label{eq:app-rs-Zdir_ref_def}
\end{equation}
The radial factor is deterministic and independent of the disorder, while the disorder dependence and the nontrivial overlap structure remain in $Z_{\rm dir}^{\rm ref}(\Dcal_N)$.
The exponent in \eqref{eq:app-rs-Zrad_ref_def} is
\begin{equation}
 \left(\frac N2-1\right)\log\Qt-\frac{N\lambda_{\mathrm{ref}}}2\Qt
 =
 N\left[\frac12\log\Qt-\frac{\lambda_{\mathrm{ref}}}2\Qt\right]-\log\Qt.
\label{eq:app-rs-radial_exponent_ref}
\end{equation}
The leading reference radial action is
\begin{equation}
 f_{\mathrm{rad}}^{\rm ref}(\Qt)
 :=
 \frac12\log\Qt-\frac{\lambda_{\mathrm{ref}}}2\Qt.
\label{eq:app-rs-f_rad_ref_def}
\end{equation}
The saddle equation and curvature are
\begin{equation}
 \frac{\partial f_{\mathrm{rad}}^{\rm ref}}{\partial\Qt}
 =
 \frac1{2\Qt}-\frac{\lambda_{\mathrm{ref}}}2,
 \qquad
 \frac{\partial^2 f_{\mathrm{rad}}^{\rm ref}}{\partial\Qt^2}
 =
 -\frac1{2\Qt^2}<0.
\label{eq:app-rs-f_rad_ref_derivatives}
\end{equation}
Thus the unique radial maximum is
\begin{equation}
 \Qt_\star=\lambda_{\mathrm{ref}}^{-1}.
\label{eq:app-rs-Qtilde_star}
\end{equation}
The radial contribution at the saddle is
\begin{equation}
 f_{\mathrm{rad}}^{\rm ref}(\Qt_\star)
 =
 -\frac12(1+\log\lambda_{\mathrm{ref}}).
\label{eq:app-rs-f_rad_ref_star}
\end{equation}

It remains to compute the directional contribution \eqref{eq:app-rs-Zdir_ref_def}.  Unlike the radial factor, this term depends on the quenched Gaussian patterns.  For integer $n\ge1$, the disorder-averaged replicated directional partition is
\begin{align}
 \E_{\Dcal_N}\left[(Z_{\rm dir}^{\rm ref})^n\right]
 &=
 \int
 \left[\prod_{a=1}^{n}\dd\sigma(\vvec^a)\right]
 \left[
 \E_{\boldsymbol\xi}
 \prod_{a=1}^{n}
 \Theta\!\left(\frac{\boldsymbol\xi\cdot\vvec^a}{\sqrt N}\right)
 \right]^M .
\label{eq:app-rs-ED_Zdir_n_start}
\end{align}
For replicas $\{\vvec^a\}_{a=1}^{n}$, define
\begin{equation}
 \htilde^a
 :=
 \frac{\boldsymbol\xi\cdot\vvec^a}{\sqrt N},
 \qquad
 a=1,\ldots,n.
\label{eq:app-rs-h_tilde_a_def}
\end{equation}
Since the pattern is Gaussian, the vector $\htilde=(\htilde^1,\ldots,\htilde^n)^T$ is also centered Gaussian with covariance
\begin{align}
 \E_{\boldsymbol\xi}[\htilde^a\htilde^b]
 &=
 \E_{\boldsymbol\xi}\left[
 \frac1{\sqrt N}\sum_{i=1}^{N}(\boldsymbol\xi)_i\vvec_i^a
 \frac1{\sqrt N}\sum_{j=1}^{N}(\boldsymbol\xi)_j\vvec_j^b
 \right]
\notag\\
 &=
 \frac1N\sum_{i,j=1}^{N}\vvec_i^a\vvec_j^b\E_{\boldsymbol\xi}[(\boldsymbol\xi)_i(\boldsymbol\xi)_j]
 =
 \frac1N\sum_{i=1}^{N}\vvec_i^a\vvec_i^b.
\label{eq:app-rs-cov_h_tilde_derivation}
\end{align}
Define the replica overlap matrix
\begin{equation}
 q_{ab}
 :=
 \frac1N\vvec^a\cdot\vvec^b
 =
 \frac1N\sum_{i=1}^{N}\vvec_i^a\vvec_i^b.
\label{eq:app-rs-qab_def}
\end{equation}
Then the vector $\htilde$ is Gaussian with covariance matrix
\begin{equation}
 \htilde\sim\Ncal(0,q),
 \qquad
 q=(q_{ab})_{a,b=1}^{n}.
\label{eq:app-rs-htilde_gaussian_q}
\end{equation}
For a positive-definite covariance matrix $q$, define the one-pattern reference-satisfaction probability
\begin{align}
 P_n(q)
 &:=
 \E_{\htilde\sim\Ncal(0,q)}
 \prod_{a=1}^{n}\Theta(\htilde^a)
\label{eq:app-rs-Pn_q_def_a}\\
 &=
 \int_{\R^n}
 \frac{\dd^n h}{(2\pi)^{n/2}(\det q)^{1/2}}
 \exp\!\left[-\frac12h^Tq^{-1}h\right]
 \prod_{a=1}^{n}\Theta(h^a).
\label{eq:app-rs-Pn_q_def_b}
\end{align}
Using \eqref{eq:app-rs-Pn_q_def_a}, \eqref{eq:app-rs-ED_Zdir_n_start} becomes
\begin{equation}
 \E_{\Dcal_N}\left[(Z_{\rm dir}^{\rm ref})^n\right]
 =
 \int
 \left[\prod_{a=1}^{n}\dd\sigma(\vvec^a)\right]
 \left[P_n(q(\mathbf V))\right]^M,
\label{eq:app-rs-ED_Zdir_n_Pn}
\end{equation}
where $\mathbf V=(\vvec^1,\ldots,\vvec^n)$ denotes the collection of replicated directions and
\begin{equation}
 q(\mathbf V)_{ab}
 =
 \frac1N\vvec^a\cdot\vvec^b.
\label{eq:app-rs-q_of_V_def}
\end{equation}

The surface measure can be represented, at exponential order, by an ambient coordinate integral with norm constraints.  For each replica,
\begin{equation}
 \dd\sigma(\vvec^a)
 \doteqN
 \left[\prod_{i=1}^{N}\dd\vvec_i^a\right]
 \delta\!\left(N-\sum_{i=1}^{N}(\vvec_i^a)^2\right).
\label{eq:app-rs-dsigma_delta_norm}
\end{equation}
The diagonal constraints in \eqref{eq:app-rs-dsigma_delta_norm} impose $q_{aa}=1$ in \eqref{eq:app-rs-qab_def}.  For compact notation, insert an overlap resolution of identity for all pairs $(a,b)$.
\begin{equation}
 1=
 \int\dd q_{ab}\,
 \delta\!\left(Nq_{ab}-\sum_{i=1}^{N}\vvec_i^a\vvec_i^b\right).
\label{eq:app-rs-overlap_delta_identity_ref}
\end{equation}
The delta representation is taken on a suitable complex contour.
\begin{equation}
 \delta(x)
 =
 \int_{\mathcal C}\frac{\dd\hat q}{2\pi i}
 \exp\!\left[-\frac12\hat q x\right].
\label{eq:app-rs-delta_conjugate_identity}
\end{equation}
Using \eqref{eq:app-rs-delta_conjugate_identity},
\begin{align}
&\prod_{a,b}
 \delta\!\left(Nq_{ab}-\sum_{i=1}^{N}\vvec_i^a\vvec_i^b\right)
\notag\\
&\quad=
 \int\left[\prod_{a,b}\frac{\dd\hat q_{ab}}{2\pi i}\right]
 \exp\!\left[
 -\frac N2\sum_{a,b}\hat q_{ab}q_{ab}
 +\frac12\sum_{i=1}^{N}\sum_{a,b}\hat q_{ab}\vvec_i^a\vvec_i^b
 \right].
\label{eq:app-rs-delta_product_conjugate}
\end{align}
Substituting \eqref{eq:app-rs-overlap_delta_identity_ref} and \eqref{eq:app-rs-delta_product_conjugate} into \eqref{eq:app-rs-ED_Zdir_n_Pn},
\begin{align}
 \E_{\Dcal_N}\left[(Z_{\rm dir}^{\rm ref})^n\right]
 &\doteqN
 \int
 \left[\prod_{a,b}\dd q_{ab}\,\dd\hat q_{ab}\right]
 \exp\!\left[-\frac N2\sum_{a,b}\hat q_{ab}q_{ab}\right]
 \left[P_n(q)\right]^M
\notag\\
&\quad\times
 \int
 \left[\prod_{i=1}^{N}\prod_{a=1}^{n}\dd\vvec_i^a\right]
 \exp\!\left[
 \frac12\sum_{i=1}^{N}
 \left(
 \sum_{a,b}\hat q_{ab}\vvec_i^a\vvec_i^b
 \right)
 \right].
\label{eq:app-rs-Zdir_replicated_before_factor}
\end{align}
The last integral factorizes over coordinates $i$,
\begin{align}
&\int
 \left[\prod_{i=1}^{N}\prod_{a=1}^{n}\dd\vvec_i^a\right]
 \exp\!\left[
 \frac12\sum_{i=1}^{N}
 \left(
 \sum_{a,b}\hat q_{ab}\vvec_i^a\vvec_i^b
 \right)
 \right]
\notag\\
&\quad=
 \prod_{i=1}^{N}
 \left[
 \int
 \left[\prod_{a=1}^{n}\dd\vt^a\right]
 \exp\!\left(
 \frac12\sum_{a,b}\hat q_{ab}\vt^a\vt^b
 \right)
 \right]
\notag\\
&\quad=
 \left[
 \int_{\R^n}\dd^n\vt\,
 \exp\!\left(\frac12\sum_{a,b}\hat q_{ab}\vt^a\vt^b\right)
 \right]^N.
\label{eq:app-rs-coordinate_factorization_ref}
\end{align}
In \eqref{eq:app-rs-coordinate_factorization_ref}, the variables $\vt^a$ in the single-site integral are scalar coordinate variables, not full $N$-dimensional vectors. 
Define the shorthand
\begin{equation}
 \int\dd q\,\dd\hat q
 :=
 \int\left[\prod_{a,b}\dd q_{ab}\,\dd\hat q_{ab}\right].
\label{eq:app-rs-dq_dqhat_shorthand}
\end{equation}
Then \eqref{eq:app-rs-Zdir_replicated_before_factor}--\eqref{eq:app-rs-coordinate_factorization_ref} give
\begin{equation}
 \E_{\Dcal_N}[(Z_{\rm dir}^{\rm ref})^n]
 \doteqN
 \int\dd q\,\dd\hat q\,
 \exp\!\left
 \{
 N\left[G_{\mathrm S}^{(n)}(q,\hat q)+\alpha G_{\mathrm E}^{(n)}(q)\right]
 \right\},
\label{eq:app-rs-Zdir_action_general}
\end{equation}
where $M/N\to\alpha$ and
\begin{equation}
 G_{\mathrm S}^{(n)}(q,\hat q)
 :=
 -\frac12\sum_{a,b}\hat q_{ab}q_{ab}
 +
 \log
 \int_{\R^n}\dd^n\vt\,
 \exp\!\left(\frac12\sum_{a,b}\hat q_{ab}\vt^a\vt^b\right),
\label{eq:app-rs-GS_n_general}
\end{equation}
\begin{equation}
 G_{\mathrm E}^{(n)}(q)
 :=
 \log P_n(q),
\label{eq:app-rs-GE_n_general}
\end{equation}
with $P_n(q)$ defined in \eqref{eq:app-rs-Pn_q_def_a}--\eqref{eq:app-rs-Pn_q_def_b}.
The shell observable \eqref{eq:app-rs-Phi_from_Zns} is not a reference free entropy by itself.  Nevertheless, the reference sector supplies two inputs to the shell saddle, the radial norm $\Qt_\star$ and the selected reference overlap $q_{\mathrm{ref}}$.  The latter is obtained from the formal quenched reference continuation
\begin{equation}
 \extr_{q,\hat q}
 \left[
 \lim_{n\to0}\frac1n \left(G_{\mathrm S}^{(n)}(q,\hat q)+\alpha G_{\mathrm E}^{(n)}(q)\right)\right].
\label{eq:app-rs-phi_dir_replica_general}
\end{equation}

\subsection{Replica-symmetric reference saddle}
\label{app:rs-GS_reference_RS}
We evaluate the reference-sector saddle with the replica-symmetric ansatz. The overlap matrix is
\begin{equation}
 q_{ab}
 =
 \delta_{ab}+(1-
\delta_{ab})q.
\label{eq:app-rs-RS_q_ref_ansatz}
\end{equation}
The conjugate RS ansatz is
\begin{equation}
 \hat q_{ab}
 =
 -\widehat Q\,\delta_{ab}
 +(1-\delta_{ab})\widehat r.
\label{eq:app-rs-RS_qhat_ref_ansatz}
\end{equation}
The variables $\widehat Q$ and $\widehat r$ are auxiliary conjugate saddle variables.
Substituting \eqref{eq:app-rs-RS_qhat_ref_ansatz} into the quadratic form in \eqref{eq:app-rs-GS_n_general},
\begin{align}
 \frac12\sum_{a,b}\hat q_{ab}\vt^a\vt^b
 &=
 \frac12\sum_a(-\widehat Q)(\vt^a)^2
 +
 \frac12\sum_{a\ne b}\widehat r\,\vt^a\vt^b
\notag\\
 &=
 -\frac12(\widehat Q+\widehat r)
 \sum_{a=1}^{n}(\vt^a)^2
 +
 \frac{\widehat r}{2}
 \left(\sum_{a=1}^{n}\vt^a\right)^2 .
\label{eq:app-rs-RS_quadratic_form_ref}
\end{align}
Use the Hubbard-Stratonovich identity
\begin{equation}
 \exp\!\left(\frac{\widehat r}{2}X^2\right)
 =
 \int Dz\,\exp\!\left(\sqrt{\widehat r}\,zX\right),
 \qquad
 Dz:=\frac{e^{-z^2/2}}{\sqrt{2\pi}}\dd z,
\label{eq:app-rs-HS_identity}
\end{equation}
and define
\begin{align}
 I_n(\widehat Q,\widehat r)
 &:={}
 \int_{\R^n}\dd^n\vt\,
 \exp\!\left[
 -\frac12(\widehat Q+\widehat r)
 \sum_{a=1}^{n}(\vt^a)^2
 +
 \frac{\widehat r}{2}
 \left(\sum_{a=1}^{n}\vt^a\right)^2
 \right]
\label{eq:app-rs-In_ref_def_a}\\
 &=
 \int Dz
 \left[
 \int_{\R}\dd\vt\,
 \exp\!\left(-\frac12(\widehat Q+\widehat r)\vt^2+\sqrt{\widehat r}\,z\vt\right)
 \right]^n
\label{eq:app-rs-In_ref_def_b}\\
 &=
 \int Dz
 \left[
 \sqrt{\frac{2\pi}{\widehat Q+\widehat r}}
 \exp\!\left(\frac{\widehat r z^2}{2(\widehat Q+\widehat r)}\right)
 \right]^n .
\label{eq:app-rs-In_ref_def_c}
\end{align}
Consequently,
\begin{equation}
 \lim_{n\to0}\frac1n\log I_n(\widehat Q,\widehat r)
 =
 \frac12\log\frac{2\pi}{\widehat Q+\widehat r}
 +
 \frac12\frac{\widehat r}{\widehat Q+\widehat r}.
\label{eq:app-rs-In_limit_ref}
\end{equation}
The conjugate term satisfies
\begin{align}
 -\frac1{2n}\sum_{a,b}\hat q_{ab}q_{ab}
 &=
 -\frac1{2n}
 \left[
 \sum_a(-\widehat Q)(1)
 +
 \sum_{a\ne b}\widehat r q
 \right]
\notag\\
 &=
 -\frac1{2n}
 \left[-n\widehat Q+n(n-1)\widehat r q\right].
\label{eq:app-rs-conjugate_limit_ref}
\end{align}
Taking $n\to0$ gives
\begin{equation}
 \lim_{n\to0}
 \left(-\frac1{2n}\sum_{a,b}\hat q_{ab}q_{ab}\right)
 =
 \frac12\widehat Q+\frac12q\widehat r.
\label{eq:app-rs-conjugate_limit_ref_final}
\end{equation}
Combining \eqref{eq:app-rs-In_limit_ref} and \eqref{eq:app-rs-conjugate_limit_ref_final}, define the $n\to0$ site entropy before eliminating the conjugates.
\begin{equation}
 G_{\mathrm S}(q,\widehat Q,\widehat r)
 :=
 \frac12\widehat Q
 +
 \frac12q\widehat r
 +
 \frac12\log\frac{2\pi}{\widehat Q+\widehat r}
 +
 \frac12\frac{\widehat r}{\widehat Q+\widehat r}.
\label{eq:app-rs-GS_q_Qhat_rhat_ref}
\end{equation}
The full reference directional saddle is obtained by imposing stationarity in $q$, $\widehat Q$, and $\widehat r$ on
\begin{equation}
 G_{\mathrm S}(q,\widehat Q,\widehat r)+\alpha G_{\mathrm E}^{\rm ref}(q).
\label{eq:app-rs-reference_total_RS_action_before_conjugates}
\end{equation}
The energetic term does not depend on $\widehat Q$ or $\widehat r$, so their stationarity conditions are
\begin{align}
 0&=\partial_{\widehat Q}G_{\mathrm S}
 =
 \frac12-\frac{1}{2(\widehat Q+\widehat r)}-\frac{\widehat r}{2(\widehat Q+\widehat r)^2},
\label{eq:app-rs-stationarity_Qhat_ref}\\
0&=\partial_{\widehat r}G_{\mathrm S}
 =
 \frac q2-\frac{1}{2(\widehat Q+\widehat r)}+\frac{\widehat Q}{2(\widehat Q+\widehat r)^2}.
\label{eq:app-rs-stationarity_rhat_ref}
\end{align}
Solving \eqref{eq:app-rs-stationarity_Qhat_ref}--\eqref{eq:app-rs-stationarity_rhat_ref} gives
\begin{equation}
 \widehat r
 =
 \frac{q}{(1-q)^2},
 \qquad
 \widehat Q
 =
 \frac{1-2q}{(1-q)^2},
 \qquad
 \widehat Q+
 \widehat r
 =
 \frac1{1-q}.
\label{eq:app-rs-qhat_solution_ref}
\end{equation}
Substituting \eqref{eq:app-rs-qhat_solution_ref} into \eqref{eq:app-rs-GS_q_Qhat_rhat_ref} gives the reference site entropy
\begin{equation}
 G_{\mathrm S}^{\rm ref}(q)
 =
 \frac1{2(1-q)}
 +
 \frac12\log(1-q)
 +
 \frac12\log(2\pi).
\label{eq:app-rs-GS_ref_final}
\end{equation}

We now evaluate the one-pattern reference factor under the RS covariance \eqref{eq:app-rs-RS_q_ref_ansatz}.  Let
\begin{equation}
 z_0,z_1,\ldots,z_n\stackrel{\rm iid}{\sim}\Ncal(0,1).
\label{eq:app-rs-z_variables_ref}
\end{equation}
Then the Gaussian field vector \eqref{eq:app-rs-htilde_gaussian_q} can be represented by
\begin{equation}
 \htilde^a
 =
 \sqrt q\,z_0+
 \sqrt{1-q}\,z_a,
 \qquad
 a=1,\ldots,n.
\label{eq:app-rs-htilde_RS_representation}
\end{equation}
Indeed,
\begin{equation}
 \E[\htilde^a\htilde^b]
 =
 q+(1-q)\delta_{ab}.
\label{eq:app-rs-htilde_RS_cov_check}
\end{equation}
Define the Gaussian upper tail and density
\begin{equation}
 \Htail(x):=\int_x^\infty Dz,
 \qquad
 \phig(x):=\frac{e^{-x^2/2}}{\sqrt{2\pi}}.
\label{eq:app-rs-Htail_phi_def}
\end{equation}
Using \eqref{eq:app-rs-htilde_RS_representation}, \eqref{eq:app-rs-Pn_q_def_a} becomes
\begin{align}
 P_n(q)
 &=
 \int Dz_0
 \prod_{a=1}^{n}
 \int Dz_a\,
 \Theta(\sqrt q\,z_0+
 \sqrt{1-q}\,z_a)
\notag\\
 &=
 \int Dz_0
 \left[
 \Htail\!\left(-\sqrt{\frac{q}{1-q}}z_0\right)
 \right]^n .
\label{eq:app-rs-Pn_RS_ref}
\end{align}
Thus
\begin{equation}
 G_{\mathrm E}^{\rm ref}(q)
 :=
 \lim_{n\to0}\frac1n\log P_n(q)
 =
 \int Dz\,
 \log\Htail\!\left(-\sqrt{\frac{q}{1-q}}z\right).
\label{eq:app-rs-GE_ref_final}
\end{equation}
This is the energetic contribution of the exact reference constraints.  For later differentiation define
\begin{equation}
 a_z
 :=
 -\sqrt{\frac{q}{1-q}}z,
 \qquad
 R(a_z)
 :=
 \frac{\phig(a_z)}{\Htail(a_z)}.
\label{eq:app-rs-az_R_def}
\end{equation}
From \eqref{eq:app-rs-GS_ref_final},
\begin{equation}
 \frac{\dd}{\dd q}G_{\mathrm S}^{\rm ref}(q)
 =
 \frac{q}{2(1-q)^2}.
\label{eq:app-rs-dGS_ref_dq}
\end{equation}
For $G_{\mathrm E}^{\rm ref}$, first note that
\begin{equation}
 \frac{\dd a_z}{\dd q}
 =
 -\frac{z}{2\sqrt q(1-q)^{3/2}}.
\label{eq:app-rs-daz_dq}
\end{equation}
Since $\Htail'(x)=-\phig(x)$,
\begin{equation}
 \frac{\dd}{\dd q}G_{\mathrm E}^{\rm ref}(q)
 =
 -\int Dz\,R(a_z)\frac{\dd a_z}{\dd q}
 =
 \frac1{2\sqrt q(1-q)^{3/2}}
 \int Dz\,zR(a_z).
\label{eq:app-rs-dGE_before_IBP}
\end{equation}
Gaussian integration by parts gives
\begin{equation}
 \int Dz\,zR(a_z)
 =
 \int Dz\,\frac{\dd}{\dd z}R(a_z).
\label{eq:app-rs-IBP_gaussian}
\end{equation}
Using
\begin{equation}
 \frac{\dd a_z}{\dd z}
 =
 -\sqrt{\frac{q}{1-q}},
 \qquad
 R'(a)=-aR(a)+R(a)^2,
\label{eq:app-rs-Rprime_def}
\end{equation}
one obtains, with $I:=\int Dz\,zR(a_z)$,
\begin{equation}
 I
 =
 -\sqrt{\frac{q}{1-q}}
 \int Dz\,[-a_zR(a_z)+R(a_z)^2]
 =
 -\frac{q}{1-q}I
 -\sqrt{\frac{q}{1-q}}
 \int Dz\,R(a_z)^2.
\label{eq:app-rs-zR_integral_intermediate}
\end{equation}
Solving for $I$ gives
\begin{equation}
 \int Dz\,zR(a_z)
 =
 -\sqrt{q(1-q)}
 \int Dz\,R(a_z)^2.
\label{eq:app-rs-zR_integral_result}
\end{equation}
Substituting \eqref{eq:app-rs-zR_integral_result} into \eqref{eq:app-rs-dGE_before_IBP} yields
\begin{equation}
 \frac{\dd}{\dd q}G_{\mathrm E}^{\rm ref}(q)
 =
 -\frac1{2(1-q)}
 \int Dz\,
 \left[\frac{\phig(a_z)}{\Htail(a_z)}\right]^2.
\label{eq:app-rs-dGE_ref_dq}
\end{equation}

Combining \eqref{eq:app-rs-Zref_rad_dir_factor}, \eqref{eq:app-rs-f_rad_ref_star}, \eqref{eq:app-rs-GS_ref_final}, and \eqref{eq:app-rs-GE_ref_final}, the RS prediction for the reference free-entropy density is
\begin{equation}
 \phi_{\rm ref}^{\rm RS}
 =
 -\frac12(1+\log\lambda_{\mathrm{ref}})
 +
 \extr_{0<q<1}
 \left[
 G_{\mathrm S}^{\rm ref}(q)+\alpha G_{\mathrm E}^{\rm ref}(q)
 \right].
\label{eq:app-rs-phi_ref_RS_final}
\end{equation}
The scalar value of \eqref{eq:app-rs-phi_ref_RS_final} is not the final shell observable.  Its role here is to fix the reference-sector order parameters used by the reference-conditioned shell calculation.

The reference radial saddle is
\begin{equation}
 \Qt_\star
 =
 \lambda_{\mathrm{ref}}^{-1}.
\label{eq:app-rs-Qtilde_star_final}
\end{equation}
The interior reference overlap saddle $q_{\mathrm{ref}}\in(0,1)$ is a selected RS stationary point of
\begin{equation}
 \frac{\dd}{\dd q}
 \left[
 G_{\mathrm S}^{\rm ref}(q)+\alpha G_{\mathrm E}^{\rm ref}(q)
 \right]
 =0.
\label{eq:app-rs-qref_stationarity_abstract}
\end{equation}
Using \eqref{eq:app-rs-dGS_ref_dq} and \eqref{eq:app-rs-dGE_ref_dq}, \eqref{eq:app-rs-qref_stationarity_abstract} is equivalently
\begin{equation}
 \frac{q}{(1-q)^2}
 =
 \frac{\alpha}{1-q}
 \int Dz\,
 \left[
 \frac{\phig(a_z)}{\Htail(a_z)}
 \right]^2,
 \qquad
 a_z=-\sqrt{\frac q{1-q}}z.
\label{eq:app-rs-qref_saddle_equation}
\end{equation}
Define the set of interior RS reference stationary points by
\begin{equation}
 \mathcal S_{\rm ref}^{\rm RS}
 :=
 \left\{
 q\in(0,1):
 \frac{q}{(1-q)^2}
 =
 \frac{\alpha}{1-q}
 \int Dz\,
 \left[
 \frac{\phig(a_z)}{\Htail(a_z)}
 \right]^2,
 \quad
 a_z=-\sqrt{\frac q{1-q}}z
 \right\}.
\label{eq:app-rs-Sref_RS_def}
\end{equation}
The selected reference branch is denoted by $q_{\mathrm{ref}}$ and satisfies
\begin{equation}
 q_{\mathrm{ref}}\in\mathcal S_{\rm ref}^{\rm RS}.
\label{eq:app-rs-qref_def_final}
\end{equation}
Within the selected RS description, $q_{\mathrm{ref}}$ is interpreted as the typical overlap of two reference directions drawn from the same hard reference ensemble.
\begin{equation}
 q_{\mathrm{ref}}
 =
 \lim_{N\to\infty}\frac1N\vvec^a\cdot\vvec^b,
 \qquad
 a\ne b,
\label{eq:app-rs-qref_overlap_interpretation}
\end{equation}
within the RS interpretation.  Equations \eqref{eq:app-rs-Qtilde_star_final} and \eqref{eq:app-rs-qref_def_final} are the reference order parameters passed to the shell calculation.  In the shell sector one sets
\begin{equation}
 \Qt_\star=\lambda_{\mathrm{ref}}^{-1},
 \qquad
 q=q_{\mathrm{ref}},
\label{eq:app-rs-reference_params_pass_to_shell}
\end{equation}
and then optimizes over the shell variables subject to their shell feasibility conditions.
\subsection{Shell geometry and order parameters}
\label{app:rs-shell_radial_decomp_distance}

The reference radial saddle imported in \eqref{eq:app-rs-reference_params_pass_to_shell} gives
\begin{equation}
 \wt^a=\sqrt{\Qt_\star}\,\vvec^a,
 \qquad
 \|\vvec^a\|_2^2=N,
 \qquad
 a=1,\ldots,n.
\label{eq:app-rs-shell_wtilde_vtilde_replicas}
\end{equation}

Before imposing any shell-replica symmetry, a fully replicated radial decomposition would assign an independent norm density to each shell replica,
\begin{equation}
 \wvec^\gamma=\sqrt{Q_\gamma}\,\uvec^\gamma,
 \qquad
 \|\uvec^\gamma\|_2^2=N,
 \qquad
 Q_\gamma>0,
 \qquad
 \gamma=1,\ldots,s.
\label{eq:app-rs-shell_w_gamma_radial_general_def}
\end{equation}

The fixed-distance constraint is imposed separately for each $\gamma$ against the same selected reference $\wt^1$.  Therefore it fixes
\begin{equation}
 \frac{\uvec^\gamma\cdot\vvec^1}{N}
 =
 c_d(Q_\gamma,r)
 :=
 \frac{Q_\gamma+\Qt_\star-r^2}{2\sqrt{Q_\gamma\Qt_\star}}.
\label{eq:app-rs-shell_cd_Qgamma_general}
\end{equation}
The distance constraint fixes each replica's distance from the selected
reference, not its distance from the origin.  Points on a sphere centered at
$\wt^1$ can therefore have different norms $Q_\gamma$.  Setting all
\(Q_\gamma=Q\) is an additional RS assumption that restores permutation
symmetry among shell replicas,
\begin{equation}
 Q_\gamma=Q,
 \qquad
 \gamma=1,\ldots,s.
\label{eq:app-rs-shell_Qgamma_RS_ansatz}
\end{equation}
After this ansatz we write, throughout the rest of the shell calculation,
\begin{equation}
 \wvec^\gamma=\sqrt{Q}\,\uvec^\gamma,
 \qquad
 \|\uvec^\gamma\|_2^2=N,
 \qquad
 \gamma=1,\ldots,s,
 \qquad
 Q>0.
\label{eq:app-rs-shell_w_gamma_radial_def}
\end{equation}
A more general radial-RSB construction would keep the full collection $(Q_1,\ldots,Q_s)$ and would impose $T_{1\gamma}=c_d(Q_\gamma,r)$ before any symmetry reduction. 

With the reference normalization represented by $n$ reference replicas and the shell logarithm by $s$ shell replicas, the replicated shell object is

\begin{align}
 Z_{n,s}(r)
 &=
 \E_{\Dcal_N}
 \int
 \left[\prod_{a=1}^{n}\dd^N\wt^a\right]
 \left[
 \prod_{a=1}^{n}
 \exp\!\left(-\frac{\lambda_{\mathrm{ref}}}2\|\wt^a\|_2^2\right)
 \prod_{\mu=1}^{M}\Theta(h^\mu(\wt^a))
 \right]
\notag\\
&\quad\times
 \int
 \left[\prod_{\gamma=1}^{s}\dd^N \wvec^\gamma\right]
 \left[
 \prod_{\gamma=1}^{s}
 \exp\!\left(
 -\beta_{\mathrm p}\sum_{\mu=1}^{M}\ell(h^\mu(\wvec^\gamma))
 -\frac{\beta_{\mathrm p}\lambda_{\mathrm{sh}}}2\|\wvec^\gamma\|_2^2
 \right)
 \right]
\notag\\
&\quad\times
 \left[
 \prod_{\gamma=1}^{s}
 \delta\!\left(Nr^2-\|\wvec^\gamma-\wt^1\|_2^2\right)
 \right].
\end{align}
We now isolate the shell-dependent radial and angular factors in the second integral.  At exponential order,
\begin{equation}
 \dd^N\wvec^\gamma
 \doteqN
 \dd Q\,\dd\sigma(\uvec^\gamma)\,Q^{N/2-1}.
\label{eq:app-rs-shell_dNw_radial_measure}
\end{equation}
Combining the shell radial Jacobian with the $L^2$ shell energy, and then imposing the RS norm ansatz \eqref{eq:app-rs-shell_Qgamma_RS_ansatz}, gives
\begin{align}
 \prod_{\gamma=1}^{s}\dd^N\wvec^\gamma
 \prod_{\gamma=1}^{s}
 \exp\!\left[-\frac{\beta_{\mathrm p}\lambda_{\mathrm{sh}}}2\|\wvec^\gamma\|_2^2\right]
 &\doteqN
 \dd Q\,
 \exp\!\left
 \{Ns\left[\frac12\log Q-\frac{\beta_{\mathrm p}\lambda_{\mathrm{sh}}}2Q\right]\right\}
 \prod_{\gamma=1}^{s}\dd\sigma(\uvec^\gamma).
\label{eq:app-rs-shell_radial_l2_factor}
\end{align}
Define
\begin{equation}
 f_{\mathrm{rad}}^{\rm shell}(Q)
 :=
 \frac12\log Q-\frac{\beta_{\mathrm p}\lambda_{\mathrm{sh}}}2Q.
\label{eq:app-rs-shell_frad_def}
\end{equation}
The physical shell field is
\begin{equation}
 h^\mu(\wvec^\gamma)
 =
 \frac{\boldsymbol\xi^\mu\cdot \wvec^\gamma}{\sqrt N}
 =
 \sqrt Q\frac{\boldsymbol\xi^\mu\cdot \uvec^\gamma}{\sqrt N}.
\label{eq:app-rs-shell_physical_field_radial}
\end{equation}
Thus $Q$ appears not only in \eqref{eq:app-rs-shell_frad_def}, but also in the loss argument in \eqref{eq:app-rs-shell_physical_field_radial}.

For the distance constraint, use \eqref{eq:app-rs-shell_wtilde_vtilde_replicas} and \eqref{eq:app-rs-shell_w_gamma_radial_def}.
\begin{align}
 \|\wvec^\gamma-\wt^1\|_2^2
 &=
 \|\sqrt Q\,\uvec^\gamma-\sqrt{\Qt_\star}\,\vvec^1\|_2^2
\notag\\
 &=
 Q\|\uvec^\gamma\|_2^2+\Qt_\star\|\vvec^1\|_2^2
 -2\sqrt{Q\Qt_\star}\,\uvec^\gamma\cdot\vvec^1
\notag\\
 &=
 N\left[Q+\Qt_\star-2\sqrt{Q\Qt_\star}\,\frac{\uvec^\gamma\cdot\vvec^1}{N}\right].
\label{eq:app-rs-shell_distance_expansion}
\end{align}
Define the distance cosine
\begin{equation}
 c_d(Q,r)
 :=
 \frac{Q+\Qt_\star-r^2}{2\sqrt{Q\Qt_\star}}.
\label{eq:app-rs-shell_cd_def}
\end{equation}
Then \eqref{eq:app-rs-shell_distance_expansion} implies
\begin{equation}
 \frac{\uvec^\gamma\cdot\vvec^1}{N}=c_d(Q,r).
\label{eq:app-rs-shell_distance_cosine_constraint}
\end{equation}
The exact delta transformation is
\begin{align}
 \delta\!\left(Nr^2-\|\sqrt Q\,\uvec^\gamma-\sqrt{\Qt_\star}\,\vvec^1\|_2^2\right)
 &=
 \frac{1}{2\sqrt{Q\Qt_\star}}
 \delta\!\left(\uvec^\gamma\cdot\vvec^1-Nc_d(Q,r)\right).
\label{eq:app-rs-shell_distance_delta_transform}
\end{align}
The prefactor in \eqref{eq:app-rs-shell_distance_delta_transform} contributes only an additive $o(N)$ term to $\log Z_{n,s}$ and is omitted in exponential-order formulae.

Substituting \eqref{eq:app-rs-shell_wtilde_vtilde_replicas}--\eqref{eq:app-rs-shell_distance_delta_transform} into $Z_{n,s}(r)$,
\begin{align}
 Z_{n,s}(r)
 &\doteqN
 \int_0^\infty \dd Q\,
 \exp\!\left[Ns f_{\mathrm{rad}}^{\rm shell}(Q)\right]
 \E_{\Dcal_N}
 \int
 \left[\prod_{a=1}^{n}\dd\sigma(\vvec^a)\right]
 \left[\prod_{\gamma=1}^{s}\dd\sigma(\uvec^\gamma)\right]
\notag\\
&\quad\times
 \prod_{\mu=1}^{M}
 \left[
 \prod_{a=1}^{n}
 \Theta\!\left(\frac{\boldsymbol\xi^\mu\cdot\vvec^a}{\sqrt N}\right)
 \right]
 \prod_{\mu=1}^{M}
 \left[
 \prod_{\gamma=1}^{s}
 \exp\!\left(-\beta_{\mathrm p}\ell\!\left(\sqrt Q\frac{\boldsymbol\xi^\mu\cdot \uvec^\gamma}{\sqrt N}\right)\right)
 \right]
\notag\\
&\quad\times
 \prod_{\gamma=1}^{s}
 \left[
 \delta\!\left(\uvec^\gamma\cdot\vvec^1-Nc_d(Q,r)\right)
 \right].
\label{eq:app-rs-shell_Zns_after_radial}
\end{align}

Define reference-reference overlaps
\begin{equation}
 R_{ab}:=\frac1N\vvec^a\cdot\vvec^b
 =
 \frac1N\sum_{i=1}^{N}\vvec_i^a\vvec_i^b,
 \qquad
 a,b=1,\ldots,n.
\label{eq:app-rs-shell_Rab_def}
\end{equation}
Define shell-shell overlaps
\begin{equation}
 P_{\gamma\delta}:=\frac1N \uvec^\gamma\cdot \uvec^\delta
 =
 \frac1N\sum_{i=1}^{N}\uvec_i^\gamma \uvec_i^\delta,
 \qquad
 \gamma,\delta=1,\ldots,s.
\label{eq:app-rs-shell_Pgd_def}
\end{equation}
Define reference-shell overlaps
\begin{equation}
 T_{a\gamma}:=\frac1N\vvec^a\cdot \uvec^\gamma
 =
 \frac1N\sum_{i=1}^{N}\vvec_i^a \uvec_i^\gamma,
 \qquad
 a=1,\ldots,n,
 \quad
 \gamma=1,\ldots,s.
\label{eq:app-rs-shell_Tag_def}
\end{equation}
The distance constraint \eqref{eq:app-rs-shell_distance_cosine_constraint} is
\begin{equation}
 T_{1\gamma}=c_d(Q,r),
 \qquad
 \gamma=1,\ldots,s.
\label{eq:app-rs-shell_T1gamma_cd_constraint}
\end{equation}
The index $a=1$ is special because $\vvec^1$ is the selected reference in $Z_{n,s}(r)$.  The indices $a$ and $\gamma$ belong to different replica sets.

Insert the following identities into \eqref{eq:app-rs-shell_Zns_after_radial}.
\begin{equation}
 1
 =
 \int\dd R_{ab}\,
 \delta\!\left(NR_{ab}-\sum_{i=1}^{N}\vvec_i^a\vvec_i^b\right),
\label{eq:app-rs-shell_R_identity}
\end{equation}
\begin{equation}
 1
 =
 \int\dd P_{\gamma\delta}\,
 \delta\!\left(NP_{\gamma\delta}-\sum_{i=1}^{N}\uvec_i^\gamma \uvec_i^\delta\right),
\label{eq:app-rs-shell_P_identity}
\end{equation}
\begin{equation}
 1
 =
 \int\dd T_{a\gamma}\,
 \delta\!\left(NT_{a\gamma}-\sum_{i=1}^{N}\vvec_i^a \uvec_i^\gamma\right).
\label{eq:app-rs-shell_T_identity}
\end{equation}

Given fixed $R,P,T$, define the joint covariance matrix
\begin{equation}
 \mathcal M
 :=
 \begin{pmatrix}
 R&T\\
 T^T&P
 \end{pmatrix}.
\label{eq:app-rs-shell_M_block_def}
\end{equation}
For one disorder pattern, define fields
\begin{equation}
 \htilde^a:=\frac{\boldsymbol\xi\cdot\vvec^a}{\sqrt N},
 \qquad
 g^\gamma:=\frac{\boldsymbol\xi\cdot \uvec^\gamma}{\sqrt N}.
\label{eq:app-rs-shell_htilde_g_fields_def}
\end{equation}
Using \eqref{eq:app-rs-shell_Rab_def}--\eqref{eq:app-rs-shell_Tag_def}, their covariances are
\begin{equation}
 \E_{\boldsymbol\xi}[\htilde^a\htilde^b]=R_{ab},
 \qquad
 \E_{\boldsymbol\xi}[g^\gamma g^\delta]=P_{\gamma\delta},
 \qquad
 \E_{\boldsymbol\xi}[\htilde^a g^\gamma]=T_{a\gamma}.
\label{eq:app-rs-shell_field_covariances}
\end{equation}
Hence
\begin{equation}
 (\htilde^1,\ldots,\htilde^n,g^1,\ldots,g^s)^T
 \sim
 \Ncal(0,\mathcal M).
\label{eq:app-rs-shell_fields_gaussian_M}
\end{equation}
Define the one-pattern energetic factor
\begin{align}
 \mathcal P_{n,s}(Q;R,P,T)
 &:={}
 \E_{(\htilde,g)\sim\Ncal(0,\mathcal M)}
 \left[
 \prod_{a=1}^{n}\Theta(\htilde^a)
 \prod_{\gamma=1}^{s}\exp\!\left(-\beta_{\mathrm p}\ell(\sqrt Q\,g^\gamma)\right)
 \right].
\label{eq:app-rs-shell_Pns_def}
\end{align}
Since the patterns are independent, the disorder-averaged contribution of the $M$ pattern factors in \eqref{eq:app-rs-shell_Zns_after_radial} is
\begin{equation}
 \left[\mathcal P_{n,s}(Q;R,P,T)\right]^M
 =
 \exp\!\left[N\alpha\log\mathcal P_{n,s}(Q;R,P,T)+o(N)\right].
\label{eq:app-rs-shell_Pns_power_M}
\end{equation}

The angular entropic integral at fixed overlaps is
\begin{align}
 \mathcal I_{n,s}(R,P,T)
 &:={}
 \int
 \left[\prod_{a=1}^{n}\dd\sigma(\vvec^a)\right]
 \left[\prod_{\gamma=1}^{s}\dd\sigma(\uvec^\gamma)\right]
\notag\\
&\quad\times
 \prod_{a,b}
 \left[
 \delta\!\left(NR_{ab}-\sum_{i=1}^{N}\vvec_i^a\vvec_i^b\right)
 \right]
 \prod_{\gamma,\delta}
 \left[
 \delta\!\left(NP_{\gamma\delta}-\sum_{i=1}^{N}\uvec_i^\gamma \uvec_i^\delta\right)
 \right]
\notag\\
&\quad\times
 \prod_{a,\gamma}
 \left[
 \delta\!\left(NT_{a\gamma}-\sum_{i=1}^{N}\vvec_i^a \uvec_i^\gamma\right)
 \right].
\label{eq:app-rs-shell_Ins_def}
\end{align}
Using \eqref{eq:app-rs-shell_Pns_def} and \eqref{eq:app-rs-shell_Ins_def}, \eqref{eq:app-rs-shell_Zns_after_radial} becomes
\begin{align}
 Z_{n,s}(r)
 &\doteqN
 \int_0^\infty\dd Q\,
 \exp\!\left[Ns f_{\mathrm{rad}}^{\rm shell}(Q)\right]
 \int \dd R\,\dd P\,\dd T
 \prod_{\gamma=1}^{s}
 \left[
 \delta\!\left(T_{1\gamma}-c_d(Q,r)\right)
 \right]
\notag\\
&\quad\times
 \mathcal I_{n,s}(R,P,T)
 \left[\mathcal P_{n,s}(Q;R,P,T)\right]^M.
\label{eq:app-rs-shell_Zns_with_I_P}
\end{align}
Here
\begin{equation}
 \dd R:=\prod_{a,b}\dd R_{ab},
 \qquad
 \dd P:=\prod_{\gamma,\delta}\dd P_{\gamma\delta},
 \qquad
 \dd T:=\prod_{a,\gamma}\dd T_{a\gamma}.
\label{eq:app-rs-shell_dR_dP_dT_def}
\end{equation}

\subsection{Angular integral and reference-conditioned shell action}
\label{app:rs-evaluate_angular_integral}

Let
\begin{equation}
 d:=n+s.
\label{eq:app-rs-shell_d_combined_def}
\end{equation}
For each coordinate $i=1,\ldots,N$, define the combined site vector
\begin{equation}
 y_i
 :=
 (\vvec_i^1,\ldots,\vvec_i^n,\uvec_i^1,\ldots,\uvec_i^s)^T
 \in\R^{d}.
\label{eq:app-rs-shell_yi_def}
\end{equation}
Then \eqref{eq:app-rs-shell_Rab_def}--\eqref{eq:app-rs-shell_Tag_def} combine into
\begin{equation}
 \mathcal M_{\alpha\kappa}
 =
 \frac1N\sum_{i=1}^{N}y_i^\alpha y_i^\kappa,
 \qquad
 \alpha,\kappa=1,\ldots,d.
\label{eq:app-rs-shell_M_combined_overlap}
\end{equation}
Thus \eqref{eq:app-rs-shell_Ins_def} can be written as
\begin{equation}
 \mathcal I_{n,s}(\mathcal M)
 =
 \int
 \left[\prod_{i=1}^{N}\dd^d y_i\right]
 \prod_{\alpha,\kappa=1}^{d}
 \left[
 \delta\!\left(N\mathcal M_{\alpha\kappa}-\sum_{i=1}^{N}y_i^\alpha y_i^\kappa\right)
 \right],
\label{eq:app-rs-shell_Ins_combined_delta}
\end{equation}
where surface-measure constants have been absorbed into exponential-order normalization.

Introduce a conjugate matrix
\begin{equation}
 \widehat{\mathcal M}
 =
 (\widehat{\mathcal M}_{\alpha\kappa})_{\alpha,\kappa=1}^{d}.
\label{eq:app-rs-shell_Mhat_def}
\end{equation}
Using the formal representation
\begin{equation}
 \delta(x)
 =
 \int \frac{\dd\hat x}{2\pi i}\,e^{-\hat x x/2},
\label{eq:app-rs-shell_delta_conjugate}
\end{equation}
we have
\begin{align}
&\prod_{\alpha,\kappa}
 \delta\!\left(N\mathcal M_{\alpha\kappa}-\sum_{i=1}^{N}y_i^\alpha y_i^\kappa\right)
\notag\\
&\quad=
 \int\dd\widehat{\mathcal M}
 \exp\!\left[
 -\frac N2\sum_{\alpha,\kappa}\widehat{\mathcal M}_{\alpha\kappa}\mathcal M_{\alpha\kappa}
 +\frac12\sum_{i=1}^{N}\sum_{\alpha,\kappa}\widehat{\mathcal M}_{\alpha\kappa}y_i^\alpha y_i^\kappa
 \right]
\notag\\
&\quad=
 \int\dd\widehat{\mathcal M}
 \exp\!\left[
 -\frac N2\Tr(\widehat{\mathcal M}\mathcal M)
 +\frac12\sum_{i=1}^{N}y_i^T\widehat{\mathcal M}y_i
 \right].
\label{eq:app-rs-shell_delta_conjugate_matrix}
\end{align}
Substitution into \eqref{eq:app-rs-shell_Ins_combined_delta} gives an auxiliary saddle over $\widehat{\mathcal M}$.
\begin{align}
 \mathcal I_{n,s}(\mathcal M)
 &=
 \int\dd\widehat{\mathcal M}
 \exp\!\left[-\frac N2\Tr(\widehat{\mathcal M}\mathcal M)\right]
 \int
 \left[\prod_{i=1}^{N}\dd^d y_i\right]
 \exp\!\left[\frac12\sum_{i=1}^{N}y_i^T\widehat{\mathcal M}y_i\right]
\notag\\
 &=
 \int\dd\widehat{\mathcal M}
 \exp\!\left[-\frac N2\Tr(\widehat{\mathcal M}\mathcal M)\right]
 \prod_{i=1}^{N}
 \left[
 \int_{\R^d}\dd^d y\,
 \exp\!\left(\frac12 y^T\widehat{\mathcal M}y\right)
 \right]
\notag\\
 &=
 \int\dd\widehat{\mathcal M}
 \exp\!\left\{
 N\left[
 -\frac12\Tr(\widehat{\mathcal M}\mathcal M)
 +\log\int_{\R^d}\dd^d y\,
 e^{y^T\widehat{\mathcal M}y/2}
 \right]
 \right\}.
\label{eq:app-rs-shell_Ins_Mhat_action}
\end{align}
The Gaussian single-site integral is
\begin{equation}
 \int_{\R^d}\dd^d y\,e^{y^T\widehat{\mathcal M}y/2}
 =
 (2\pi)^{d/2}\det(-\widehat{\mathcal M})^{-1/2},
\label{eq:app-rs-shell_gaussian_y_integral}
\end{equation}
with the contour chosen so that $-\widehat{\mathcal M}$ has positive real part at the saddle.  Define
\begin{equation}
 \Psi(\widehat{\mathcal M};\mathcal M)
 :=
 -\frac12\Tr(\widehat{\mathcal M}\mathcal M)
 +\frac d2\log(2\pi)
 -\frac12\log\det(-\widehat{\mathcal M}).
\label{eq:app-rs-shell_Psi_Mhat_def}
\end{equation}
The auxiliary matrix $\widehat{\mathcal M}$ appears only in the angular factor \eqref{eq:app-rs-shell_Ins_Mhat_action}.  Therefore it can be eliminated by its own saddle before the remaining saddle over the physical order parameters is taken.  The saddle equation for $\widehat{\mathcal M}$ is
\begin{equation}
 0=\partial_{\widehat{\mathcal M}}\Psi
 \quad\Longleftrightarrow\quad
 -\widehat{\mathcal M}=\mathcal M^{-1}.
\label{eq:app-rs-shell_Mhat_saddle}
\end{equation}
At \eqref{eq:app-rs-shell_Mhat_saddle},
\begin{equation}
 \Psi(\widehat{\mathcal M}_\star;\mathcal M)
 =
 \frac12\log\det\mathcal M+\frac d2\log(2\pi e).
\label{eq:app-rs-shell_Ins_final_logdet}
\end{equation}
Hence
\begin{equation}
 \mathcal I_{n,s}(\mathcal M)
 \doteqN
 \exp\!\left[N\left(\frac12\log\det\mathcal M+\frac{n+s}{2}\log(2\pi e)\right)\right].
\label{eq:app-rs-shell_Ins_exp_result}
\end{equation}

Define
\begin{equation}
 G_{\mathrm S,{\rm full}}^{n,s}(\mathcal M)
 :=
 \frac12\log\det\mathcal M+\frac{n+s}{2}\log(2\pi e).
\label{eq:app-rs-shell_GS_full_def}
\end{equation}
Define
\begin{equation}
 G_{\mathrm E,{\rm full}}^{n,s}(Q;\mathcal M)
 :=
 \log\mathcal P_{n,s}(Q;R,P,T),
\label{eq:app-rs-shell_GE_full_def}
\end{equation}
where $\mathcal P_{n,s}$ is defined in \eqref{eq:app-rs-shell_Pns_def}.  With \eqref{eq:app-rs-shell_Ins_exp_result} and \eqref{eq:app-rs-shell_Pns_power_M}, \eqref{eq:app-rs-shell_Zns_with_I_P} becomes
\begin{align}
 Z_{n,s}(r)
 &\doteqN
 \int \dd Q\,\dd R\,\dd P\,\dd T
 \prod_{\gamma=1}^{s}
 \left[
 \delta\!\left(T_{1\gamma}-c_d(Q,r)\right)
 \right]
\notag\\
&\quad\times
 \exp\!\left\{
 N\left[
 s f_{\mathrm{rad}}^{\rm shell}(Q)
 +G_{\mathrm S,{\rm full}}^{n,s}(\mathcal M)
 +\alpha G_{\mathrm E,{\rm full}}^{n,s}(Q;\mathcal M)
 \right]
 \right\}.
\label{eq:app-rs-shell_Zns_full_G_action}
\end{align}

At $s=0$, the Gram matrix \eqref{eq:app-rs-shell_M_block_def} reduces to $R$.  Define the reference-only entropic part
\begin{equation}
 G_{\mathrm S,{\rm ref}}^{n}(R)
 :=
 G_{\mathrm S,{\rm full}}^{n,0}(R)
 =
 \frac12\log\det R+\frac n2\log(2\pi e).
\label{eq:app-rs-shell_GS_ref_def}
\end{equation}
Define the reference-only one-pattern factor
\begin{align}
 \mathcal P_{n,0}(R)
 &:={}
 \E_{\htilde\sim\Ncal(0,R)}
 \prod_{a=1}^{n}\Theta(\htilde^a),
\label{eq:app-rs-shell_Pn0_ref_def}\\
 G_{\mathrm E,{\rm ref}}^n(R)
 &:={}
 \log\mathcal P_{n,0}(R).
\label{eq:app-rs-shell_GE_ref_def}
\end{align}
Then
\begin{equation}
 G_{\mathrm S,{\rm full}}^{n,s}(\mathcal M)
 =
 G_{\mathrm S,{\rm ref}}^{n}(R)+G_{\mathrm S,{\rm cond}}^{n,s}(\mathcal M;R),
\label{eq:app-rs-shell_GS_split}
\end{equation}
where
\begin{align}
 G_{\mathrm S,{\rm cond}}^{n,s}(\mathcal M;R)
 &:={}
 G_{\mathrm S,{\rm full}}^{n,s}(\mathcal M)-G_{\mathrm S,{\rm ref}}^{n}(R)
\notag\\
 &=
 \frac12\log\frac{\det\mathcal M}{\det R}
 +\frac s2\log(2\pi e).
\label{eq:app-rs-shell_GS_cond_def}
\end{align}
Similarly,
\begin{equation}
 G_{\mathrm E,{\rm full}}^{n,s}(Q;\mathcal M)
 =
 G_{\mathrm E,{\rm ref}}^{n}(R)+G_{\mathrm E,{\rm cond}}^{n,s}(Q;\mathcal M,R),
\label{eq:app-rs-shell_GE_split}
\end{equation}
where
\begin{align}
 G_{\mathrm E,{\rm cond}}^{n,s}(Q;\mathcal M,R)
 &:={}
 G_{\mathrm E,{\rm full}}^{n,s}(Q;\mathcal M)-G_{\mathrm E,{\rm ref}}^{n}(R)
\notag\\
 &=
 \log\frac{\mathcal P_{n,s}(Q;R,P,T)}{\mathcal P_{n,0}(R)}.
\label{eq:app-rs-shell_GE_cond_def}
\end{align}

Using \eqref{eq:app-rs-shell_GS_split} and \eqref{eq:app-rs-shell_GE_split}, \eqref{eq:app-rs-shell_Zns_full_G_action} may be rewritten as
\begin{align}
 Z_{n,s}(r)
 &\doteqN
 \int \dd R\,
 \exp\!\left\{N\left[G_{\mathrm S,{\rm ref}}^n(R)+\alpha G_{\mathrm E,{\rm ref}}^n(R)\right]\right\}
\notag\\
&\quad\times
 \int \dd Q\,\dd P\,\dd T
 \prod_{\gamma=1}^{s}
 \left[
 \delta\!\left(T_{1\gamma}-c_d(Q,r)\right)
 \right]
\notag\\
&\quad\times
 \exp\!\left\{
 N\left[
 s f_{\mathrm{rad}}^{\rm shell}(Q)
 +G_{\mathrm S,{\rm cond}}^{n,s}(\mathcal M;R)
 +\alpha G_{\mathrm E,{\rm cond}}^{n,s}(Q;\mathcal M,R)
 \right]
 \right\}.
\label{eq:app-rs-shell_Zns_conditional_form}
\end{align}
The ratio in \eqref{eq:app-rs-shell_GS_cond_def} can be simplified using the Schur complement.  Since
\begin{equation}
 \mathcal M=
 \begin{pmatrix}
 R&T\\
 T^T&P
 \end{pmatrix},
\label{eq:app-rs-shell_M_block_repeat}
\end{equation}
block determinant factorization gives
\begin{equation}
 \det\mathcal M
 =
 \det R\,\det(P-T^TR^{-1}T).
\label{eq:app-rs-shell_block_det_identity}
\end{equation}
Define
\begin{equation}
 S:=P-T^TR^{-1}T.
\label{eq:app-rs-shell_S_schur_def}
\end{equation}
Then
\begin{equation}
 \frac{\det\mathcal M}{\det R}
 =
 \det S,
\label{eq:app-rs-shell_det_ratio_schur}
\end{equation}
and
\begin{equation}
 G_{\mathrm S,{\rm cond}}^{n,s}
 =
 \frac12\log\det S+\frac s2\log(2\pi e).
\label{eq:app-rs-shell_GS_cond_schur}
\end{equation}

Equation \eqref{eq:app-rs-shell_Zns_conditional_form} is the exponential-order replicated action after the reference-only factor has been separated.  Equivalently, up to the already selected reference normalization, the shell-dependent part can be written as
\begin{align}
 Z^{\rm shell}_{n,s}(r\mid R)
 &:={}
 \frac{Z_{n,s}(r)}{Z_{n,0}}
\notag\\
 &\doteqN
 \int_0^\infty\dd Q\,\dd P\,\dd T
 \prod_{\gamma=1}^{s}
 \delta\!\left(T_{1\gamma}-c_d(Q,r)\right)
\notag\\
&\quad\times
 \exp\!\left\{
 N\mathcal A_{n,s}^{\rm shell}(Q;R,P,T,r)
 \right\},
\label{eq:app-rs-shell_Zns_shell_action_before_delta_resolution}
\end{align}
where
\begin{equation}
 \mathcal A_{n,s}^{\rm shell}(Q;R,P,T,r)
 :=
 s f_{\mathrm{rad}}^{\rm shell}(Q)
 +G_{\mathrm S,{\rm cond}}^{n,s}(\mathcal M;R)
 +\alpha G_{\mathrm E,{\rm cond}}^{n,s}(Q;\mathcal M,R).
\label{eq:app-rs-shell_A_ns_shell_def}
\end{equation}
The distance constraint is now carried entirely by the explicit deltas $T_{1\gamma}=c_d(Q,r)$. Since they do not create an additional disorder-average term, the same shell object may be written after resolving the deltas as
\begin{align}
 Z^{\rm shell}_{n,s}(r\mid R)
 &\doteqN
 \int_0^\infty\dd Q\,\dd P\,\dd T_{\rm red}
 \exp\!\left\{
 N\mathcal A_{n,s}^{\rm shell}
 \left(Q;R,P,T_\star(Q,r),r\right)
 \right\},
\label{eq:app-rs-shell_Zns_after_T1gamma_resolution}
\end{align}
where
\begin{equation}
 (T_\star)_{1\gamma}=c_d(Q,r),
 \qquad
 (T_\star)_{a\gamma}=T_{a\gamma}\quad(a\ge2),
 \qquad
 \dd T_{\rm red}:=\prod_{a=2}^{n}\prod_{\gamma=1}^{s}\dd T_{a\gamma}.
\label{eq:app-rs-shell_Tstar_after_delta_resolution}
\end{equation}
Resolving the distance deltas fixes the entire selected-reference row
$a=1$.  The reduced measure therefore integrates only the nonselected rows
$a=2,\ldots,n$.

Finally, importing the reference saddle fixes the reference block to
\begin{equation}
 R=R_\star(q_{\mathrm{ref}}),
 \qquad
 (R_\star)_{ab}=\delta_{ab}+(1-\delta_{ab})q_{\mathrm{ref}},
 \qquad
 \Qt_\star=\lambda_{\mathrm{ref}}^{-1}.
\label{eq:app-rs-shell_import_Rstar_Qstar}
\end{equation}
Therefore the actual shell saddle target is an exponential-order integral over the remaining shell order parameters,
\begin{align}
 Z^{\rm shell}_{n,s}(r\mid q_{\mathrm{ref}},\Qt_\star)
 &\doteqN
 \int_0^\infty\dd Q\,\dd P\,\dd T_{\rm red}
 \exp\!\left\{
 N\mathcal A_{n,s}^{\rm shell}
 \left(Q;R_\star(q_{\mathrm{ref}}),P,T_\star(Q,r),r\right)
 \right\}.
\label{eq:app-rs-shell_Zns_final_pre_RS_target}
\end{align}
It is still a functional of $P$ and of the unfixed rows of $T$, with $Q$ also optimized.  The next step is to impose the shell RS ansatz for $P$ and $T$, reducing \eqref{eq:app-rs-shell_Zns_final_pre_RS_target} to a scalar saddle over $(Q,p,t)$.

\subsection{Replica-symmetric shell functional}
\label{app:rs-RS_ansatz_GS_shell}

The reference block is fixed to the selected RS reference branch.
\begin{equation}
 R_{ab}=\delta_{ab}+(1-\delta_{ab})q_{\mathrm{ref}}.
\label{eq:app-rs-shell_RS_R}
\end{equation}
The shell-shell block is assigned the RS form
\begin{equation}
 P_{\gamma\delta}=\delta_{\gamma\delta}+(1-\delta_{\gamma\delta})p.
\label{eq:app-rs-shell_RS_P}
\end{equation}
The mixed block is
\begin{equation}
 T_{a\gamma}=c_d\delta_{a1}+t(1-\delta_{a1}),
 \qquad
 c_d=c_d(Q,r).
\label{eq:app-rs-shell_RS_T}
\end{equation}
Define
\begin{equation}
 \mathbf{1}_s:=(1,\ldots,1)^T\in\R^s,
 \qquad
 \mathbf{1}_n:=(1,\ldots,1)^T\in\R^n,
\label{eq:app-rs-shell_one_vectors}
\end{equation}
\begin{equation}
 b:=(c_d,t,t,\ldots,t)^T\in\R^n.
\label{eq:app-rs-shell_b_vector_def}
\end{equation}
Then
\begin{equation}
 T=b\mathbf{1}_s^T,
\label{eq:app-rs-shell_T_b_form}
\end{equation}
and
\begin{equation}
 T^TR^{-1}T=(b^TR^{-1}b)\mathbf{1}_s\mathbf{1}_s^T.
\label{eq:app-rs-shell_TRT_b}
\end{equation}
The RS reference block is
\begin{equation}
 R=(1-q_{\mathrm{ref}})I_n+q_{\mathrm{ref}}\mathbf{1}_n\mathbf{1}_n^T.
\label{eq:app-rs-shell_R_matrix_RS}
\end{equation}
Its inverse is
\begin{equation}
 R^{-1}
 =
 \frac1{1-q_{\mathrm{ref}}}I_n-
 \frac{q_{\mathrm{ref}}}{(1-q_{\mathrm{ref}})(1-q_{\mathrm{ref}}+nq_{\mathrm{ref}})}\mathbf{1}_n\mathbf{1}_n^T.
\label{eq:app-rs-shell_R_inv_RS}
\end{equation}
For \eqref{eq:app-rs-shell_b_vector_def},
\begin{equation}
 b^Tb=c_d^2+(n-1)t^2,
 \qquad
 \mathbf{1}_n^Tb=c_d+(n-1)t.
\label{eq:app-rs-shell_b_norm_sum}
\end{equation}
Define
\begin{equation}
 m:=\lim_{n\to0}b^TR^{-1}b.
\label{eq:app-rs-shell_m_def}
\end{equation}
Using \eqref{eq:app-rs-shell_R_inv_RS} and \eqref{eq:app-rs-shell_b_norm_sum},
\begin{align}
 m
 &=
 \frac{c_d^2-t^2}{1-q_{\mathrm{ref}}}
 -
 \frac{q_{\mathrm{ref}}(c_d-t)^2}{(1-q_{\mathrm{ref}})^2}
\notag\\
 &=
 \frac{c_d^2(1-2q_{\mathrm{ref}})+2q_{\mathrm{ref}}c_dt-t^2}{(1-q_{\mathrm{ref}})^2}.
\label{eq:app-rs-shell_m_explicit}
\end{align}
Thus
\begin{align}
 1-m
 &=
 \frac{(1-c_d^2)(1-2q_{\mathrm{ref}})+q_{\mathrm{ref}}^2-2q_{\mathrm{ref}}c_dt+t^2}{(1-q_{\mathrm{ref}})^2}.
\label{eq:app-rs-shell_one_minus_m}
\end{align}
The shell block is
\begin{equation}
 P=(1-p)I_s+p\mathbf{1}_s\mathbf{1}_s^T.
\label{eq:app-rs-shell_P_matrix_RS}
\end{equation}
Combining \eqref{eq:app-rs-shell_S_schur_def}, \eqref{eq:app-rs-shell_TRT_b}, and \eqref{eq:app-rs-shell_P_matrix_RS},
\begin{equation}
 S=(1-p)I_s+(p-m)\mathbf{1}_s\mathbf{1}_s^T.
\label{eq:app-rs-shell_S_RS_form}
\end{equation}
The eigenvalues of \eqref{eq:app-rs-shell_S_RS_form} are
\begin{equation}
 \lambda_\perp=1-p,
 \qquad
 \operatorname{mult}(\lambda_\perp)=s-1,
\label{eq:app-rs-shell_S_perp_eigen}
\end{equation}
\begin{equation}
 \lambda_\parallel=1-p+s(p-m),
 \qquad
 \operatorname{mult}(\lambda_\parallel)=1,
\label{eq:app-rs-shell_S_parallel_eigen}
\end{equation}
where $\operatorname{mult}(\lambda)$ denotes the multiplicity of the eigenvalue $\lambda$.

Hence
\begin{equation}
 \det S=(1-p)^{s-1}[1-p+s(p-m)].
\label{eq:app-rs-shell_detS_RS}
\end{equation}
Using \eqref{eq:app-rs-shell_GS_cond_schur}, the shell entropic contribution is
\begin{align}
 G_{\mathrm S}^{\rm shell}(p,t;c_d,q_{\mathrm{ref}})
 &:={}
 \lim_{n\to0}\left.\partial_s G_{\mathrm S,{\rm cond}}^{n,s}\right|_{s=0}
\notag\\
 &=
 \frac12\left[\log(1-p)+\frac{p-m}{1-p}\right]
 +\frac12\log(2\pi e)
\notag\\
 &=
 \frac12\log(2\pi)+\frac12\log(1-p)+\frac{1-m}{2(1-p)}.
\label{eq:app-rs-shell_GS_shell_m_form}
\end{align}
Substituting \eqref{eq:app-rs-shell_one_minus_m},
\begin{equation}
 G_{\mathrm S}^{\rm shell}(p,t;c_d,q_{\mathrm{ref}})
 =
 \frac{(1-c_d^2)(1-2q_{\mathrm{ref}})+q_{\mathrm{ref}}^2-2q_{\mathrm{ref}}c_dt+t^2}{2(1-p)(1-q_{\mathrm{ref}})^2}
 +\frac12\log(2\pi)+\frac12\log(1-p).
\label{eq:app-rs-shell_GS_shell_final}
\end{equation}
The domain required by \eqref{eq:app-rs-shell_GS_shell_final} includes
\begin{equation}
 p<1.
\label{eq:app-rs-shell_p_less_than_one}
\end{equation}

Under \eqref{eq:app-rs-shell_RS_R}, a selected reference field is represented by
\begin{equation}
 \htilde^1=\sqrt{q_{\mathrm{ref}}}\,z_0+\sqrt{1-q_{\mathrm{ref}}}\,z_1,
\label{eq:app-rs-shell_h1_representation}
\end{equation}
where
\begin{equation}
 z_0,z_1\sim\Ncal(0,1),
 \qquad
 z_0\perp z_1.
\label{eq:app-rs-shell_z0_z1_indep}
\end{equation}
A nonselected reference field is
\begin{equation}
 \htilde^a=\sqrt{q_{\mathrm{ref}}}\,z_0+\sqrt{1-q_{\mathrm{ref}}}\,z_a,
 \qquad
 a\ne1,
\label{eq:app-rs-shell_ha_nonselected}
\end{equation}
with $z_a$ independent standard Gaussians.

A representative shell field is written as
\begin{equation}
 g
 =
 \frac{t}{\sqrt{q_{\mathrm{ref}}}}z_0+\frac{c_d-t}{\sqrt{1-q_{\mathrm{ref}}}}z_1+
 \sqrt A\,z_2+\sqrt{1-p}\,z_3,
\label{eq:app-rs-shell_g_representation}
\end{equation}
where
\begin{equation}
 z_2,z_3\sim\Ncal(0,1),
 \qquad
 z_0,z_1,z_2,z_3\text{ independent},
\label{eq:app-rs-shell_z_independence}
\end{equation}
and
\begin{equation}
 A:=p-\frac{t^2}{q_{\mathrm{ref}}}-\frac{(c_d-t)^2}{1-q_{\mathrm{ref}}}.
\label{eq:app-rs-shell_A_def}
\end{equation}
The representation \eqref{eq:app-rs-shell_g_representation} satisfies
\begin{equation}
 \begin{aligned}
 \Cov(g,\htilde^1)&=c_d,
 &\qquad
 \Cov(g,\htilde^a)&=t\quad(a\ne1),
 \\
 \Cov(g^\gamma,g^\delta)&=p\quad(\gamma\ne\delta),
 &\qquad
 \Var(g)&=1.
 \end{aligned}
\label{eq:app-rs-shell_cov_matching_summary}
\end{equation}
It is real-valued only if
\begin{equation}
 q_{\mathrm{ref}}\in(0,1),
 \qquad
 p<1,
 \qquad
 A\ge0.
\label{eq:app-rs-shell_cov_feasibility_basic}
\end{equation}

At fixed $z_0$, the selected-reference feasibility probability is
\begin{align}
 H_0(z_0)
 &:={}
 \int Dz_1\,
 \Theta\!\left(\sqrt{q_{\mathrm{ref}}}\,z_0+\sqrt{1-q_{\mathrm{ref}}}\,z_1\right)
\notag\\
 &=
 \Htail\!\left(-\sqrt{\frac{q_{\mathrm{ref}}}{1-q_{\mathrm{ref}}}}z_0\right).
\label{eq:app-rs-shell_H0_def}
\end{align}
Using \eqref{eq:app-rs-shell_h1_representation}--\eqref{eq:app-rs-shell_H0_def}, the prelimit one-pattern factor \eqref{eq:app-rs-shell_Pns_def} becomes
\begin{align}
 \mathcal P_{n,s}(Q;R,P,T)
 &=
 \int Dz_0\,H_0(z_0)^{n-1}
 \int Dz_1\,
 \Theta\!\left(\sqrt{q_{\mathrm{ref}}}\,z_0+\sqrt{1-q_{\mathrm{ref}}}\,z_1\right)
\notag\\
&\quad\times
 \int Dz_2
 \left[
 \int Dz_3\,
 \exp\!\left(-\beta_{\mathrm p}\ell(\sqrt Q\,g)\right)
 \right]^s.
\label{eq:app-rs-shell_Pns_RS_prelimit}
\end{align}
At $s=0$,
\begin{equation}
 \mathcal P_{n,0}(R)
 =
 \int Dz_0\,H_0(z_0)^n.
\label{eq:app-rs-shell_Pn0_RS}
\end{equation}
The shell energetic term is
\begin{align}
 G_{\mathrm E}^{\rm shell}(Q,p,t;r)
 &:={}
 \lim_{n\to0}\left.\partial_s G_{\mathrm E,{\rm cond}}^{n,s}(Q;\mathcal M,R)\right|_{s=0}
\label{eq:app-rs-shell_GE_shell_def}\\
 &=
 \lim_{n\to0}\left.\partial_s\log\frac{\mathcal P_{n,s}(Q;R,P,T)}{\mathcal P_{n,0}(R)}\right|_{s=0}.
\label{eq:app-rs-shell_GE_shell_ratio_derivative}
\end{align}
Since \eqref{eq:app-rs-shell_Pn0_RS} is independent of $s$,
\begin{align}
 G_{\mathrm E}^{\rm shell}(Q,p,t;r)
 &=
 \int Dz_0\,
 \frac{
 \int Dz_1\,
 \Theta\!\left(\sqrt{q_{\mathrm{ref}}}\,z_0+\sqrt{1-q_{\mathrm{ref}}}\,z_1\right)
 \int Dz_2\,
 \log\int Dz_3\,e^{-\beta_{\mathrm p}\ell(\sqrt Q\,g)}
 }{H_0(z_0)}.
\label{eq:app-rs-shell_GE_shell_final}
\end{align}
Equations \eqref{eq:app-rs-shell_g_representation}, \eqref{eq:app-rs-shell_A_def}, and \eqref{eq:app-rs-shell_H0_def} are part of the definition of \eqref{eq:app-rs-shell_GE_shell_final}.

Define the feasible shell domain
\begin{align}
 \Dcal_{\rm shell}(r)
 :=
 \bigg\{(Q,p,t):
 &\ Q>0,
 \quad
 |c_d(Q,r)|\le1,
 \quad
 p<1,
\notag\\
&\quad
 A(Q,p,t;r)
 :=
 p-\frac{t^2}{q_{\mathrm{ref}}}-\frac{(c_d(Q,r)-t)^2}{1-q_{\mathrm{ref}}}
 \ge0
 \bigg\}.
\label{eq:app-rs-shell_domain_def}
\end{align}
The shell RS functional is
\begin{equation}
 \mathcal F_{\mathrm{RS}}(Q,p,t;r)
 :=
 f_{\mathrm{rad}}^{\rm shell}(Q)
 +G_{\mathrm S}^{\rm shell}(p,t;c_d(Q,r),q_{\mathrm{ref}})
 +\alpha G_{\mathrm E}^{\rm shell}(Q,p,t;r),
\label{eq:app-rs-shell_Fshell_def}
\end{equation}
with $f_{\mathrm{rad}}^{\rm shell}$ from \eqref{eq:app-rs-shell_frad_def}, $G_{\mathrm S}^{\rm shell}$ from \eqref{eq:app-rs-shell_GS_shell_final}, and $G_{\mathrm E}^{\rm shell}$ from \eqref{eq:app-rs-shell_GE_shell_final}.  The RS shell local entropy density is
\begin{equation}
 \Phi_{\mathrm{RS}}(r)
 =
 \extr_{(Q,p,t)\in\Dcal_{\rm shell}(r)}
 \mathcal F_{\mathrm{RS}}(Q,p,t;r).
\label{eq:app-rs-shell_Phi_RS_final}
\end{equation}
If $(Q_\star(r),p_\star(r),t_\star(r))$ is the selected branch, then
\begin{equation}
 \Phi_{\mathrm{RS}}(r)
 =
 \mathcal F_{\mathrm{RS}}(Q_\star(r),p_\star(r),t_\star(r);r).
\label{eq:app-rs-shell_Phi_selected_branch}
\end{equation}

\subsection{Saddle equations and constrained branches}
\label{app:rs-interior_saddle_eqs}

The equations in this section apply only in the interior domain
\begin{equation}
 Q>0,
 \qquad
 |c_d(Q,r)|<1,
 \qquad
 p<1,
 \qquad
 A(Q,p,t;r)>0.
\label{eq:app-rs-shell_interior_domain}
\end{equation}
Let
\begin{equation}
 c:=c_d(Q,r),
 \qquad
 D_{\mathrm{ref}}:=(1-q_{\mathrm{ref}})^2,
\label{eq:app-rs-shell_c_Dq_def}
\end{equation}
\begin{equation}
 B(c,t)
 :=
 (1-c^2)(1-2q_{\mathrm{ref}})+q_{\mathrm{ref}}^2-2q_{\mathrm{ref}}ct+t^2.
\label{eq:app-rs-shell_B_def}
\end{equation}
Then
\begin{equation}
 G_{\mathrm S}^{\rm shell}
 =
 \frac{B(c,t)}{2(1-p)D_{\mathrm{ref}}}
 +\frac12\log(2\pi)+\frac12\log(1-p).
\label{eq:app-rs-shell_GS_saddle_notation}
\end{equation}
The distance derivative is
\begin{equation}
 c_Q:=\frac{\partial c_d(Q,r)}{\partial Q}
 =
 \frac{Q-\Qt_\star+r^2}{4Q^{3/2}\sqrt{\Qt_\star}}.
\label{eq:app-rs-shell_cQ_def}
\end{equation}
The entropic derivatives are
\begin{equation}
 \partial_pG_{\mathrm S}^{\rm shell}
 =
 \frac{B(c,t)}{2D_{\mathrm{ref}}(1-p)^2}-\frac1{2(1-p)},
\label{eq:app-rs-shell_dGS_dp}
\end{equation}
\begin{equation}
 \partial_tG_{\mathrm S}^{\rm shell}
 =
 \frac{t-q_{\mathrm{ref}}c}{(1-p)D_{\mathrm{ref}}},
\label{eq:app-rs-shell_dGS_dt}
\end{equation}
\begin{equation}
 \partial_cG_{\mathrm S}^{\rm shell}
 =
 -\frac{c(1-2q_{\mathrm{ref}})+q_{\mathrm{ref}}t}{(1-p)D_{\mathrm{ref}}},
\label{eq:app-rs-shell_dGS_dc}
\end{equation}
\begin{equation}
 \partial_QG_{\mathrm S}^{\rm shell}
 =
 c_Q\partial_cG_{\mathrm S}^{\rm shell}.
\label{eq:app-rs-shell_dGS_dQ}
\end{equation}

For the energetic derivatives, define
\begin{equation}
 \mathcal Z_3(Q,p,t,c;z_0,z_1,z_2)
 :=
 \int Dz_3\,e^{-\beta_{\mathrm p}\ell(\sqrt Q\,g)},
\label{eq:app-rs-shell_Z3_def}
\end{equation}
where $g$ is given by \eqref{eq:app-rs-shell_g_representation}.  Define the inner average
\begin{equation}
 \langle X\rangle_3
 :=
 \frac{\int Dz_3\,X(z_3)e^{-\beta_{\mathrm p}\ell(\sqrt Q\,g)}}{\mathcal Z_3(Q,p,t,c;z_0,z_1,z_2)}.
\label{eq:app-rs-shell_inner_average_def}
\end{equation}
Define the outer selected-reference average
\begin{equation}
 \langle F\rangle_+
 :=
 \int Dz_0\,
 \frac{
 \int Dz_1\,\Theta\!\left(\sqrt{q_{\mathrm{ref}}}\,z_0+\sqrt{1-q_{\mathrm{ref}}}\,z_1\right)
 \int Dz_2\,F(z_0,z_1,z_2)
 }{H_0(z_0)}.
\label{eq:app-rs-shell_outer_average_def}
\end{equation}
Then
\begin{equation}
 G_{\mathrm E}^{\rm shell}
 =
 \left\langle \log\mathcal Z_3\right\rangle_+.
\label{eq:app-rs-shell_GE_average_notation}
\end{equation}
For $x\in\{p,t,c\}$,
\begin{equation}
 \partial_xG_{\mathrm E}^{\rm shell}
 =
 -\beta_{\mathrm p}
 \left\langle
 \left\langle
 \ell'(\sqrt Q\,g)\sqrt Q\,\partial_xg
 \right\rangle_3
 \right\rangle_+.
\label{eq:app-rs-shell_dGE_dx_general}
\end{equation}
The derivatives of $A$ are
\begin{equation}
 \partial_pA=1,
\label{eq:app-rs-shell_dA_dp}
\end{equation}
\begin{equation}
 \partial_tA=-\frac{2t}{q_{\mathrm{ref}}}+\frac{2(c-t)}{1-q_{\mathrm{ref}}},
\label{eq:app-rs-shell_dA_dt}
\end{equation}
\begin{equation}
 \partial_cA=-\frac{2(c-t)}{1-q_{\mathrm{ref}}}.
\label{eq:app-rs-shell_dA_dc}
\end{equation}
Thus
\begin{equation}
 \partial_pg
 =
 \frac{z_2}{2\sqrt A}-\frac{z_3}{2\sqrt{1-p}},
\label{eq:app-rs-shell_dg_dp}
\end{equation}
\begin{equation}
 \partial_tg
 =
 \frac{z_0}{\sqrt{q_{\mathrm{ref}}}}-\frac{z_1}{\sqrt{1-q_{\mathrm{ref}}}}
 +\frac{1}{2\sqrt A}\left[-\frac{2t}{q_{\mathrm{ref}}}+\frac{2(c-t)}{1-q_{\mathrm{ref}}}\right]z_2,
\label{eq:app-rs-shell_dg_dt}
\end{equation}
\begin{equation}
 \partial_cg
 =
 \frac{z_1}{\sqrt{1-q_{\mathrm{ref}}}}-\frac{c-t}{(1-q_{\mathrm{ref}})\sqrt A}z_2.
\label{eq:app-rs-shell_dg_dc}
\end{equation}
The $Q$ derivative is
\begin{equation}
 \partial_Q(\sqrt Q\,g)
 =
 \frac{g}{2\sqrt Q}+\sqrt Q\,c_Q\partial_cg.
\label{eq:app-rs-shell_d_sqrtQg_dQ}
\end{equation}
Therefore
\begin{equation}
 \partial_QG_{\mathrm E}^{\rm shell}
 =
 -\beta_{\mathrm p}
 \left\langle
 \left\langle
 \ell'(\sqrt Q\,g)
 \left[\frac{g}{2\sqrt Q}+\sqrt Q\,c_Q\partial_cg\right]
 \right\rangle_3
 \right\rangle_+.
\label{eq:app-rs-shell_dGE_dQ}
\end{equation}
The derivative of the shell radial term with respect to $Q$ is
\begin{equation}
 \partial_Q f_{\mathrm{rad}}^{\rm shell}(Q)
 =
 \frac1{2Q}-\frac{\beta_{\mathrm p}\lambda_{\mathrm{sh}}}2.
\label{eq:app-rs-shell_dfrad_dQ}
\end{equation}

The interior saddle equations are
\begin{equation}
 \partial_Q\mathcal F_{\mathrm{RS}}=0,
 \qquad
 \partial_p\mathcal F_{\mathrm{RS}}=0,
 \qquad
 \partial_t\mathcal F_{\mathrm{RS}}=0.
\label{eq:app-rs-shell_saddle_compact}
\end{equation}
Using \eqref{eq:app-rs-shell_Fshell_def}, the $p$-equation is
\begin{align}
 0
 &=
 \frac{B(c,t)}{2D_{\mathrm{ref}}(1-p)^2}-\frac1{2(1-p)}
\notag\\
&\quad-
 \alpha\beta_{\mathrm p}
 \left\langle
 \left\langle
 \ell'(\sqrt Q\,g)\sqrt Q
 \left[
 \frac{z_2}{2\sqrt A}-\frac{z_3}{2\sqrt{1-p}}
 \right]
 \right\rangle_3
 \right\rangle_+.
\label{eq:app-rs-shell_p_saddle_expanded}
\end{align}
The $t$-equation is
\begin{align}
 0
 &=
 \frac{t-q_{\mathrm{ref}}c}{(1-p)D_{\mathrm{ref}}}
\notag\\
&\quad-
 \alpha\beta_{\mathrm p}
 \left\langle
 \left\langle
 \ell'(\sqrt Q\,g)\sqrt Q
 \left[
 \frac{z_0}{\sqrt{q_{\mathrm{ref}}}}-\frac{z_1}{\sqrt{1-q_{\mathrm{ref}}}}
 +\frac{1}{2\sqrt A}\left(-\frac{2t}{q_{\mathrm{ref}}}+\frac{2(c-t)}{1-q_{\mathrm{ref}}}\right)z_2
 \right]
 \right\rangle_3
 \right\rangle_+.
\label{eq:app-rs-shell_t_saddle_expanded}
\end{align}
The $Q$-equation is
\begin{align}
 0
 &=
 \frac1{2Q}-\frac{\beta_{\mathrm p}\lambda_{\mathrm{sh}}}2
 -c_Q\frac{c(1-2q_{\mathrm{ref}})+q_{\mathrm{ref}}t}{(1-p)D_{\mathrm{ref}}}
\notag\\
&\quad-
 \alpha\beta_{\mathrm p}
 \left\langle
 \left\langle
 \ell'(\sqrt Q\,g)
 \left[
 \frac{g}{2\sqrt Q}+\sqrt Q\,c_Q\left(\frac{z_1}{\sqrt{1-q_{\mathrm{ref}}}}-\frac{c-t}{(1-q_{\mathrm{ref}})\sqrt A}z_2\right)
 \right]
 \right\rangle_3
 \right\rangle_+.
\label{eq:app-rs-shell_Q_saddle_expanded}
\end{align}
Equations \eqref{eq:app-rs-shell_p_saddle_expanded}--\eqref{eq:app-rs-shell_Q_saddle_expanded} are the interior full-feasible saddle equations for $(Q,p,t)$.

The feasible domain \eqref{eq:app-rs-shell_domain_def} contains the inequality
\begin{equation}
 A(Q,p,t;r)\ge0.
\label{eq:app-rs-shell_A_geq_zero_boundary}
\end{equation}
The equations \eqref{eq:app-rs-shell_p_saddle_expanded}--\eqref{eq:app-rs-shell_Q_saddle_expanded} assume $A>0$.  

On the boundary,
\begin{equation}
 A(Q,p,t;r)=0,
\label{eq:app-rs-shell_A_zero_boundary}
\end{equation}
derivatives containing $1/\sqrt A$ are not valid as unconstrained interior
derivatives.

A useful explicit parametrization of the $A=0$ boundary is, for $|c_d|\le1$ and boundary parameter $s_b$ chosen so that $p<1$,
\begin{equation}
 p=c_d^2+s_b^2(1-c_d^2),
\label{eq:app-rs-shell_boundary_p}
\end{equation}
\begin{equation}
 t=q_{\mathrm{ref}}c_d+s_b\sqrt{q_{\mathrm{ref}}(1-q_{\mathrm{ref}})(1-c_d^2)}.
\label{eq:app-rs-shell_boundary_t}
\end{equation}
On the $\zeta=0$ face of the full feasible map, $s_b=\chi$.  We screened this
face during branch selection, but the retained
production branch was interior throughout the 42-point grid, with
$A\geq 7.04\times10^{-7}$.

\subsection{Summary of the final shell result}
\label{app:rs-shell_summary}

The full shell local entropy density, after inserting the reference saddle data \eqref{eq:app-rs-reference_params_pass_to_shell}, is
\begin{equation}
 \Phi_{\mathrm{RS}}(r)
 =
 \extr_{(Q,p,t)\in\Dcal_{\rm shell}(r)}
 \left[
 \frac12\log Q-\frac{\beta_{\mathrm p}\lambda_{\mathrm{sh}}}2Q
 +G_{\mathrm S}^{\rm shell}(p,t;c_d(Q,r),q_{\mathrm{ref}})
 +\alpha G_{\mathrm E}^{\rm shell}(Q,p,t;r)
 \right].
\label{eq:app-rs-shell_summary_phi}
\end{equation}
The entropic term is
\begin{equation}
 G_{\mathrm S}^{\rm shell}(p,t;c_d,q_{\mathrm{ref}})
 =
 \frac{(1-c_d^2)(1-2q_{\mathrm{ref}})+q_{\mathrm{ref}}^2-2q_{\mathrm{ref}}c_dt+t^2}{2(1-p)(1-q_{\mathrm{ref}})^2}
 +\frac12\log(2\pi)+\frac12\log(1-p).
\label{eq:app-rs-shell_summary_GS}
\end{equation}
The energetic term is
\begin{align}
 G_{\mathrm E}^{\rm shell}(Q,p,t;r)
 &=
 \int Dz_0\,
 \frac{
 \int Dz_1\,
 \Theta\!\left(\sqrt{q_{\mathrm{ref}}}\,z_0+\sqrt{1-q_{\mathrm{ref}}}\,z_1\right)
 \int Dz_2\,
 \log\int Dz_3\,e^{-\beta_{\mathrm p}\ell(\sqrt Q\,g)}
 }{\Htail\!\left(-\sqrt{\frac{q_{\mathrm{ref}}}{1-q_{\mathrm{ref}}}}z_0\right)},
\label{eq:app-rs-shell_summary_GE}
\end{align}
where
\begin{equation}
 g
 =
 \frac{t}{\sqrt{q_{\mathrm{ref}}}}z_0+\frac{c_d-t}{\sqrt{1-q_{\mathrm{ref}}}}z_1+
 \sqrt{p-\frac{t^2}{q_{\mathrm{ref}}}-\frac{(c_d-t)^2}{1-q_{\mathrm{ref}}}}\,z_2+
 \sqrt{1-p}\,z_3.
\label{eq:app-rs-shell_summary_g}
\end{equation}
The feasible set is
\begin{align}
 \Dcal_{\rm shell}(r)
 =
 \bigg\{(Q,p,t):
 &\ Q>0,
 \quad
 |c_d(Q,r)|\le1,
 \quad
 p<1,
\notag\\
&\quad
 p-\frac{t^2}{q_{\mathrm{ref}}}-\frac{(c_d(Q,r)-t)^2}{1-q_{\mathrm{ref}}}\ge0
 \bigg\}.
\label{eq:app-rs-shell_summary_domain}
\end{align}
The distance cosine is
\begin{equation}
 c_d(Q,r)=\frac{Q+\lambda_{\mathrm{ref}}^{-1}-r^2}{2\sqrt{Q\lambda_{\mathrm{ref}}^{-1}}}.
\label{eq:app-rs-shell_summary_cd}
\end{equation}
Interior saddles satisfy \eqref{eq:app-rs-shell_p_saddle_expanded}, \eqref{eq:app-rs-shell_t_saddle_expanded}, and \eqref{eq:app-rs-shell_Q_saddle_expanded}.

\endgroup

\end{document}